\documentclass[twocolumn,dvipsnames]{aastex701}
\usepackage{subfigure}
\usepackage{rotating}
\usepackage{amsmath}
\usepackage{array}
\usepackage{xcolor}

\newcolumntype{P}[1]{>{\centering\arraybackslash}p{#1}}

\newcommand{\msun}{\hbox{${ M}_{\odot}$}}
\newcommand{\mstar}{\hbox{${ M}_{\star}$ }}

\begin{document}

\title{Resolved mass assembly and star formation in Milky Way Progenitors since $z = 5$ from JWST/CANUCS: From clumps and mergers to well-ordered disks}

\shorttitle{Resolved MWA mass assembly since $z = 5$}
\shortauthors{Tan et al.}

\author[0000-0002-3503-8899]{Vivian Yun Yan Tan}
\affiliation{Department of Physics and Astronomy, York University, 4700 Keele Street, Toronto, ON, M3J 1P3, Canada}
\email{tanvivia@yorku.ca}

\author[0000-0002-9330-9108]{Adam Muzzin}
\affiliation{Department of Physics and Astronomy, York University, 4700 Keele Street, Toronto, ON, M3J 1P3, Canada}
\email{muzzin@yorku.ca}

\author[0000-0001-8830-2166]{Ghassan T. E. Sarrouh}
\affiliation{Department of Physics and Astronomy, York University, 4700 Keele Street, Toronto, ON, M3J 1P3, Canada}
\email{gsarrouh@yorku.ca}

\author[0000-0002-0243-6575]{Jacqueline Antwi-Danso}
\affiliation{David A. Dunlap Department of Astronomy and Astrophysics, University of Toronto, 50 St. George Street, Toronto, Ontario, M5S 3H4, Canada}
\altaffiliation{Banting Postdoctoral Fellow}
\email{j.antwidanso@utoronto.ca}

\author[0000-0003-0780-9526]{Visal Sok}
\affiliation{Department of Physics and Astronomy, York University, 4700 Keele Street, Toronto, ON, M3J 1P3, Canada}
\email{sokvisal@yorku.ca}

\author[0009-0009-9848-3074]{Naadiyah Jagga}
\affiliation{Department of Physics and Astronomy, York University, 4700 Keele Street, Toronto, ON, M3J 1P3, Canada}
\email{jagga@yorku.ca}

\author[0009-0009-4388-898X]{Gregor Rihtar\v{s}i\v{c}}
\affiliation{Faculty of Mathematics and Physics, Jadranska ulica 19, SI-1000 Ljubljana, Slovenia}
\email{gregor.rihtarsic@fmf.uni-lj.si}

\author[0000-0002-6741-078X]{Westley Brown}
\affiliation{Department of Physics and Astronomy, York University, 4700 Keele Street, Toronto, ON, M3J 1P3, Canada}
\email{westleyb@yorku.ca}

\author[0000-0002-4542-921X]{Roberto Abraham}
\affiliation{David A. Dunlap Department of Astronomy and Astrophysics, University of Toronto, 50 St. George Street, Toronto, Ontario, M5S 3H4, Canada}
\email{abraham@astro.utoronto.ca}

\author[0000-0003-3983-5438]{Yoshihisa Asada}
\affiliation{Department of Astronomy and Physics and Institute for Computational Astrophysics, Saint Mary's University, 923 Robie Street, Halifax, Nova Scotia B3H 3C3, Canada}
\affiliation{Department of Astronomy, Kyoto University, Sakyo-ku, Kyoto 606-8502, Japan}
\email{asada@kusastro.kyoto-u.ac.jp}

\author[0000-0001-8325-1742]{Guillaume Desprez}
\affiliation{Kapteyn Astronomical Institute, University of Groningen, P.O. Box 800, 9700AV Groningen, The Netherlands}
\email{guillaume.desprez@protonmail.com}

\author[0000-0001-9298-3523]{Kartheik Iyer}
\affiliation{Columbia Astrophysics Laboratory, Columbia University, 550 West 120th Street, New York, NY 10027, USA}
\email{kgi2103@columbia.edu}

\author[0000-0003-3243-9969]{Nicholas S. Martis}
\affiliation{Faculty of Mathematics and Physics, Jadranska ulica 19, SI-1000 Ljubljana, Slovenia}
\email{nicholas.martis@fmf.uni-lj.si}

\author[0000-0001-8115-5845]{Rosa M. M\'erida}
\affiliation{Department of Astronomy and Physics and Institute for Computational Astrophysics, Saint Mary's University, 923 Robie Street, Halifax, Nova Scotia B3H 3C3, Canada}
\email{rosa.meridagonzalez@smu.ca}

\author[0000-0002-8530-9765]{Lamiya A. Mowla}
\affiliation{Whitin Observatory, Department of Physics and Astronomy, Wellesley College, 106 Central Street, Wellesley, MA 02481, USA}
\email{lmowla@wellesley.edu}

\author{Ga\"el Noirot}
\affiliation{Space Telescope Science Institute, 3700 San Martin Drive, Baltimore, Maryland 21218, USA}
\email{gnoirot@stsci.edu}

\author[0000-0002-8432-6870]{Kiyoaki Christopher Omori}
\affiliation{Department of Astronomy and Physics and Institute for Computational Astrophysics, Saint Mary's University, 923 Robie Street, Halifax, Nova Scotia B3H 3C3, Canada}
\email{kiyoaki.omori@smu.ca}

\author[0000-0002-7712-7857]{Marcin Sawicki}
\affiliation{Department of Astronomy and Physics and Institute for Computational Astrophysics, Saint Mary's University, 923 Robie Street, Halifax, Nova Scotia B3H 3C3, Canada}
\email{marcin.sawicki@smu.ca}

\author[0000-0002-9877-3491]{Roberta Tripodi}
\affiliation{Faculty of Mathematics and Physics, Jadranska ulica 19, SI-1000 Ljubljana, Slovenia}
\email{roberta.tripodi@fmf.uni-lj.si}

\author[0000-0002-4201-7367]{Chris J. Willott}
\affiliation{National Research Council of Canada, Herzberg Astronomy \& Astrophysics Research Centre, 5071 West Saanich Road, Victoria, BC, V9E 2E7, Canada}
\email{chris.willott@nrc.ca}




\begin{abstract}
We present a resolved study of 877 progenitors of Milky Way Analogs (MWAs) at $0.3<z<5$ selected with abundance matching in the ten fields of the Canadian NIRISS Unbiased Cluster Survey (CANUCS). Utilizing 18-21 bands of deep NIRCam, NIRISS, and HST photometry, we create resolved stellar mass maps and star formation rate maps via spectral energy distribution fitting with Dense Basis. We examine their resolved stellar mass and specific star formation rate (sSFR) profiles as a function of galactocentric radius, and find clear evidence for inside-out mass assembly. The total $\mstar$ of the inner 2 kpc regions of the progenitors remain roughly constant ($10^{9.3-9.4}\msun$) at $2<z<5$, while the total $\mstar$ of the regions beyond 2 kpc increases by 0.8 dex, from $10^{7.5}\msun$ to $10^{8.3}\msun$. Additionally, the sSFR of the outer regions increase with decreasing redshift, until $z\sim 2$. The median S\'ersic index of the MWA progenitors stays nearly constant at $n \sim 1$ at $2<z<5$, while the half-mass radii of their stellar mass profiles double. We perform additional morphological measurements on the stellar mass maps via the Gini-M20 plane and asymmetry parameters. They show that the rate of double-peak mergers and disturbances to galaxy structure also increase with redshift, with $\sim50\%$ of galaxies at $4<z<5$ classified as disturbed, and $\sim20\%$ classified as ongoing mergers. Overall, the early evolution of MWAs is revealed as chaotic, with significant mergers and high SFRs. Mass growth is primarily inside-out and galaxies become more disk-like after $z=3$.
\end{abstract}

\keywords{galaxies:evolution, galaxies:high-redshift, galaxies:interactions, milky way}


\section{Introduction} \label{sec:intro}

The present-day Milky Way is a large spiral galaxy with a total stellar mass of $\sim 5\times10^{10} \msun$ (see \citealt{Bland-Hawthorn:2016, Helmi:2020} for reviews). It contains a thin disk with ongoing star formation, a dynamically hotter thick disk with older stars of lower metallicity (e.g. \citealt{Bensby:2003, Fuhrmann:2011,Licquia:2015a,Kilic:2017}), and a central region with a bar and a pseudobulge (e.g. \citealt{Shen:2010, Wegg:2013, Wegg:2015, Portail:2017, Barbuy:2018}). Stellar streams in the Milky Way's halo reveal remnants of galaxies that merged with the Milky Way \citep{Helmi:2018, Belokurov:2018, Horta:2024}.

Compared to its peers of similar stellar mass, the Milky Way (MW) can either be typical, or somewhat abnormal. Work from \cite{Licquia:2016} shows that the MW lies on the Tully-Fisher relation for galaxies \citep{Tully:1977}. \cite{Fraser-McKelvie:2019} finds the MW lies within two standard deviations of the star-forming main sequence (see \citealt{Speagle:2014,Schreiber:2015,Santini:2017}), however its star formation rate is more consistent with a galaxy transitioning from star-forming to quiescent, green valley galaxy \citep{Mutch:2011}. Its disk length scale is also more compact physically compared to other spirals of the same stellar mass \citep{Bovy:2013, Licquia:2016}.
Depending on selection criteria, there exists dozens to potentially hundreds of Milky Way Analogs (MWAs) in the local universe, in terms of the galaxies' mass, morphology, and star formation rate (e.g.\citealt{Fraser-McKelvie:2019, Kormendy:2019, Boardman:2020a, Zhou:2023}). By comparing the physical characteristics of the MW to its analogs (e.g. \citealt{Bovy:2013, Licquia:2016, Boardman:2020b}), we gain a better understanding of the MW as it exists in its current state. However, the galaxies of the present-day universe had complex formation histories that we have only begun to glimpse from the current era of high-redshift observations.

Recent studies of gravitationally lensed arcs via multiband photometry and spectroscopy from JWST reveal very high redshift, reionization era galaxies that are progenitors of many of the galaxies that we find in the local universe. Among these are the Sunrise arc \citep{Vanzella:2023} at $z \sim 5.9$ with a total stellar mass of $10^6 - 10^7 \msun$,  the Firefly Sparkle \citep{Mowla:2024} at $z\sim8.3$ with a total stellar mass of $10^{5.3} - 10^{6.3}\msun$ , and the Cosmic Gems arc \citep{Adamo:2024, Bradley:2024} at $z\sim10.2$ with a total stellar mass of $10^{7.38} - 10^{7.75}\msun$ These correspond to a lookback time of $\sim$12.7 Gyr,  $\sim$13.1 Gyr and $\sim$13.3 Gyr respectively. These galaxies are gravitationally bound star clusters with extremely high star formation rates. Although these high-z protogalaxies are a far cry from the ``grand-design" spiral disks observed in the local universe, they represent the earliest stages of galaxy formation, and fall within the stellar mass range of possible progenitors of MWAs. 

A recent simulation study from \cite{Rusta:2024} has shown that the Firefly Sparkle's stellar mass and star formation history are consistent with that of a MW type galaxy. They provide evidence for the hierarchical assembly model of structure formation in the early universe \citep{White:1978}, supported by studies of the chemical and kinematic properties of stars in the galactic halo of the MW (e.g. \citealt{Tumlinson:2010, Salvadori:2010, Belokurov:2018, Belokurov:2022, Xiang:2022}). Hierarchical assembly also fits with the most widely accepted model of the MW's formation, the so-called ``inside-out" model, which posits the bulge formed first, with the disk and halo coalescing around it (see \citealt{vandenBosch:1998, Abadi:2003b, Munoz-Mateos:2007, Wang:2011, Licquia:2015b, Goddard:2017}).

With the advent of next generation high resolution hydrodynamical galaxy simulations such as TNG50 \citep{Pillepich:2019,Pillepich:2023}, FIRE \citep{Hopkins:2015, Onorbe:2015, Wetzel:2016, Hopkins:2018, Garrison-Kimmel:2018}, EAGLE \citep{Schaye:2015, Bignone:2019}, and NIHAO-UHD \citep{Buck:2020}, we are now able to observe the formation of MW and M31-sized galaxies within these simulations on the sub-kpc scale, along with more complex interactions, such as the survival of the galactic disk after major and minor mergers \citep{Sotillo-Ramos:2022}, or satellite galaxies that are gravitationally bound to MW and M31 analogs \citep{Engler:2023}. It is now possible to verify the results from these state-of-the-art hydrodynamical simulations by comparison with spatially resolved observations from the high redshift universe (i.e. \citealt{Gimenez-Arteaga:2023,Estrada-Carpenter:2024}). As we have an enormous wealth of data from JWST of many epochs from the early universe, we also have a unique opportunity to link the high-z and low-z universe via an \textit{observational} approach. Now in the JWST era, with deep, high-resolution, multiwavelength surveys, one can trace the evolution of disk galaxies further back in time. For example, \cite{LeConte:2024} used the CEERS survey to trace the evolution of resolved galaxy structures such as bars in disks from $1 < z <3$.


Previous studies which use observations to study MWA progenitors used rest-frame optical and NIR flux from HST as a proxy for stellar mass (\citealt{vanDokkum:2013, Patel:2013}, and \citealt{Papovich:2015}), and the maximum redshift reached was $z \sim 3$. \cite{Tan:2024} performed the first resolved mass assembly study but was limited to only $z\sim 2$. In all of these studies, the main disks have already formed by $z\sim 2.5$. A recent study using JWST photometry from \cite{Costantin:2023} has also found a MWA at $z\sim3$ with its disk already formed. Thus, the results from these studies favor neither inside-out nor outside-in models of formation, and the effect of galaxy interactions and mergers on mass assembly was not well explored. Galaxy mergers are known to play a role in the evolution of the MW (i.e. the last major merger of our galaxy, Gaia-Enceladus, see \citealt{Belokurov:2018, Helmi:2018}), but how common mergers are at different epochs of cosmic time, and how they may drive other processes in addition to mass assembly, such as starbursts or morphological disturbances are less well understood. 

In this paper, by utilizing the data from the Canadian NIRISS Unbiased Cluster Survey (CANUCS), we extend the study of spatially resolved galaxy mass assembly to $z \sim 5$. In doing so, we are able to provide a more holistic picture of MWA formation from its earliest steps to the present day. This paper is organized as follows: Section \ref{sec:data} provides a summary of the catalogs used and the selection criteria for MWA progenitors via abundance matching. Section \ref{sec:sed-fitting} describes the process of creating resolved stellar mass and star formation rate maps of our sample of MWA progenitors with SED-fitting using Dense Basis. Section \ref{sec:mass-growth} demonstrates the evidence for inside-out growth of MWA progenitors from $2 <z < 5$, and lockstep growth at $0 < z < 2$ from mass density and specific star formation rate densities. Section \ref{sec:morphology} examines the morphological evolution of MWA progenitors from chaotic, disturbed systems to rotationally supported disks.
For this work we assume a $\Lambda$CDM cosmology with $\Omega_\Lambda = 0.7$, $\Omega_m = 0.3$, and $H_0 = 70$ km/s/Mpc. We assume a \cite{Chabrier:2003} IMF.

\begin{table*}[t]
\caption{The catalog version and list of all NIRCam and HST filters used for each respective CANUCS field in this work. Note that the cluster fields (CLU) also contain additional NIRISS photometry in the bands \textit{F115WN, F150WN,} and \textit{F200WN}, which is included in the count of total number of filters.}\label{table1}
\hspace{2em}
\renewcommand{\arraystretch}{1.5}
\centering
\begin{tabular}{l P{2cm} P{4cm} P{4cm} P{2cm} }
\hline
Field &
 Catalog version &
  NIRCam &
  HST &
  Total number of filters \\ 
  \hline
MACS0417CLU &
  v1p0.3 &
  F090W, F115W, F150W, F200W, F277W, F356W, F410M, F444W &
  F435W, F606W, F814W, F105W, F125W, F140W, F160W &
  18\\ 
Abell370CLU &
  v1p1.1 &
  F090W, F115W, F150W, F200W, F277W, F356W, F410M, F444W &
 F435W, F606W, F814W, F105W, F110W, F125W,  F140W, F160W &
  19\\ 
MACS0416CLU &
  v1p1.1 &
 F090W, F115W,F150W, F200W, F277W, F356W, F410M, F444W &
  F435W, F606W, F814W,F105W, F110W, F125W,F140W, F160W &
  19\\ 
MACS1423CLU &
  v1p0.1 &
  F090W, F115W,F150W, F200W, F277W, F356W, F410M, F444W &
  F435W, F606W, F814W,F105W, F110W, F125W,F140W, F160W &
  19\\ 
MACS1149CLU &
  v1p2.1 &
  F090W, F115W, F150W, F200W, F277W, F356W, F410M, F444W &
  F435W, F606W, F814W,F105W, F110W, F125W,F140W, F160W &
  19 \\ 
  \hline
MACS0417NCF &
  v1p0.2 &
  F090W, F115W, F140M, F150W, F160M, F182M, F210M, F250M, F277W, F300M, F335M, F360M, F410M, F444W &
  F438WU, F606WU&
  16\\ 
Abell370NCF &
  v2p0.1 &
  F090W, F115W, F140M,F150W, F160M, F182M,F210M, F250M, F277W, F300M, F335M, F360M, F410M, F444W &
  F435W, F606W, F814W, F105W, F125W, F140W, F160W &
  21\\ 
MACS0416NCF &
  v1p1.1 &
  F090W, F115W, F140M, F150W, F160M, F182M, F210M, F250M, F277W, F300M, F335M, F360M, F410M, F444W &
  F435W, F606W, F814W, F105W, F125W, F140W, F160W &
  21\\ 
MACS1423NCF &
  v1p0.1 &
  F090W, F115W, F140M, F150W, F160M, F182M, F210M, F250M, F277W, F300M, F335M, F360M, F410M, F444W &
  F438WU, F606WU, F125W, F160W &
  18\\ 
MACS1149NCF &
  v1p2.1 &
  F090W, F115W, F140M, F150W, F160M, F182M, F210M, F250M, F277W, F300M,F335M, F360M, F410M, F444W &
  F435W, F606W, F814W, F105W, F125W, F140W, F160W &
  19\\ 
  \hline
\end{tabular}
\end{table*}

\section{Data \& Sample Selection} \label{sec:data}
\subsection{Survey Design}
We use the observations from the Canadian NIRISS Unbiased Cluster Survey (CANUCS, see \citealt{Willott:2022}, and \citealt{Sarrouh:2025} for its first data release) for its excellent depth at high-$z$ and extensive photmetric sampling. There are two fields for each cluster in CANUCS: CLU denotes the cluster field, and NCF denotes the ``NIRCam Flanking" field. For the CLU fields, the JWST NIRCam filters used are F090W, F115W, F150W, F200W, F277W, F256W, F410M, and F444W. For the NCF fields, the same wide band filters are also used, \textit{except for} F200W. Additionally, NCF fields are imaged by the medium band NIRCam filters F140M, F162M, F182M, F210M, F250M, F300M, F335M, F360M, and F410M. All imaging information for each field is described in Table \ref{table1}. Details of PSF-matching and catalog creation for CANUCS are described in \cite{Willott:2024} and \cite{Sarrouh:2024}. Source detection and aperture photometry was done using the Photutils package from astropy \citep{Bradley:2016, Bradley:2023} as described in \cite{Asada:2024}. See \S \ref{sec:vorbin} for typical aperture size used for each object in this work, and additional details on resolved photometry.

For this work, we only make use of the photometric data. Additional imaging from HST/ACS and HST/WFC3 are provided in the filters F435W, F606W, F814W, F105W, F110W, F125W, and F160W for fields that overlap with the Hubble Frontier Fields survey \citep{Lotz:2017}, i.e. both CLU and NCF for Abell 370, MACS 0416, MACS 1149, as well as MACS0417CLU and MACS1423CLU. However, for MACS0417NCF and MACS1423NCF, there is no ACS imaging, so imaging in the bluest bands is provided by WFC3/UVIS in the bands F438WU and F606WU. For exact usage of bands, and catalog version, see Table \ref{table1}.

\subsection{Data Reduction}
Image processing and catalog construction will be discussed in detail in Sarrouh and Asada et al. (in prep.), which we briefly describe here. For more information on the data reduction process for CANUCS, we refer the reader to \cite{Noirot:2023}. The full CANUCS dataset includes NIRCam photometry, with spectroscopy from NIRISS and NIRSpec of five strong lensing clusters Abell 370, MACS J0416.1-2403 (hereafter MACS 0416), MACS J0417.5-1154 (hereafter MACS 0417), MACS J1149.5+2223 (hereafter MACS 1149), and MACS J1423.8+2404 (hereafter MACS 1423). 

In general, the NCF fields are deeper than the cluster fields by 0.3 to 0.6 magnitudes for the same band. The shallowest median depth ($3\sigma$ in 0.3'' apertures) is 27.2 mag in the HST bands of F125W and F140W for MACS 0417 CLU, and the deepest median depth is the JWST NIRCam band of F277W in MACS 1149 NCF, at 30.8 mag ($3\sigma$ in 0.3'', see \citealt{Desprez:2024, Asada:2024}, and \citealt{Willott:2024} for more information). The cluster fields are affected by gravitational lensing, albeit at different strengths for different clusters. Any object in the catalogs with a best-fit model lensing magnification of $\mu > 2.5$ are excluded from the selection. We also correct for the magnification in all stellar mass and SFR calculations, whether resolved or integrated. The process of accounting for the effects of lensing in radial profiles is outlined in \S \ref{sec:density-profiles}. The Abell 370 cluster field is the field impacted the most by lensing overall, and thus our sample is the smallest in that field. The lensing models for Abell 370 are provided in \cite{Gledhill:2024}, for MACS0416 in \cite{Rihtarsic:2025}, MACS1149 in {Rihtar{\v{s}}i{\v{c}}} et al. (in prep), and MACS 0417 and MACS 1423 in Desprez et al. (in prep).

The full CANUCS catalog is complete to AB magnitude $27$ in the reddest filter F444W at all redshifts (including this work's redshift limit of $z\sim5$). However, the signal-to-noise cutoff necessary for successful Voronoi binning is $S/N > 30$ in F444W. This signal-to-noise threshold is set to ensure that the individual spatial bins have $S/N > 5$. This means that out of all possible MWAs in the CANUCS fields, only the brightest at high-$z$ can be retained in our sample. More information on Voronoi binning is provided in \S \ref{sec:vorbin}. Photometric redshifts in the CANUCS catalog \citep{Asada:2024,Asada:2024b} were obtained with an implementation of EAZY in Python known as \texttt{eazy-py} \citep{Brammer:2008}. During fitting, a systematic flux error of $5\%$ was added in quadrature to the nominal photometric uncertainty. 

\subsection{Sample Selection via Abundance Matching}\label{sec:abundance-matching}

\begin{figure*}
   \centering
    \begin{subfigure}
    \centering
        \includegraphics[width=0.6\textwidth]{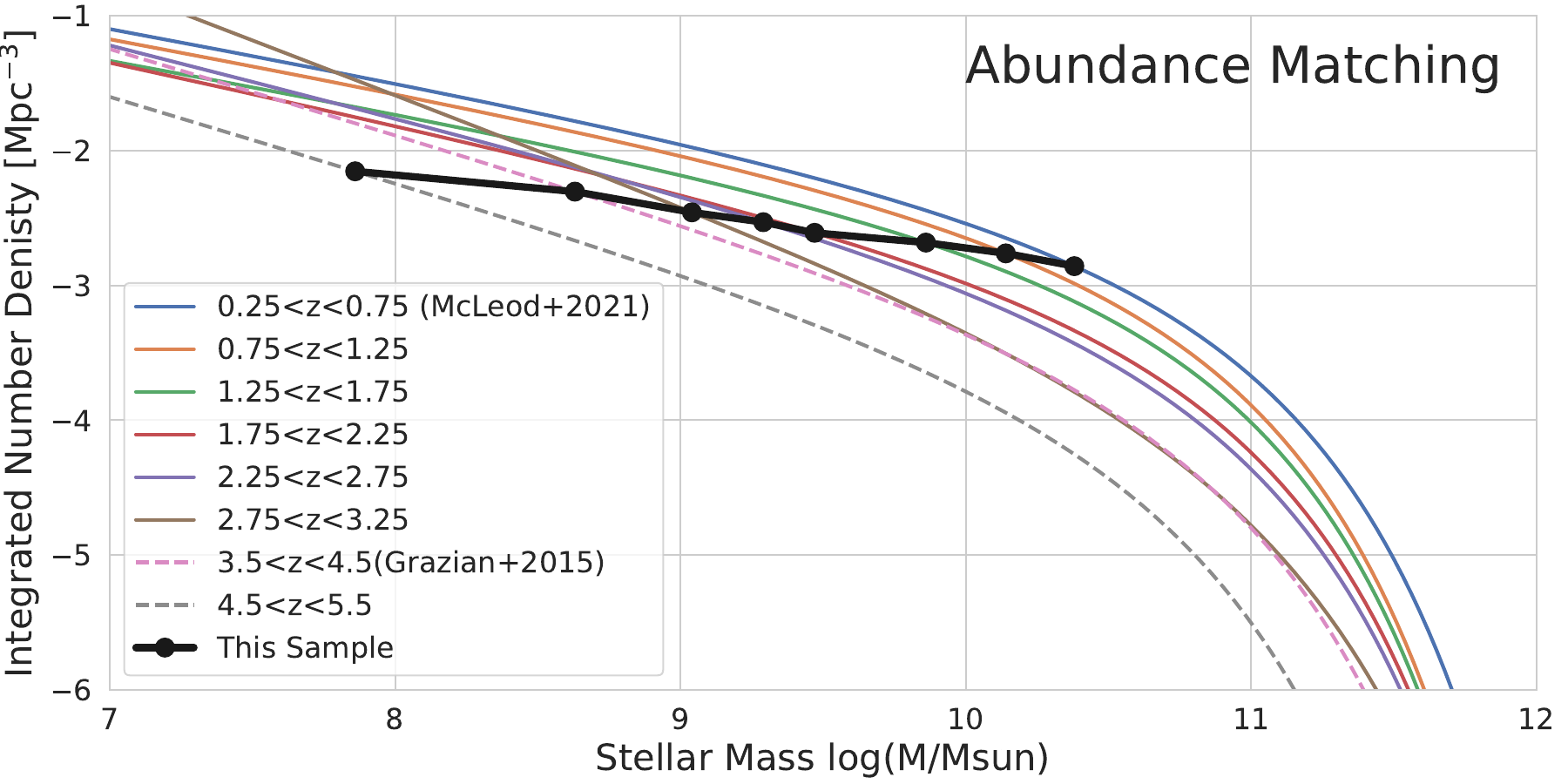}
    \end{subfigure}
    \begin{subfigure}
    \centering
        \includegraphics[width=0.8\textwidth]{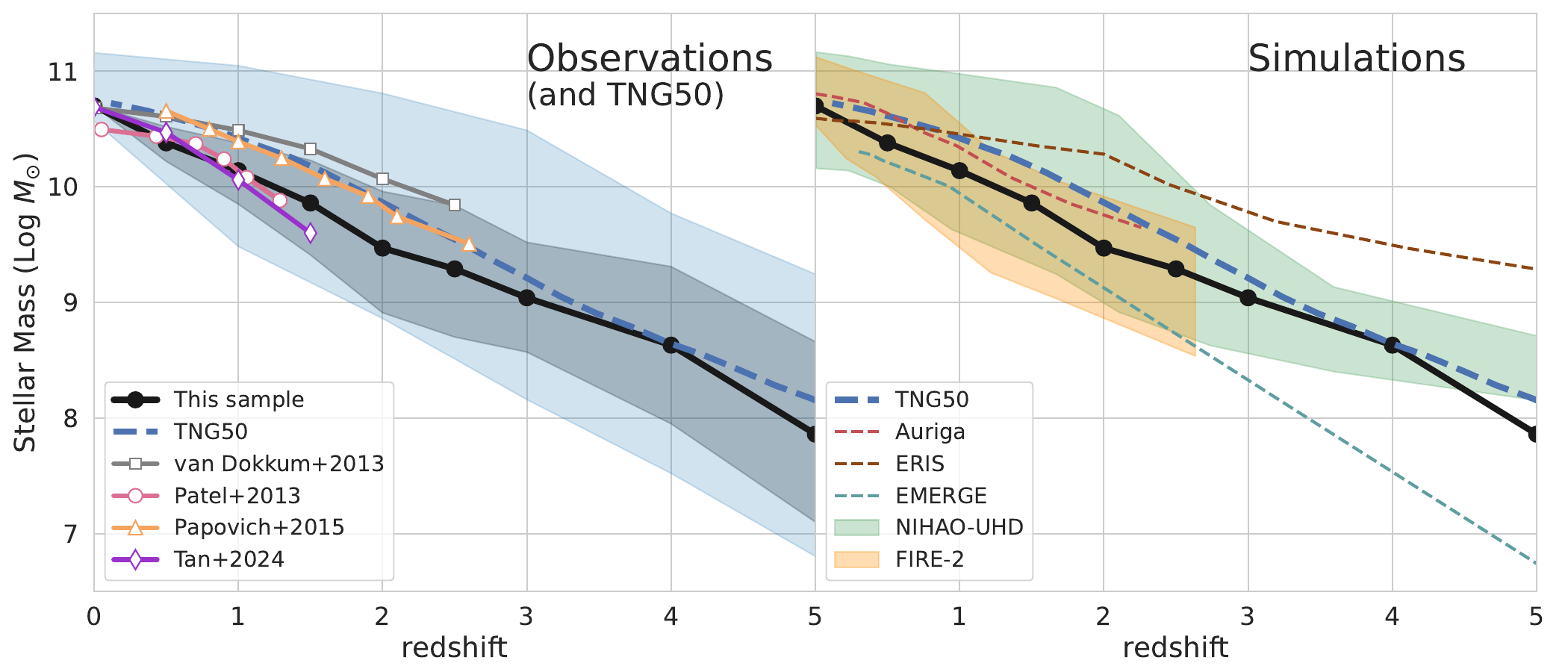}
    \end{subfigure}
    \caption{\emph{Top Panel:} Each black point represents the intersection between the cumulative number density predicted by abundance matching \citep{Behroozi:2013a} for MWA progenitors, and the stellar mass corresponding to that number density given by the stellar mass function (SMF) at that redshift interval. The starting point is a number density of 0.0011 Mpc$^{-3}$ at $z = 0$, corresponding to $\mstar = 10^{10.7}\msun$. Solid SMFs are from \cite{McLeod:2021}, dashed SMFs are from \cite{Grazian:2015}. 
    \emph{Bottom left:} Median stellar mass versus redshift for MWA progenitors plotted in black points. The shaded grey region is the $1\sigma$ deviation (16th to 84th percentile) of our sample, which comes from the $1\sigma$ range in number density from the abundance matching code. We compare our median stellar mass to the median stellar mass from TNG50 \citep{Sotillo-Ramos:2022, Pillepich:2023} represented with the blue dashed line. The blue shaded region is the TNG50 sample's 10th to 90th percentile. Previous \textit{observation-based} studies \citep{vanDokkum:2013, Patel:2013, Papovich:2015, Tan:2024} are also plotted in grey with white points. 
    \emph{Right panel:} Comparison of this sample's stellar mass evolution for MW progenitors versus predictions from various simulations, such as TNG50, again in blue, Auriga \citep{Grand:2017}, ERIS \citep{Guedes:2011}, EMERGE \citep{Moster:2018}, FIRE-2 \citep{Garrison-Kimmel:2018}, and NIHAO-UHD \citep{Buck:2020}.} \label{fig:numdensity}
\end{figure*}

Selecting MWA progenitors may be done via abundance matching, or using properties from simulations (such as stellar masses, halo masses, star-formation history, etc.) as a standard. We choose to use abundance matching, but we demonstrate that both methods are consistent in Figure \ref{fig:numdensity}. To estimate the range of stellar masses of progenitors of MW-mass galaxies at higher redshift, we assume an evolving co-moving number density with redshift, as determined in \cite{Behroozi:2013a}, with a present day number density of $\log(n/\text{Mpc}^{3}) =-2.95$ for MWAs. This number density corresponds to a MW stellar mass of $10^{10.7} \msun$ at $z=0$. The code calculates a past median galaxy number density at $z_2$, given an initial number density at $z_1$, via peak halo mass functions. Since the merger rate per unit halo per unit $\Delta z$ is roughly constant, the evolution of cumulative number density of progenitors of any given galaxy is a power law, with the change described by $0.16\times\Delta z$ dex. \cite{Behroozi:2013a} uses peak halo mass functions for which the resultant median number densities are less affected by the scatter in stellar mass and luminosity. However, this scatter does affect the $1\sigma$ errors in cumulative number density. The $1\sigma$ or 68 percentile range grows with increasing redshift. The $1\sigma$ values grow by $\sim 0.16$ dex per 0.5 change in redshift.

\cite{Behroozi:2013a} does not assign stellar masses to the number densities, but rather determines the number density of the median \textit{halo} mass. Thus, we obtain the stellar mass ranges of the MWA progenitors using the same number densities from the stellar mass functions (SMFs) from two studies, \cite{Grazian:2015} and \cite{McLeod:2021}. Both studies used a combination of ground-based and HST data: \cite{Grazian:2015} CANDELS-UDS, GOODS-South, and HUDF09 and HUDF12. The surveys used by \cite{McLeod:2021} are a combination of UKIDSS Ultra Deep Survey, UltraVISTA, CFHTLS-D1, and all five CANDELS fields. The result of this matching of cumulative number densities for MWAs to median stellar masses is shown in the top panel of Figure \ref{fig:numdensity}. A similar analysis is done in \cite{Tan:2024} for the Hubble Frontier Fields but with SMFs from \cite{Muzzin:2013b}. See \cite{Mowla:2024} for the extension of this abundance matching in CANUCS to $5 < z < 9$. 

\begin{figure*}
\centering
   \includegraphics[width=0.75\textwidth]{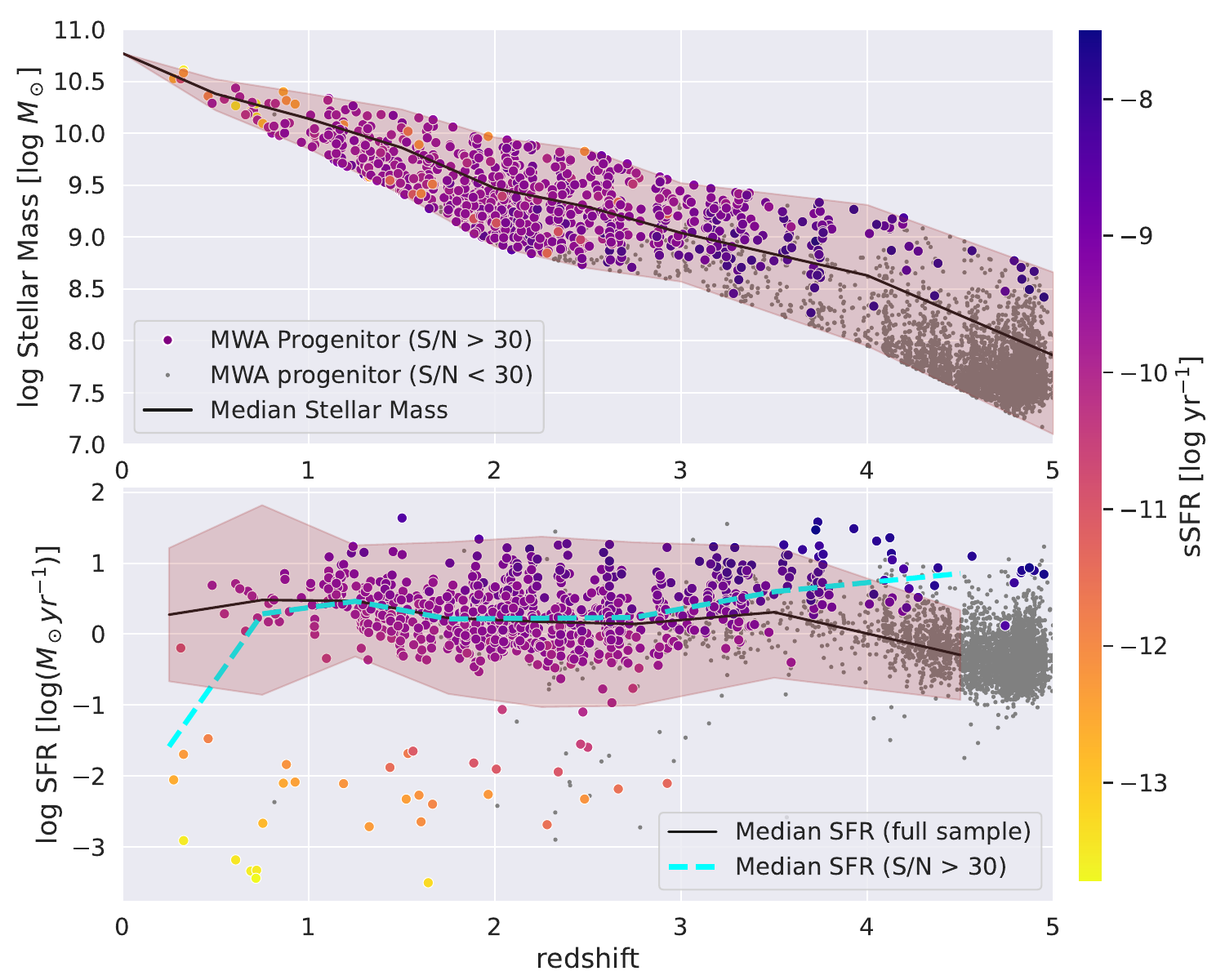}
    \caption{\emph{Top panel: }Integrated stellar masses (obtained with Dense Basis) versus photometric redshift of MWA progenitors in the CANUCS fields selected by abundance matching and SMFs. The solid black line is the rolling median of the stellar mass for each redshift range, and corresponds to the solid black line in Figure \ref{fig:numdensity}, while the shaded red region represents the $1\sigma$ scatter, same as the shaded grey region in Figure \ref{fig:numdensity}. Some of the stellar masses are scattered below the $1\sigma$ region due to lensing magnification corrections. Large colored points are galaxies that have $S/N > 30$, and their color represents the specific SFR. The small grey points have too little signal-to-noise for resolved photometry, and are excluded from our final sample. 
    \emph{Bottom panel:} Catalog log SFR vs Redshift, same selection criteria as above. The shaded red region represents the $1\sigma$ scatter of SFR. The color of each point indicates specific SFR. Note that at higher redshifts, galaxies with $S/N > 30$ skew towards the higher mass and higher SFR end of the full distribution of possible MWAs in that redshift range. Note that quiescent galaxies are not used for the main analysis.}
    \label{fig:mass_fullsamp}
\end{figure*}

\begin{figure*}
    \centering
    \includegraphics[width=\textwidth]{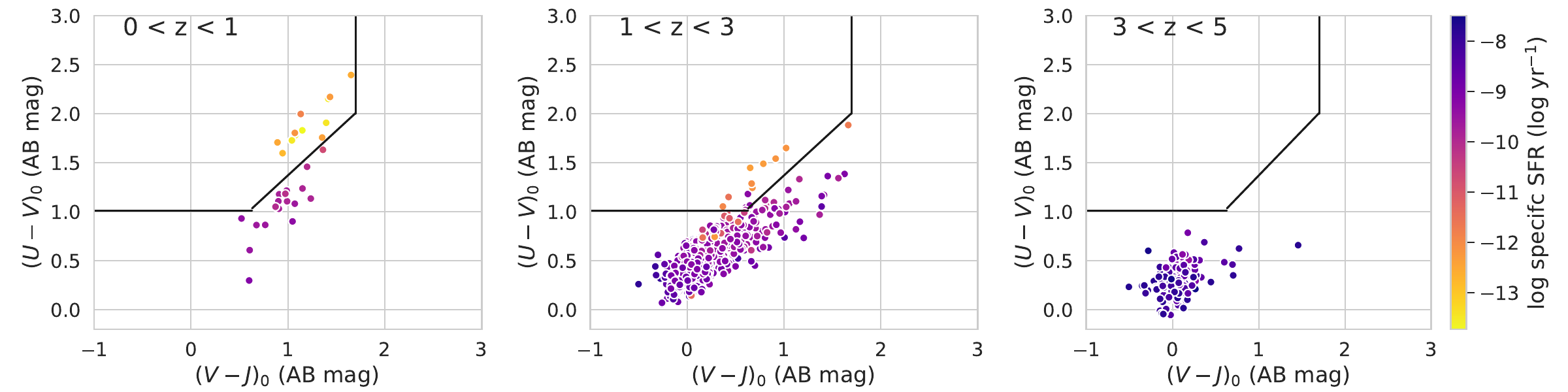}
    \caption{The distribution of UVJ colors for the MWA progenitors. Since the quiescent fraction of MWAs is small ($\sim0.03\%$), The UVJ boundaries are defined with a bimodal distribution of 9615 galaxies in CANUCS at $0 < z < 5$, with a 25\% quiescent fraction (See Appendix \ref{app:uvj}). The color calibration is adapted from \cite{Antwi-Danso:2023}. The color of the points represent the sSFR of the MWA progenitor sample.}\label{fig:uvj}
\end{figure*}

We take the median cumulative number densities at each $\Delta z$, to find the stellar mass associated with that number density from the corresponding SMF. In addition, the $1\sigma$ errors on the given number density for each redshift are used to determine the maximum and minimum boundaries on the stellar mass of the progenitors at each epoch. Galaxies with redshifts within $\pm0.05$ of the cluster's redshift are removed from our sample to prevent contamination from cluster galaxies. The median stellar mass at each redshift range is plotted in the bottom panels of Figure \ref{fig:numdensity}. Also shown in Figure \ref{fig:numdensity} are comparisons of this sample's median stellar mass and $1\sigma$ stellar mass range to past works that have studied progenitors of MWAs. The left panel focuses primarily on studies that used observational data  \citep{vanDokkum:2013,Patel:2013, Papovich:2015, Tan:2024}, while the right panel primarily focuses on studies with simulations \citep{Guedes:2011, Grand:2017, Moster:2018, Garrison-Kimmel:2018, Buck:2020, Pillepich:2023}. The TNG50 results from \cite{Pillepich:2023} are plotted in both panels for comparison. Our stellar mass range for each redshift bin for MWA progenitors agree with the literature, for both observations and simulations. 

In Figure \ref{fig:mass_fullsamp}, we apply the selection criteria as described above to the CANUCS catalogs, and plot the catalog stellar mass versus redshift in the top panel, and the catalog instantaneous SFR versus redshift in the bottom panel (unlike the photometric catalogs, the mass catalogs are obtained with Dense Basis, see \S \ref{sec:sed-fitting} for more details). Note the CANUCS catalog stellar mass and instantaneous SFR (henceforth simply called `SFR') are integrated over the whole galaxy. The color of the points show their integrated sSFR (SFR/stellar mass). The vast majority of galaxies have log(sSFR) $> -11$. Due to the relationship between stellar mass and redshift, lower-$z$ galaxies tend to be more massive but have lower sSFR than higher-$z$ galaxies with the same total SFR.

There are 5307 potential MWA progenitor galaxies at $0 < z < 5$ that fall within the $1\sigma$ boundaries as defined by applying the abundance matching code to the given SMFs. In this paper, we SED-fit galaxies in 2D, and an integrated $S/N > 30$ is required to get meaningful, spatially-resolved SED fits. Therefore, the colored dots represent the subsample of these potential MWA progenitors that have a $S/N > 30$ in their integrated F444W flux. This ensures that each spatial bin will have at least $S/N = 5$ at higher redshifts, and ensures there are at least 50 spatial bins per galaxy. We remove the small grey points from our sample, which have integrated $S/N < 30$. After additionally removing 23 more galaxies on the basis of missing photometry in all JWST bands bluer than F250M (which would make the SED fit untrustworthy for resolved modeling), our full sample of MWA progenitors for 2D SED modeling is a total of 877 galaxies.

A result of this signal-to-noise cutoff is that the higher mass galaxies at higher-$z$ are selected over lower mass galaxies. As shown in Figure \ref{fig:mass_fullsamp}, our sample is complete to $z=3$. Above $z=3$, we become increasingly biased towards more massive progenitors. In addition, the median SFR, as plotted in the bottom panel of Figure \ref{fig:mass_fullsamp}, no longer peaks at $z \sim 1.5$, but continues to rise to $z\sim 5$. Ultimately, the CANUCS dataset is among the deepest JWST surveys to date, and we continue the analysis to $z=5$. Note that at $z > 3$, there is an increasing degree of progenitor bias in the results. 
In order to be complete at $z=5$ would require an order of magnitude more depth, which can only be obtained with two orders of magnitude greater exposure time. Given that the CANUCS integration times are $\sim 1-3$ hours, this is unlikely to be obtained within the mission (see \citealt{Willott:2022}). Strong lensing samples, such as the Firefly and Cosmic Gems, are necessary for even higher redshifts, though they are limited in number for each lensing cluster. Thus, our sample shows a ``first look" at evolution of progenitors up to $z=5$.

\subsection{Quiescent fraction in the MWA sample}\label{sec:q-frac-uvj}

A limitation of abundance matching is that not all selected progenitors will evolve into an MWA at $z=0$. Since the MW is still star-forming (albeit with a much lower star formation rate that places it within the Green Valley, see \citealt{Mutch:2011}), we only select actively star-forming galaxies within our MWA progenitor sample for an accurate picture of the mass assembly relevant to the MW itself. Even though any actively star-forming higher-$z$ may become quiescent before reaching $z=0$, quiescent galaxies can certainly be discounted as MWAs. Therefore, we use rest-frame UVJ colors in combination with the sSFR of this sample to find and remove quiescent galaxies.

In Figure \ref{fig:uvj}, we plot the rest-frame colors of our progenitor sample. In order to determine the proper color-magnitude cuts, we utilize the code presented in \cite{Antwi-Danso:2023}. The code searches for a bimodality in color space by calculating the distance of each galaxy from the UVJ diagonal line. We refit the UVJ slope for the color calibration with the redshift range changed to $0<z<5$. For more information on UVJ color calibration, see Appendix \ref{app:uvj}. The UVJ boundaries are:

\begin{align}
& &&(V-J) < 1.7; \ (U-V) > 1.01;  \\
& &&(U-V) > (V-J) \times 0.91 + 0.63 . 
\end{align}

We lowered the $(U-V)$ boundary by 0.23 to accommodate a simple stellar population at $z \sim 6$ with an age of $\sim$800 Myr. This corresponds to the age of a typical post-starburst galaxy at $0.5 < z < 2.5$ \citep{Belli:2019, Wild:2020}. The divisions presented above are also shown in Figure \ref{fig:uvj}, in each panel, as solid black lines. The color of the points represent the galaxy's sSFR. 

After the signal-to-noise cut, and separating the sample into star-forming and quiescent, there are 885 star-forming galaxies, and 24 quiescent galaxies. 12 out of the 24 quiescent galaxies are at $z < 1$.  Additionally, there are only star-forming galaxies at $3 < z < 5$.

\section{Spatially resolved SED-fitting on JWST photometry} \label{sec:resolved-phot}

In this section, we explain the methods of creating spatially resolved maps of stellar mass and instantaneous SFR distributions for our sample of MWA progenitor galaxies. All resolved photometry is applied to the science images PSF-matched to F444W provided by CANUCS. The size of the PSF FWHM for the F444W filter is $0.15$'' or 2.3 pixels.

\begin{turnpage}
  \begin{figure*}
  \centering      
  \includegraphics[scale=0.5]{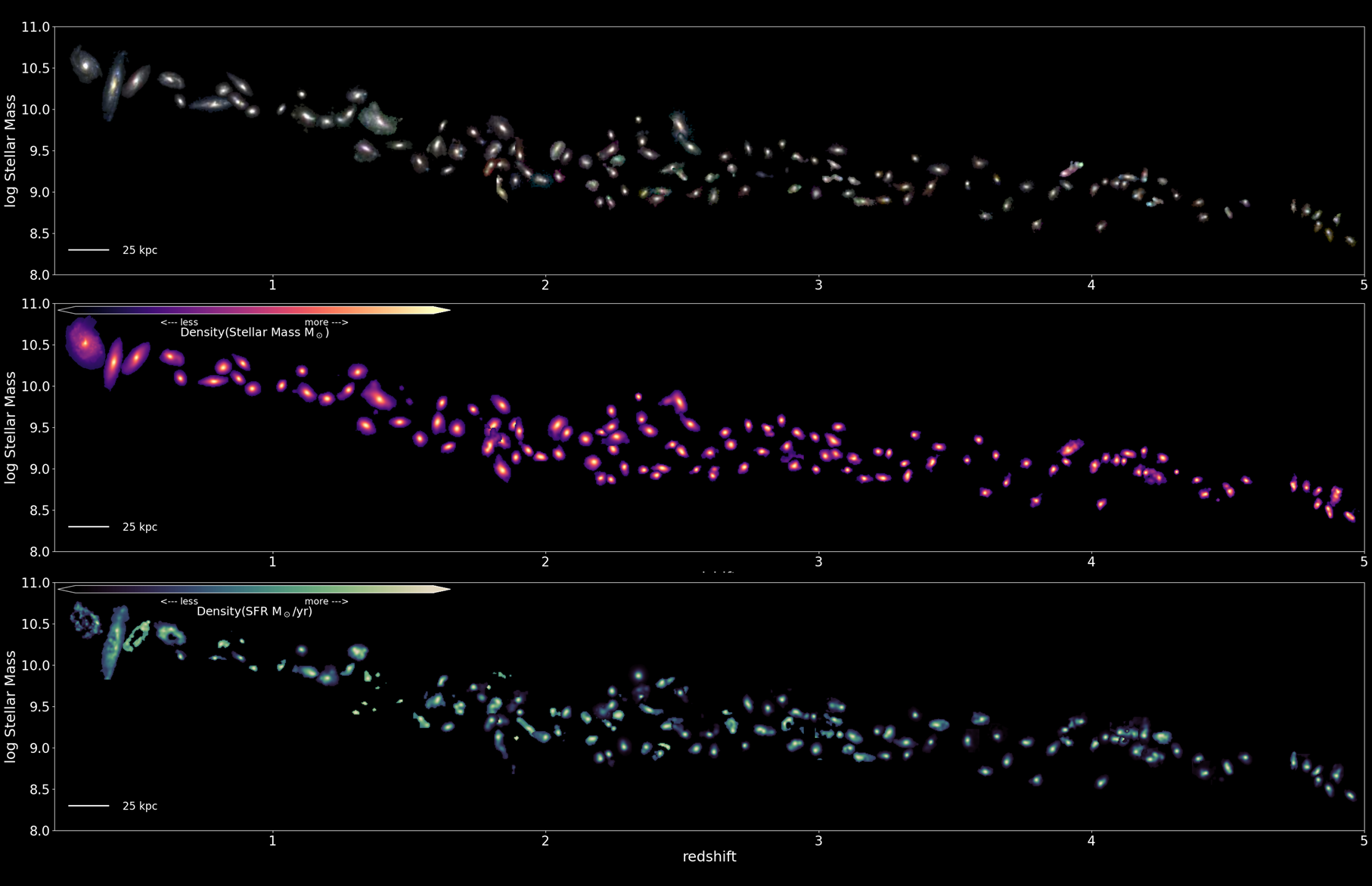}
  \caption{Color images (top panel), resolved stellar mass maps (middle panel), and resolved SFR maps (bottom panel) of a randomly selected subsample of 120 MWA progenitors. The plots are inset into a total stellar mass versus redshift plot to demonstrate the evolutionary trend of the progenitors. Each image has the same physical scale in kpc, displayed in the bottom-left corner. Certain galaxy images are offset by no more than $\pm0.25z$ or $\pm0.25$ log $\msun$ to lessen overlap. Note that at higher redshifts, the morphologies shown are not fully representative of the progenitor population.}       
  \label{fig:insetplots-maps} 
  \end{figure*}
\end{turnpage}

\begin{figure*}
\begin{tabular}{ll}
\centering
    \includegraphics[width=0.45\textwidth]{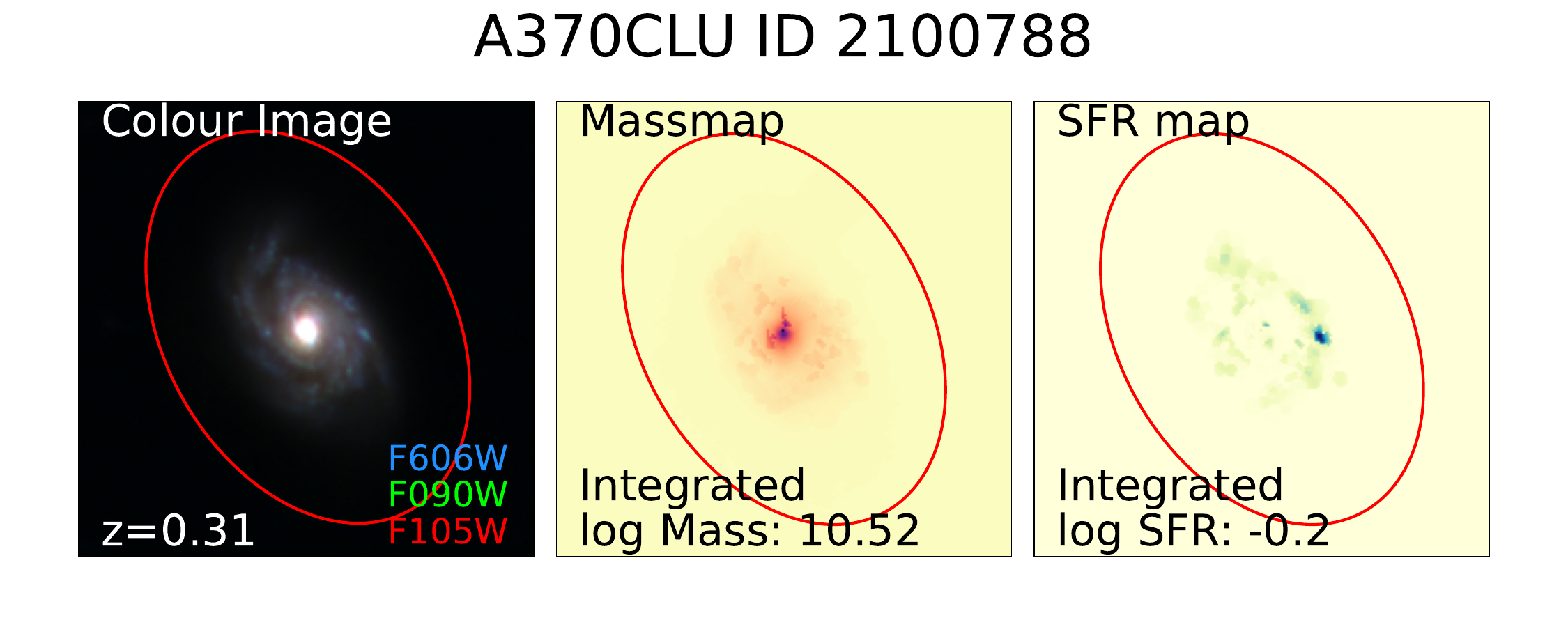}& 
    \includegraphics[width=0.45\textwidth]{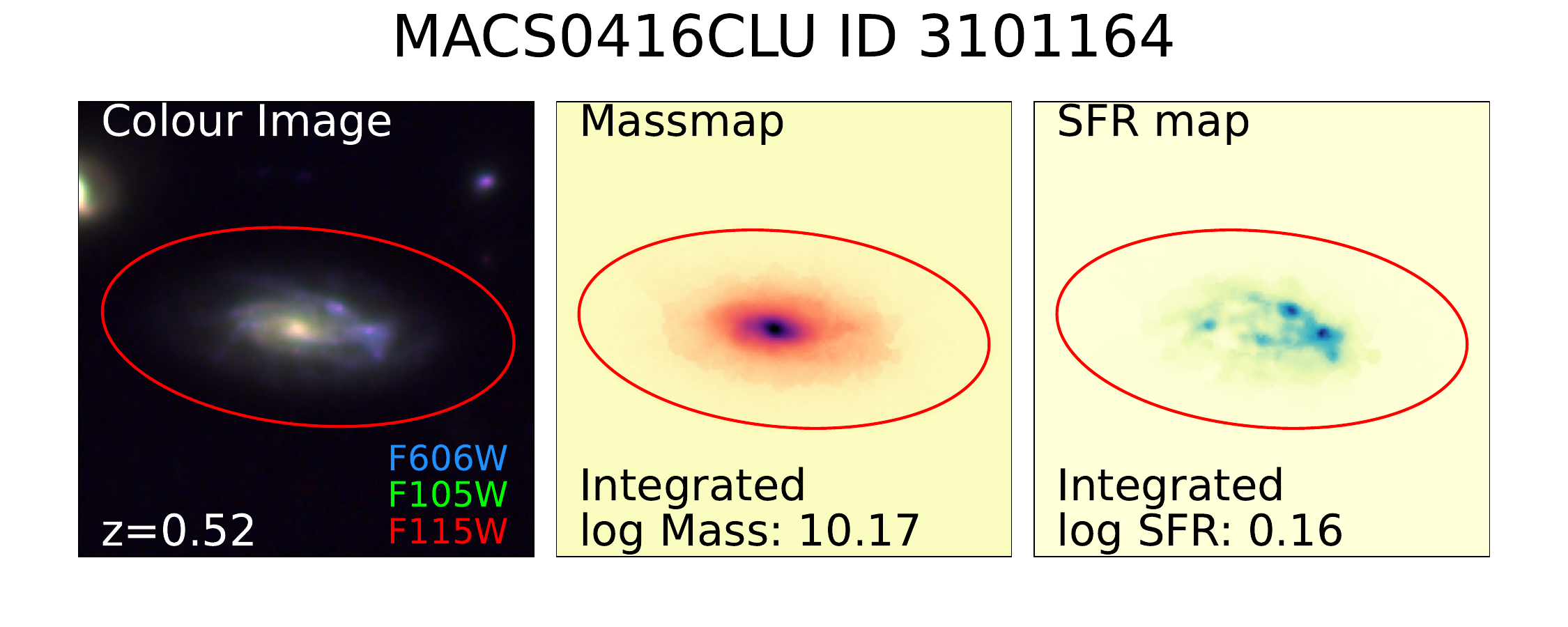}\\ 
    \includegraphics[width=0.45\textwidth]{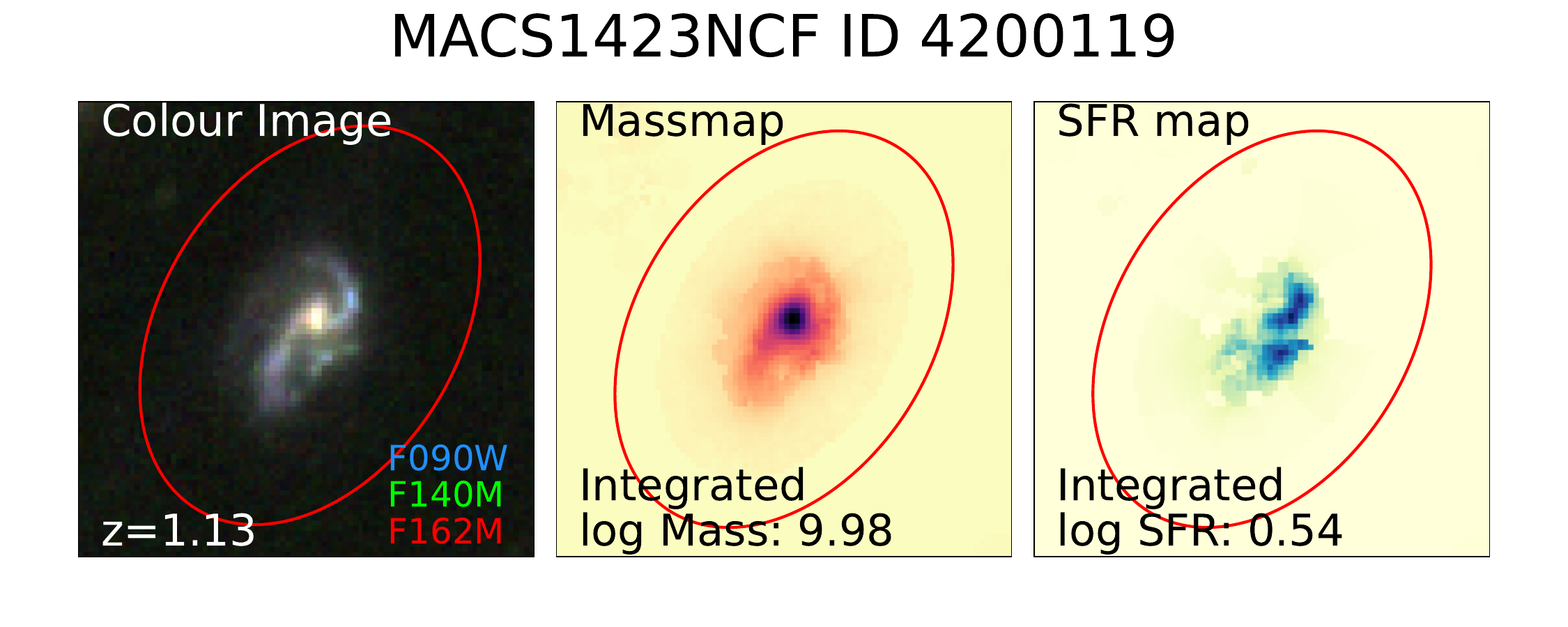} & 
    \includegraphics[width=0.45\textwidth]{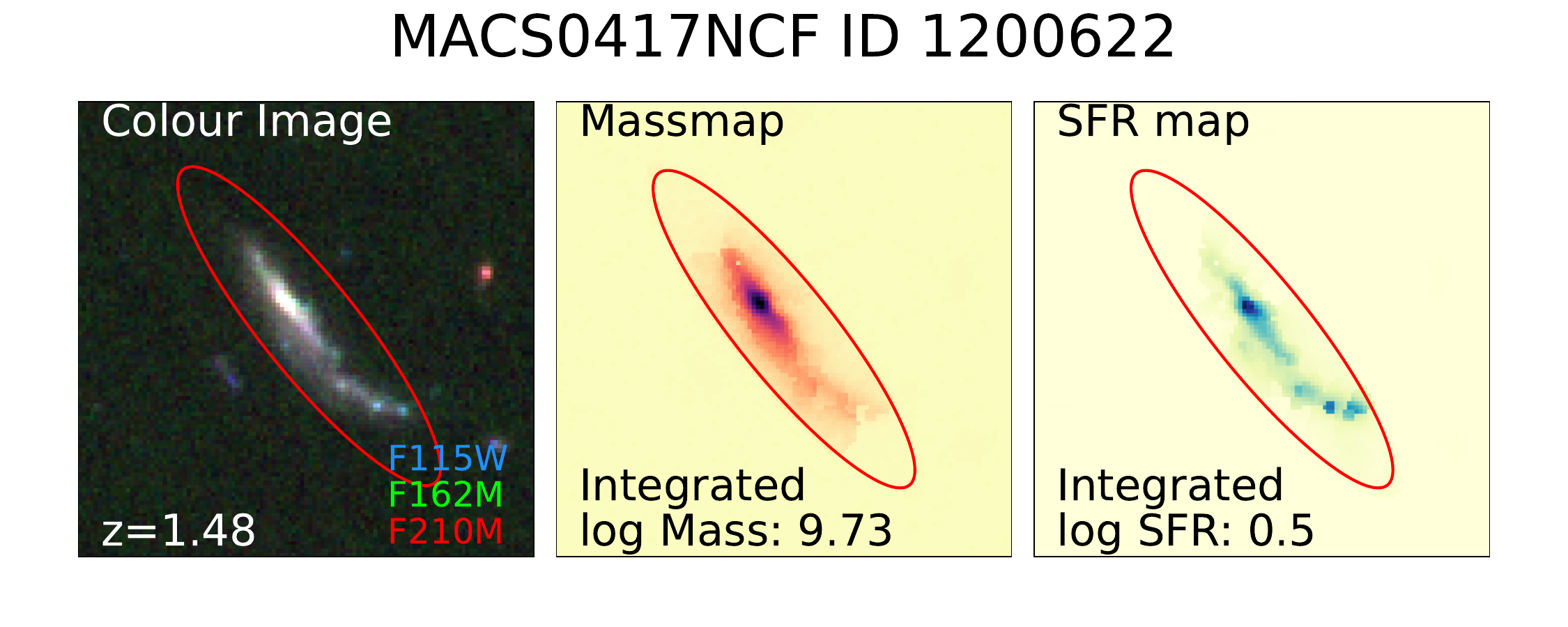} \\
    \includegraphics[width=0.45\textwidth]{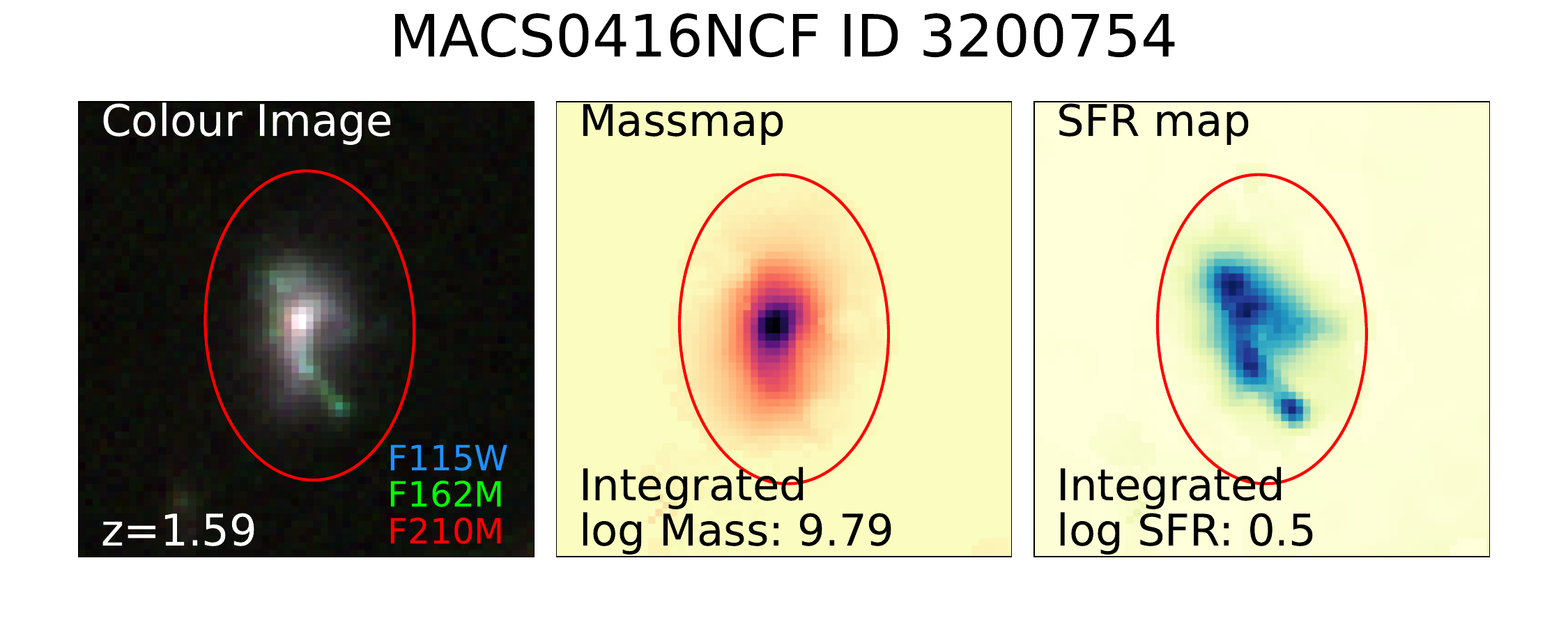} & 
    \includegraphics[width=0.45\textwidth]{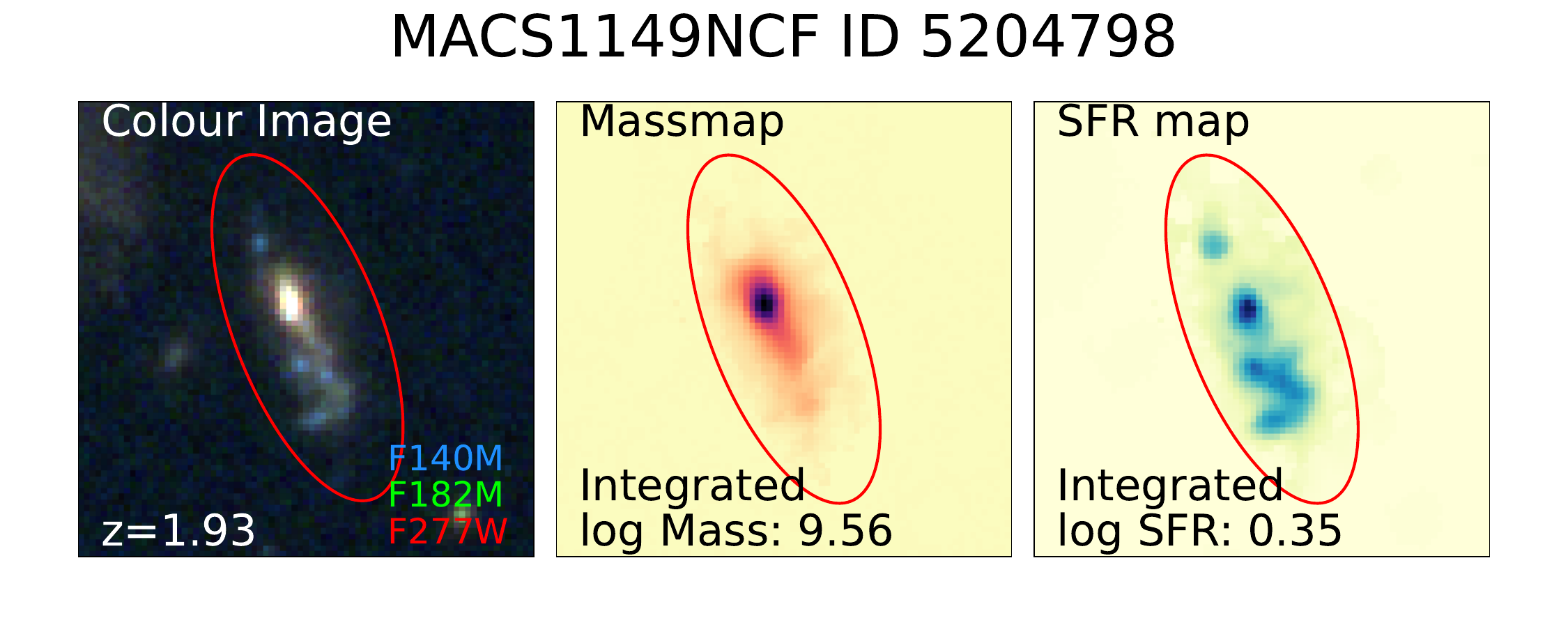} \\
    \includegraphics[width=0.45\textwidth]{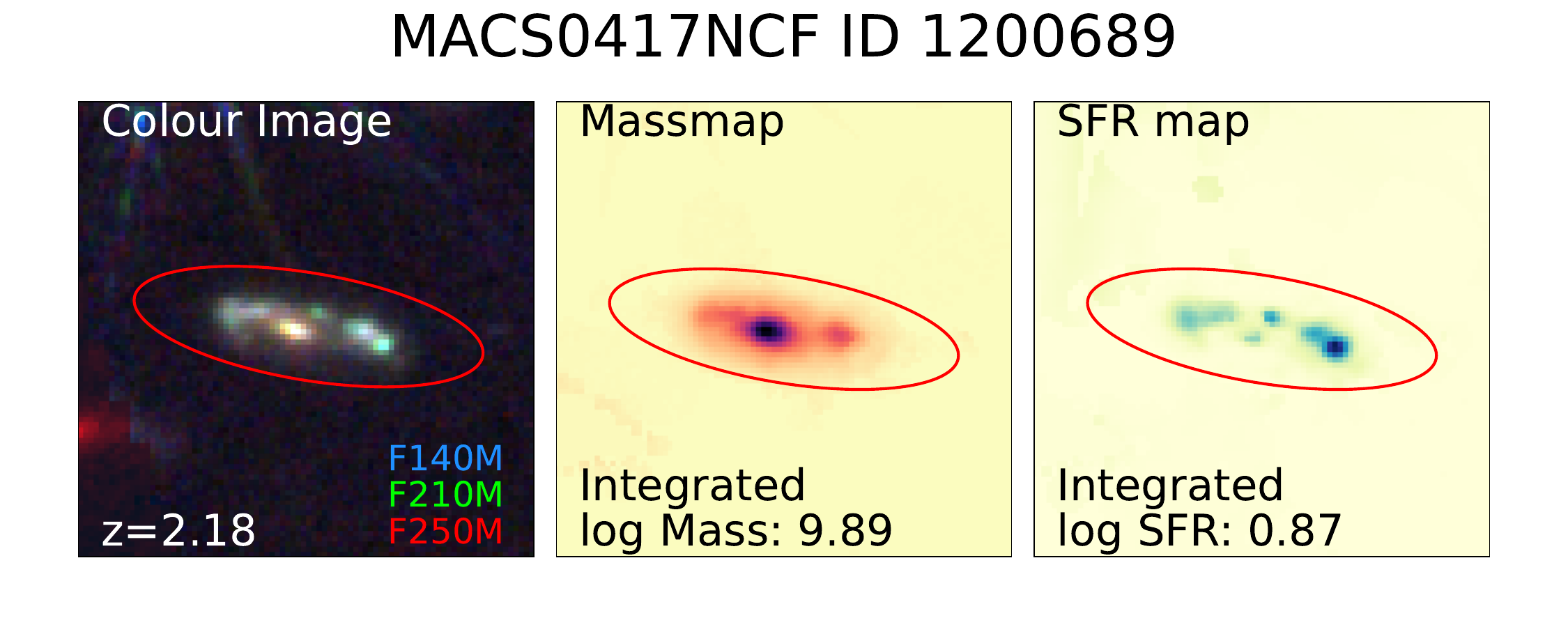} & 
    \includegraphics[width=0.45\textwidth]{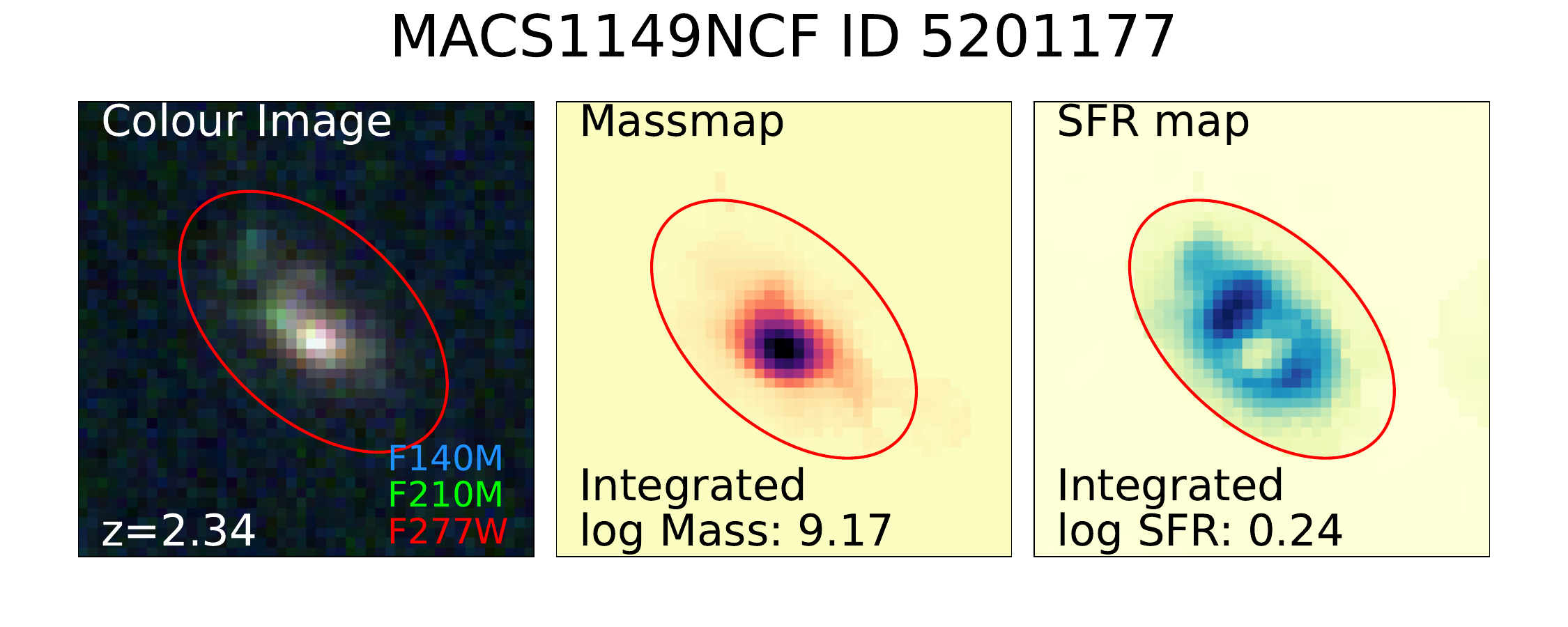} \\
     \includegraphics[width=0.45\textwidth]{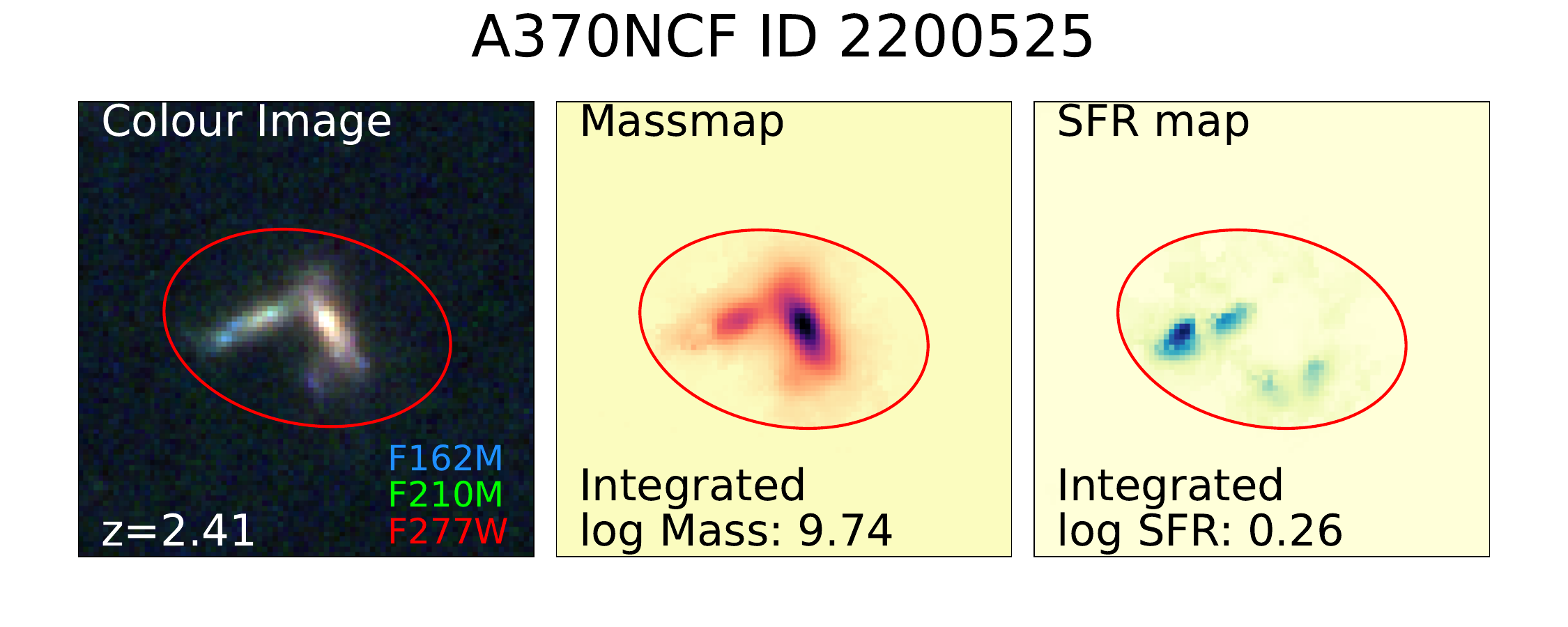}& 
    \includegraphics[width=0.45\textwidth]{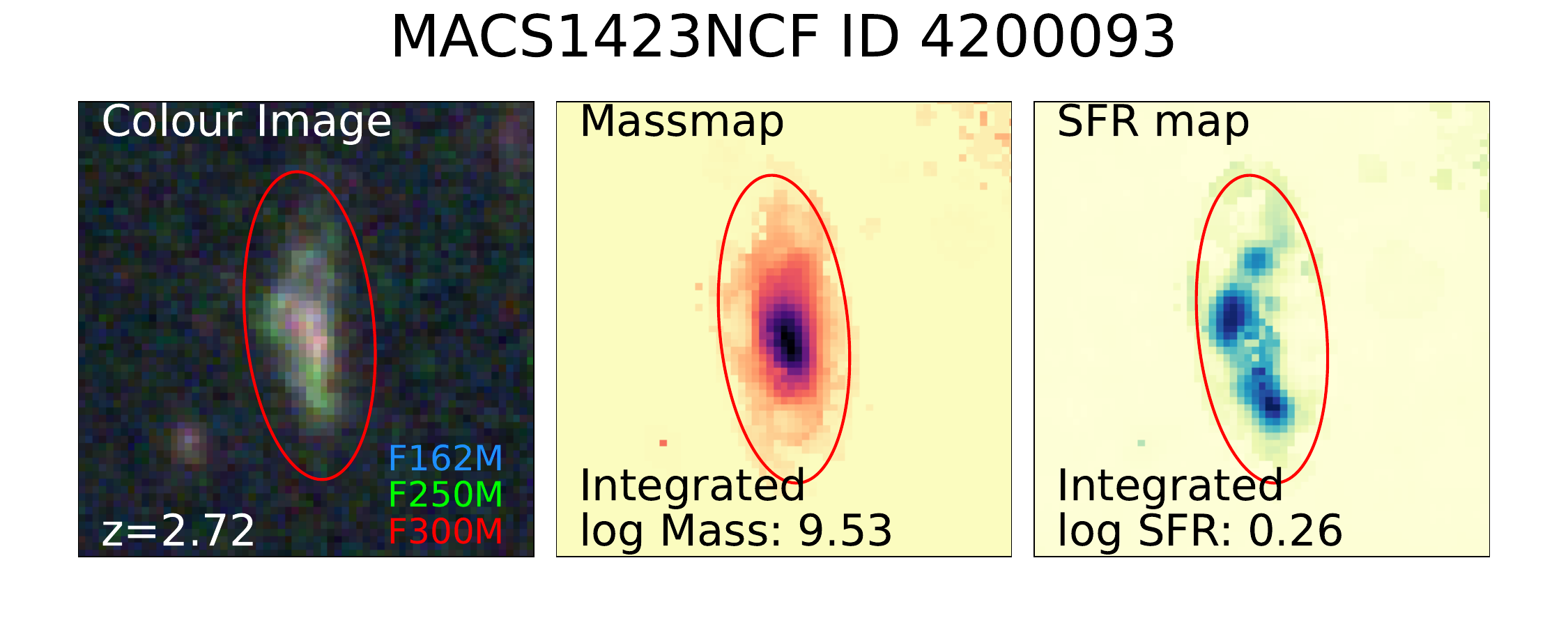} \\
     \includegraphics[width=0.45\textwidth]{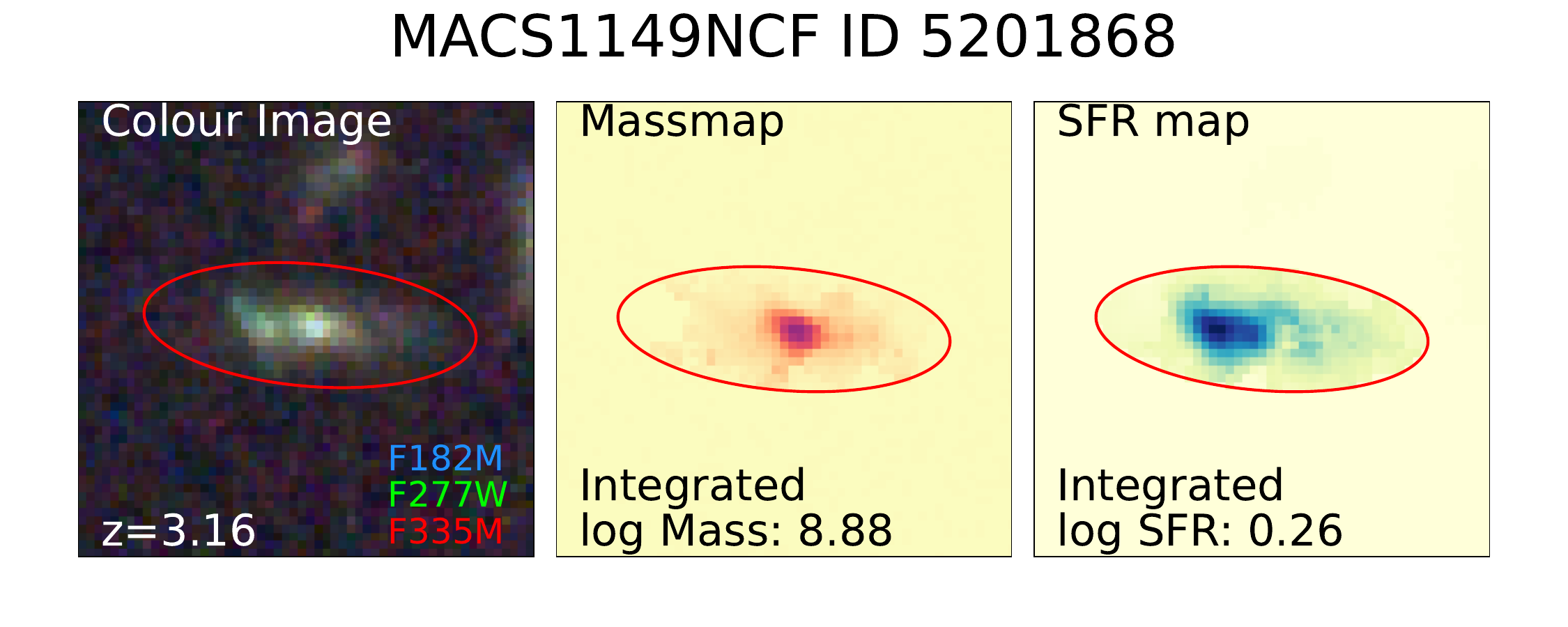} & 
    \includegraphics[width=0.45\textwidth]{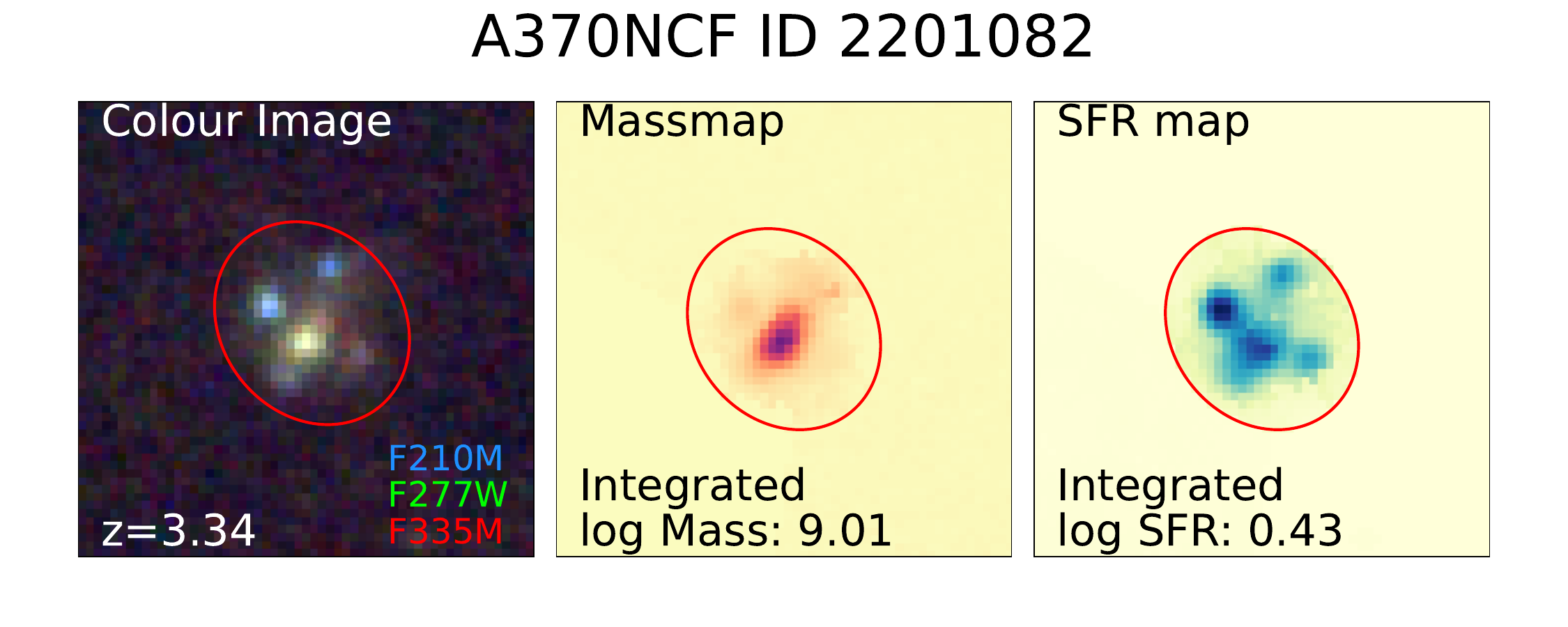} \\
     \includegraphics[width=0.45\textwidth]{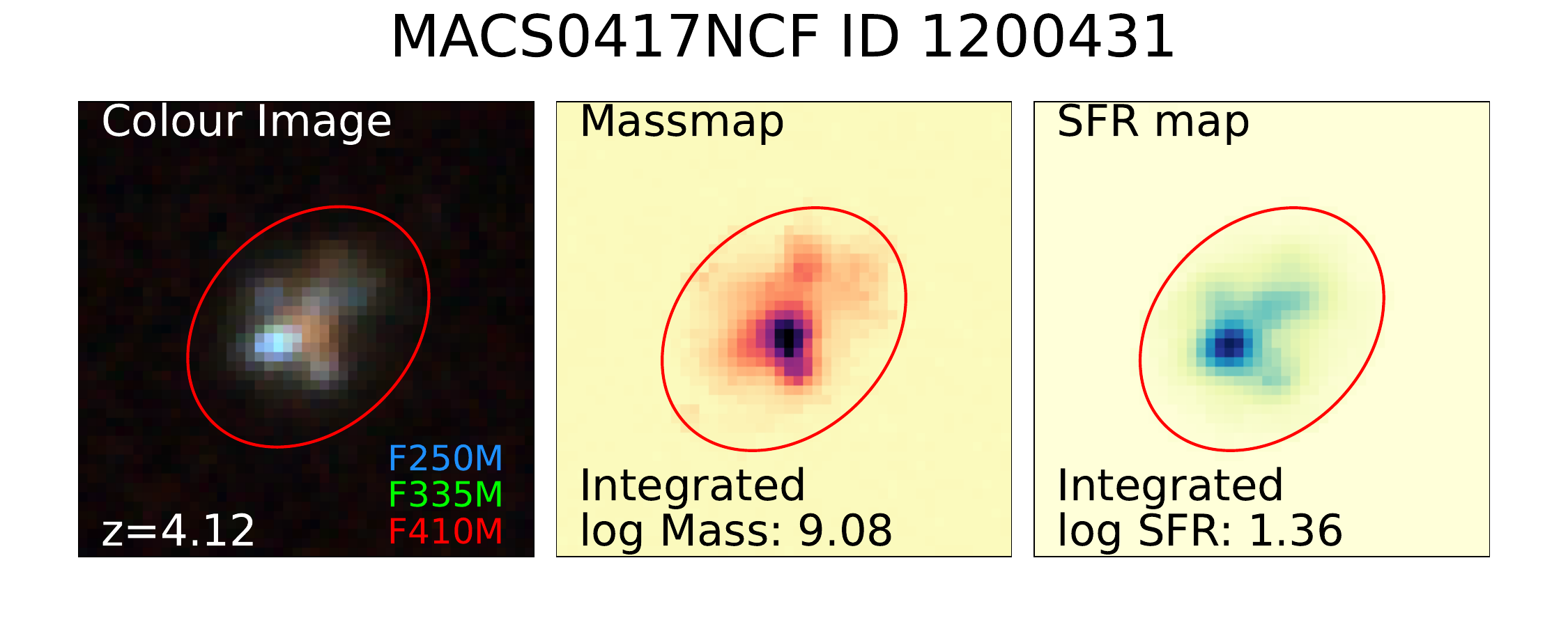} & 
    \includegraphics[width=0.45\textwidth]{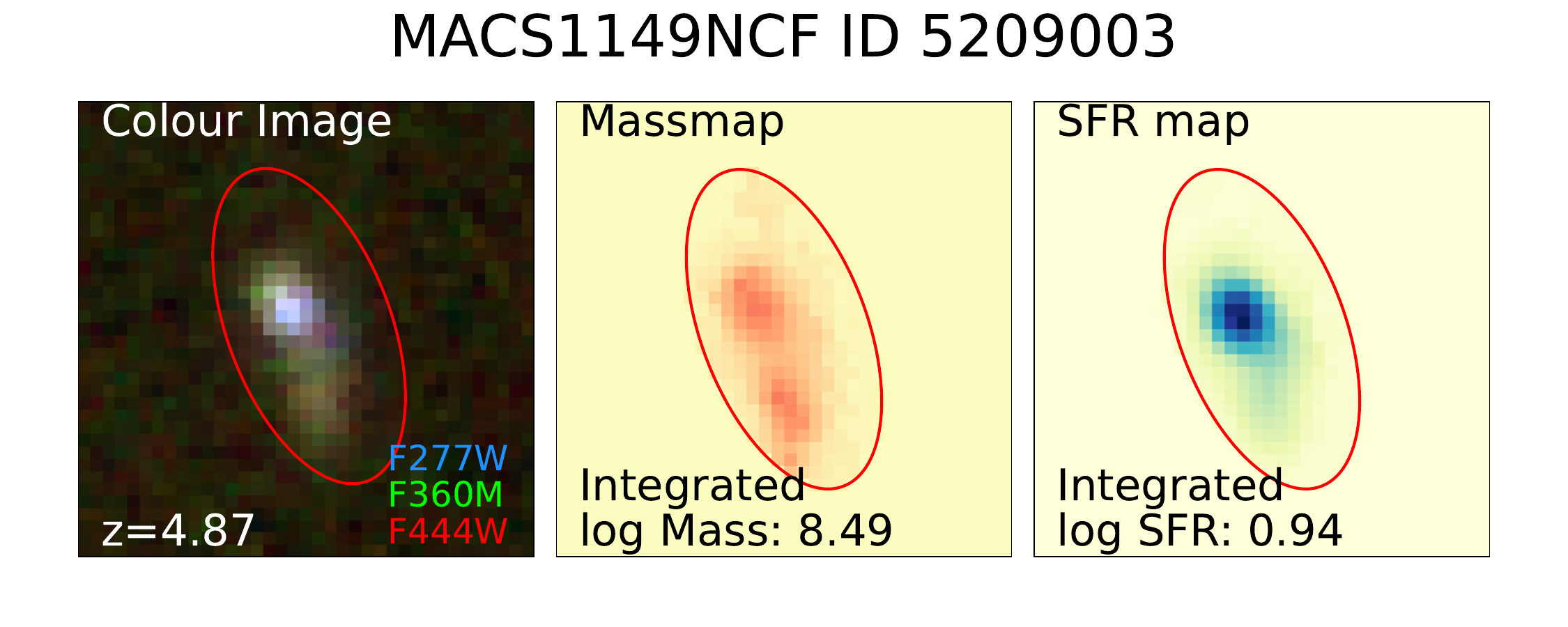} \\
\end{tabular}
\caption{Rest frame $gri$ color image, stellar mass map, and SFR map of select MWA progenitors from CANUCS. Units for integrated stellar mass are in $\log(\msun)$ and for integrated SFR are in $\log(\msun/\text{yr})$. Red ellipses are defined by Equation \ref{eqn:kron}.}\label{fig:detail-images}
\end{figure*}

\subsection{Voronoi binning of galaxy images}\label{sec:vorbin}
In preparation for resolved SED-fitting, we use Voronoi binning \citep{Cappellari:2003} on cutouts of each galaxy. The overall process is similar to the process outlined in previous works \cite{Tan:2022} and \cite{Tan:2024} using Hubble photometry, however, instead of Voronoi tessellating across the whole of the cutout, the cutouts were separated into two areas: the galaxy, and the background.

The area of the image defined to be the galaxy are the pixels of the cutout contained in an elliptical aperture scaled to the Kron radius of each object via their major and minor axes $a$ and $b$:
\begin{equation}
    R_{model,a} = 2.5 R_{kron} a \quad , \quad R_{model,b} = 2.5 R_{kron} b \quad .\label{eqn:kron}
\end{equation}

Note that this is the \textit{modified} definition of Kron radius, as defined in Source Extractor and Photutils (see \citealt{Bertin:1996} and \citealt{Bradley:2016} for details). As mentioned in \S \ref{sec:data}, the CANUCS photometric catalogs were created with the Photutils package, so $a$ and $b$ multiplied by Kron radius $R_{kron}$ would define an elliptical aperture that contains roughly 68\% of the galaxy's light. Therefore, at 2.5 times the Kron radii, the aperture would capture roughly 97\% of the galaxy's light within it.

For each field, from the preliminary sample chosen via stellar mass limits from abundance matching, we only select objects to perform pixel-binning if the total integrated flux in the F444W band as defined by the Kron radius of the object has $S/N = 30$ or higher.

In preparation for Voronoi binning, we set the minimum signal-to-noise threshold for the binning within the aperture to be $S/N = 10$ for galaxies at $z < 1.8$, and $S/N = 5$ for galaxies at $z \geq 1.8$. This is to ensure that higher-$z$ galaxies have smaller spatial bins for better resolution. For the pixels outside the aperture in the cutout, the minimum signal-to-noise threshold is set at a value of $S/N = 3$. If binning fails for the background (the pixels outside the aperture), the signal-to-noise is lowered by 1 until a minimum of $S/N = 1$. 

\subsection{Accurate flux errors for photometric catalogs of Voronoi bins}

In order to have resolved stellar mass maps, we must first create a new photometric catalog for the total flux and flux errors in each spatial bin of each cutout of a MWA progenitor in our sample. This ensures that resolved SED-fitting is performed in the same manner for this study as it was done on the integrated photometry of the entire CANUCS catalog. 

The flux errors for each spatial bin of each galaxy were derived by approximating the size of the bin to a circularized aperture of a specific radius. The longest distance between two pixels in a single bin is taken to be the ``diameter" of the aperture. This is because the Voronoi binning algorithm was set to run with both ``weighted Voronoi tesselation" (WVT) and ``centroidal Voronoi tesselation" (CVT) modes, and thus binned regions are close enough to circular for this distance to be a good approximation. Afterwards, the flux error of the circularized aperture is set as the background flux error for the specified bin. The flux error for the circularized apertures are calculated by taking the FWHM of the number of counts from $\sim$2000 empty apertures of the same size placed on empty regions across each respective CANUCS field. This means that every photometric band of every field has its own specific set of background flux errors. Similar to the photometric catalogs and mass catalogs, a systematic error of $3\%$ was added in quadrature to each spatial bin. We chose $3\%$ to be consistent with the mass catalogs of CANUCS, which are also fit with Dense Basis. (See \citealt{Sarrouh:2024} for more details related to this flux error in application to CANUCS catalog construction. See \citealt{Sok:2025a} and Jagga et al., in prep. for a similar treatment of flux errors in Voronoi binning.) In addition, the MW foreground extinction correction factors from the catalog were also applied to the total flux of the pixel bins for each respective galaxy, to ensure that the SED fit is as accurate as possible.

\subsection{Creating stellar mass and SFR maps}\label{sec:sed-fitting}

Similar to the construction of the full CANUCS mass catalogs, we use Dense Basis \citep{Iyer:2017, Iyer:2019} with the FSPS libraries \citep{Conroy:2010} to perform SED-fitting to the resolved photometric catalogs created for each individual galaxy in our MWA progenitor sample. Dense Basis is a non-parametric SED-fitting code that uses a Bayesian nested sampler and Gaussian process kernels to fit smooth star formation histories (SFHs) independent of functional form. This flexibility makes Dense Basis an ideal SED-fitting code for both obtaining the stellar mass density maps, and creating resolved SFR maps for specific sub-regions of a galaxy. Since higher-$z$ galaxies contain more star-forming clumps, SFR maps can reveal a wealth of information about the mass assembly history of MWA progenitors. 

The relevant information for both SED-fitting and SFH reconstruction is saved into an atlas of template SEDs, also called a pregrid, and was done for each catalog for the CANUCS fields. By default, Dense Basis uses a Chabrier IMF \citep{Chabrier:2003}, and the Calzetti dust law \citep{Calzetti:2000}. 
We choose a flat sSFR prior for the star-formation rate. We use an exponential dust attenuation prior with values of $A_V$ allowed to vary between 0 and 4. We also use a flat metallicity prior with a range of $-1.5 < Z < 0.25$.
For SFH reconstruction, the pregrid is given a set of timescales $\{t_x\}$ on which different SFRs are computed (i.e. $\{t_X\} = t_{10}, t_{100}, t_{300}, t_{1000}$ for the CANUCS catalogs). For consistency, we use the instantaneous SFR, which is the average SFR over the time scale at which the galaxy formed the last 1\% of its mass.

In order to be consistent with the integrated stellar mass and SFR of the CANUCS catalogs, we use the same pregrid atlases for each respective CANUCS field. When Dense Basis is run on each galaxy, it is run on each spatial bin of the galaxy cutout, treating the bin as an ``aperture" itself, with flux and errors distributed accordingly. We fix the redshift for each pixel bin to the photometric redshift of the galaxy. We take the best-fit median stellar mass and median SFR as the resultant stellar mass and SFR for each respective pixel bin. For better resolution in larger bins, we distribute the total stellar mass or SFR of that bin among the pixels depending on how much flux that individual pixel contributed to the total flux of the bin in the relevant band. For stellar mass maps, the filter used to determine distribution of the stellar mass within a single bin is the reddest filter, F444W, whereas for the SFR, the filter used is the closest broadband or medium band filter to rest-frame $u-$band (365 nm).

In Figure \ref{fig:insetplots-maps}, we display 119 randomly chosen examples of the original 877 MWAs and plot their color images, stellar mass maps, and SFR maps inset into a plot of the total stellar mass versus redshift to demonstrate the evolution of the mass density and SFR density over our entire redshift range. The images are scaled to the same angular resolution to demonstrate how much progenitors have grown since $z=5$. MWA progenitors grow significantly in total mass and size, since $z =5$, with disk structures emerging around $z\sim 2.5$. In Figure \ref{fig:detail-images}, we present more detailed color images, stellar mass maps, and SFR maps of 16 select galaxies from this sample across a variety of redshifts.  What is notable about the resolved SFR maps in Figure \ref{fig:detail-images} is that regions of high star formation are closer to and even overlap with regions of high stellar mass density at earlier times, but the star-forming regions migrate further out at later times, decoupled from regions where stellar mass is concentrated.  We further investigate these evolutionary changes in stellar mass and SFR quantitatively in \S \ref{sec:mass-growth} and \S\ref{sec:morphology}.

\section{Resolved mass assembly and star formation of MWA progenitors}\label{sec:mass-growth}
\begin{figure*}
\centering
        \begin{subfigure}
       \centering
        \includegraphics[height= 5.8cm]{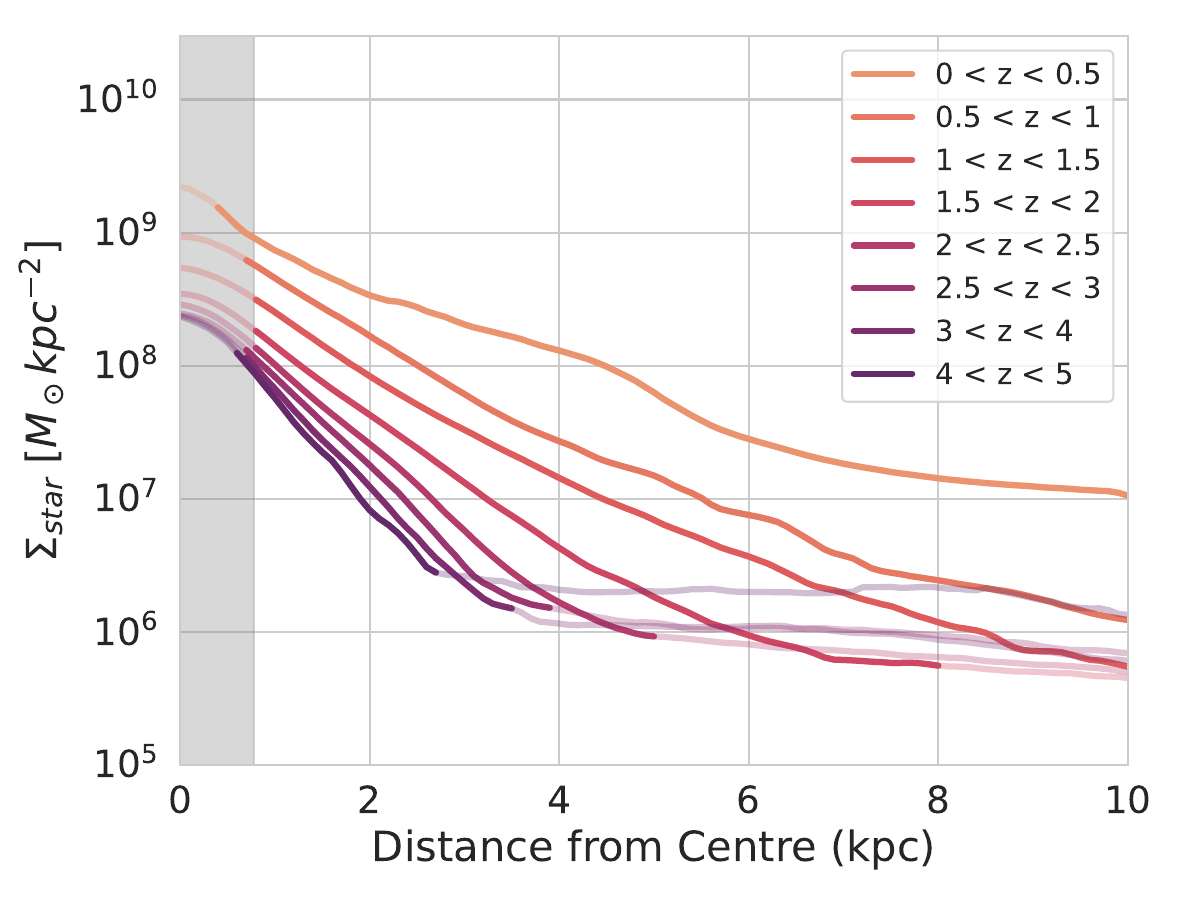}
    \end{subfigure}
    \begin{subfigure}
        \centering
        \includegraphics[height=5.8cm]{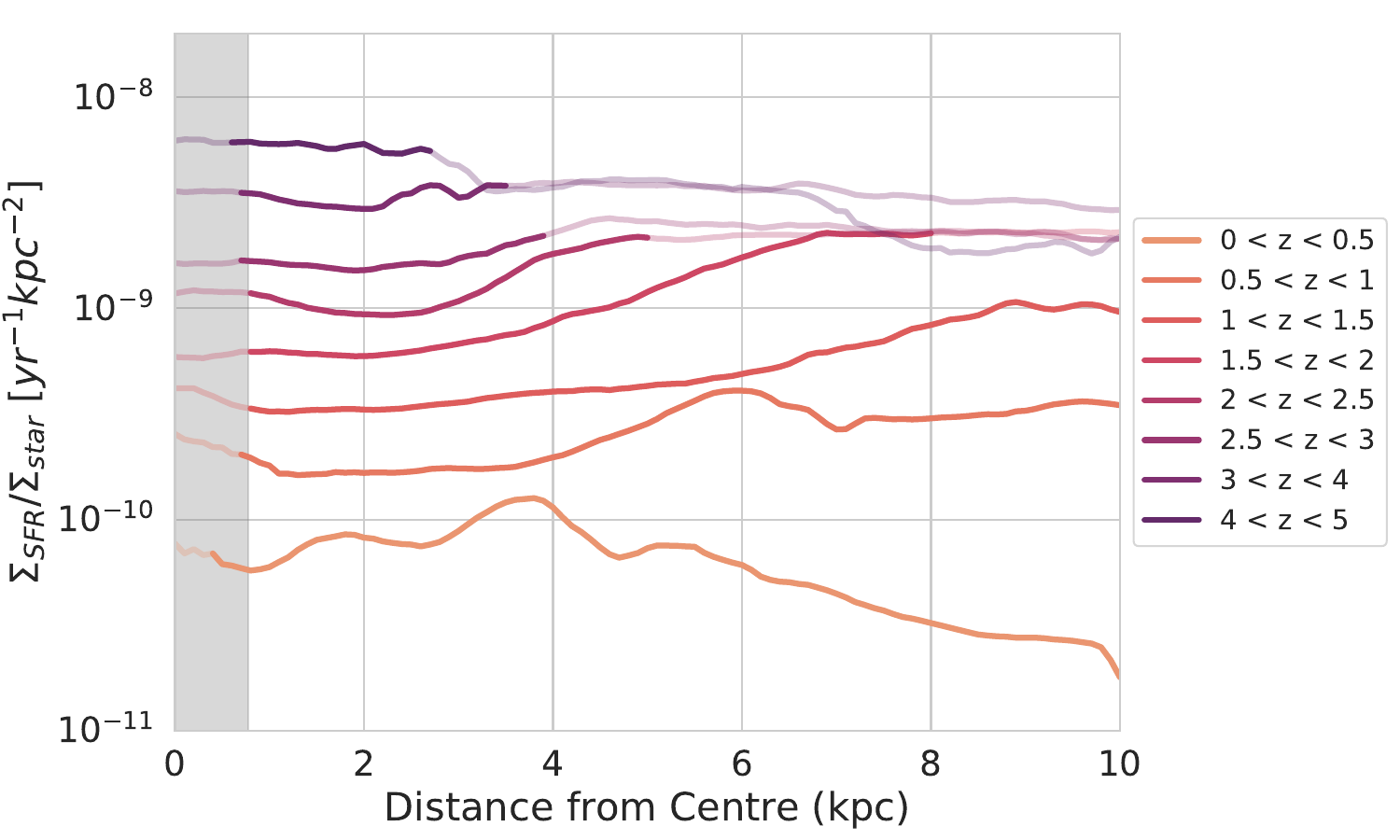}
    \end{subfigure}
    \caption{\emph{Left Panel:} Stacked and normalized stellar mass profiles for each redshift epoch. Profiles are truncated at the boundary where the noise dominates. The grey region indicates the maximum extent of the central regions affected by the PSF. \emph{Right Panel:} Stacked and normalized sSFR (specific star formation rate) profiles for each redshift epoch. Each sSFR profile was obtained by dividing each normalized SFR profile by the normalized mass profile of the same redshift (see Appendix \ref{app:profiles-extra}). The truncation in radial profile is the same as the left panel.}
    \label{fig:all_profiles}
\end{figure*}

\begin{figure*}
\centering
\begin{subfigure}
        \centering
        \includegraphics[width=\textwidth]{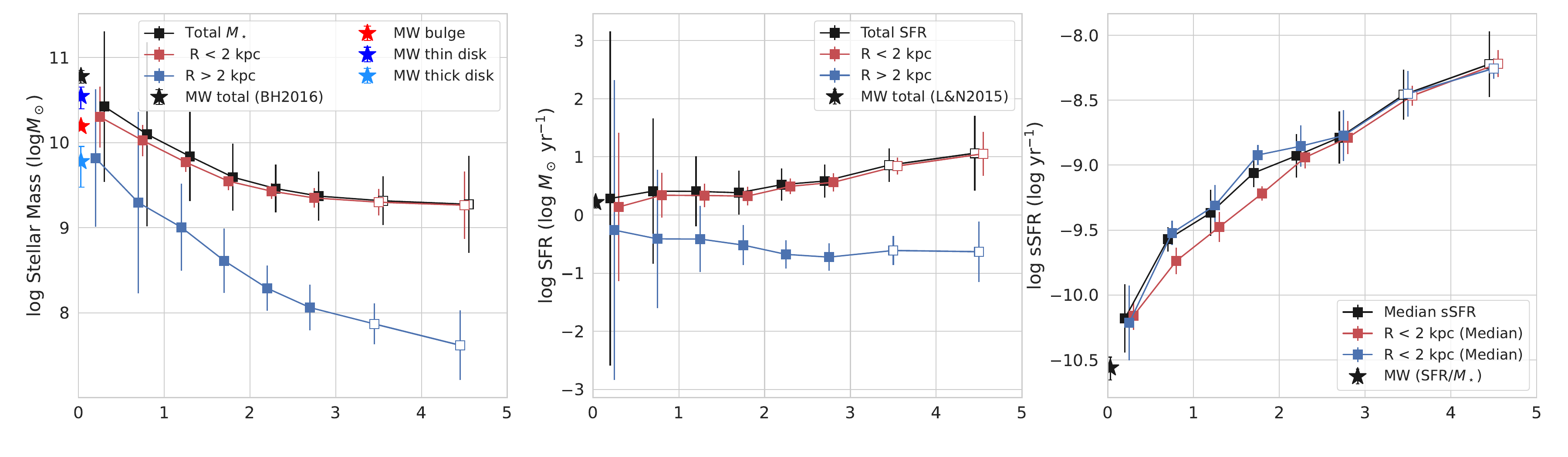}
    \end{subfigure}
    \begin{subfigure}
        \centering
        \includegraphics[width=\textwidth]{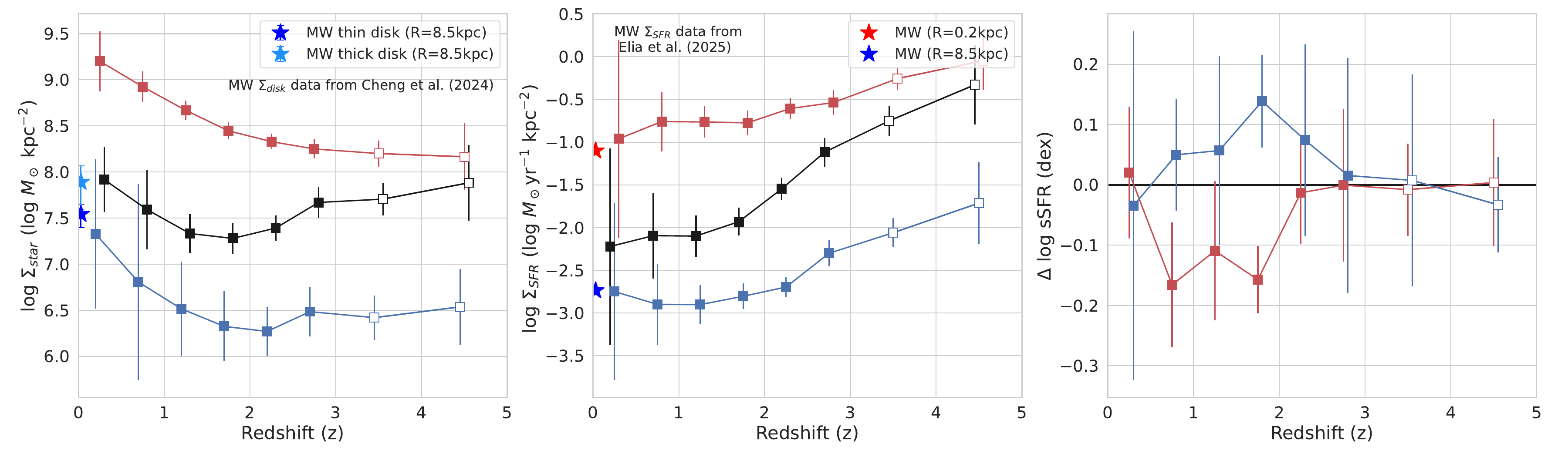}
    \end{subfigure}
    \caption{\emph{Top Panels:} Median stellar mass growth (left), median SFR (center), and median sSFR (right) as a function of redshift. Red and blue show the change in stellar mass, SFR, or sSFR in the inner ($R < 2$ kpc) and outer ($R > 2$ kpc) regions of the galaxies. Empty points represent redshifts affected by completeness of the sample. MW values from \cite{Bland-Hawthorn:2016} for stellar mass and \cite{Licquia:2015a} for SFR. \emph{Bottom Panels:} Median stellar mass and SFR densities (left and center) as a function of redshift. Right panel is the difference in sSFR between inner/outer regions and the overall median sSFR. Error bars in all plots are standard error. MW mass density and SFR density values are from \cite{Cheng:2024} and \cite{Elia:2025} respectively.}\label{fig:inner-outer}
\end{figure*}
In this section, we show measurements of the growth of MWA progenitors to better understand the mass growth history of the MW. We utilize both mass and sSFR density profiles to track the overall mass assembly since $z =5$, which spans a time period of 12.5 Gyr. Thus, we group the MWA progenitors into redshift bins according to the redshift ranges represented by the SMFs of \cite{McLeod:2021} and \cite{Grazian:2015} in \S \ref{sec:abundance-matching}. The stellar mass boundaries of each redshift bin are represented by the grey shaded region in the bottom left panel of Figure \ref{fig:numdensity}, as well as the brown shaded region in the top panel of Figure \ref{fig:mass_fullsamp}. For each redshift bin, a stacked and normalized mass and sSFR radial density profile is created from the individual radial profile (both $\Sigma_{star}$ and $\Sigma_{sSFR}$) of each galaxy. We also further separate the profiles into inner and outer regions, as defined by the positional relation to a galactocentric radius of $R = 2$ kpc. We track the inner, outer, and overall mass growth, sSFR, as well as their respective density over time to quantify the difference in mass growth over cosmic time.

\subsection{Density profiles as a function of galactocentric radius}\label{sec:density-profiles}

The 1-D mass and SFR density profiles as a function of distance from galactic center $R$ are obtained by placing elliptical annuli centered on the galaxy cutout with a width of 0.1 kpc on the stellar mass and SFR maps. The total stellar mass or SFR within each annulus is then divided by the total area of the annulus.
The shape of the annuli are defined by the galaxy's $b/a$ ratio as found in the CANUCS catalogs. 

In order to counter the effects of magnification on size measurements and radial density profiles, the conversion factor between observed angular size and the physical size in kpc were adjusted in the direction of the position angle of the galaxy (the angle of the semi-major axis to the horizontal) as well as in the direction perpendicular to the position angle. To make sure the adjustment in resolution is as close as possible to the true physical scale, we define a matrix which describes the ellipse of galaxy on the observed plane using $a$, $b$, and $\theta \equiv$ position angle. We also define a transformation matrix based on lensing shear orientation $\phi$, the angle at which the lensing distortion is the strongest, $\mu_{tan}\equiv$ the magnification in the $\phi$ direction, and $\mu_{rad} \equiv$ the magnification in the direction perpendicular to $\phi$. Applying the inverse of the transformation on the matrix that defines the ellipse and then obtaining the eigenvalues will return the de-magnified lengths of $a$ and $b$, as well as the de-lensed position angle. However, since we are keeping to the observed plane and not the source plane in order to preserve resolution and the measured PSF, we scale the angular resolution by the change in length of $a$ and $b$ respectively from the source plane to the observed plane.

Each individual galaxy's mass density or SFR density profile is then sorted according to their redshift bin, then stacked and normalized to obtain the density profile of that redshift. These normalized density profiles are shown in Figure \ref{fig:all_profiles}, with distances from the center measured along the semi-major axis. See Appendix \ref{app:profiles-extra} for the individual mass and SFR density profiles for each galaxy, divided into redshift bins.

The solid lines of the mass density profiles in the top panel of Figure \ref{fig:all_profiles} reveal that at higher redshifts ($2< z< 5$), mass density increases between different redshift steps more at greater distance from the galactic centre. At $R < 1$ kpc, the increase in mass density is roughly $\sim 0.1 - 0.2$ dex, from $4 < z < 5$ to $2<z<2.5$. At a further distance of $R \sim 2.5$ kpc, the increase in mass density from one redshift epoch to the next is $\sim 1$ dex, an entire order of magnitude greater. This trend is reflected in the sSFR density profiles as well, as the sSFR starts off relatively flat across the entire normalized profile at $4<z<5$, but the highest sSFR consistently occurs at the furthest point from the centre of the galaxy, until $z \sim 0.5$ when it flattens out once more.

At higher redshifts, the mass density profiles retain a similar shape in the center, but their outskirts increase in mass. This might best be shown by the comparison of the growth in the centre ($R < 2$kpc), to the growth at 4.5 kpc, which is approximately the effective radius of MWA at $z\sim 0$ (see \S\ref{sec:sersic-params}). Over the redshift range covered by our data the central part of the MWA grows by about an order of magnitude in mass, however, the growth at $R \sim 4.5$ kpc is two orders of magnitude, implying ten times more growth at the effective radius. Likewise, in the sSFR density, while the sSFR is continuously decreasing at later times, the highest sSFR density is usually near the limits of the density profile's boundary, implying that the outskirts of the galaxies are where most of the sSFR is concentrated. This shows that towards lower redshift, galaxies are increasingly growing their disks, which is evidence for inside-out growth. 

We note that this trend may be affected by PSF smearing, because the physical kpc scale of the PSF increases up to $z\sim 1.6$ (the affected regions of the density profile have greater transparency in Figure \ref{fig:all_profiles}). However, since the size of the PSF is never greater than 1 kpc even at the redshift bin with the worst physical resolution, PSF smearing cannot account for all of the density profile changes within $R = 2$kpc, which is defined as the inner regions of the profiles. Additionally, the total stellar mass and total sSFR in the inner regions should be less affected by the PSF, and how those quantities change with redshift is discussed in the next section.

\subsection{Inner regions versus outskirts}\label{sec:inside-out-growth}

With the stacked and normalized 1-D profiles representative of each redshift epoch in Figure \ref{fig:all_profiles}, we divide them into inner and outer regions with a 2 kpc cutoff point. While the 2 kpc cutoff point is most of the galaxy at higher redshift bins, especially $z > 4$, the MW in the current universe has a bulge extent of roughly 2 kpc, and thus this region represents the regions which will eventually \textit{become} the bulge for present day MWAs. 

In Figure \ref{fig:inner-outer}, the outer regions, represented by the blue line on the plots, are shown in the top panels to increase in both stellar mass and in sSFR much faster than the inner regions at higher-$z$. In fact, the epoch between $ 2.5 < z < 5$ shows very little stellar mass growth in the inner 2 kpc region -- only $\sim0.1$ dex -- while the outer region increased $\sim 1$ dex, an order of magnitude from the highest redshift bin to $z\sim2.25$.  The stellar masses of the outer regions at the lowest redshift bin appear to be directly in between the \mstar of the MW's thin and thick disks, while the inner regions are already more massive than the \mstar of the MW's bulge. However, one must keep in mind that the MW has an unusually small bulge for a galaxy of its size. \citep{Shen:2010}. The overall SFR and the SFR of the inner regions decline as redshift decreases, but the SFR of the outer regions remains constant at earlier times at $\sim0.2 \msun$/yr ($2<z<5$) and increase at later times ($z < 2$), up to $\sim0.6\msun$/yr at the lowest redshifts. For the median sSFR plotted in the top-right panel, we again see a consistent decline in the inner regions as redshift decreases, but from $1.5<z<3$ the sSFR of the outer regions are much higher, and decrease less sharply. From $z\sim 2.5$ onwards to lower redshift, the outer regions have consistently higher sSFR, until the lowest redshift bin. This demonstrates the rapid and sustained inside-out growth of this sample of progenitors.

In the bottom-left panel of Figure \ref{fig:inner-outer}, the total, inner, and outer stellar mass density ($\Sigma_{star}$) is plotted as a function of redshift. From $z=5$ to $z=2$ the total $\Sigma_{star}$ is decreasing, the outer $\Sigma_{star}$ is roughly constant, while the inner $\Sigma_{star}$ grows slowly. While this seems counter to the stellar mass growth in the top-left panel, recall that the galaxies themselves are also increasing in size.  Therefore if the stellar mass is added to the outskirts of the galaxy, and increasing its scale-length, the decrease in mass density is expected. The stellar mass density evolution also shows evidence of MWA progenitors growing inside-out at $2<z<5$. It is important to note that the amount of time between $2<z<5$ is roughly 2 Gyr.  In that span of time, the abundance matching implies galaxies have grown ten times in stellar mass, and increased in size from $R_e\sim1.0$ kpc to $R_e \sim1.8$ kpc (see \S \ref{sec:sersic-params} for $R_e$ measurements and more information on general morphological evolution). After that growth, however, the inner and outer regions increase in stellar mass density at roughly the same rate (i.e. the slope of the inner and outer mass density lines are very similar). Therefore, the mass is growing proportionally at $z < 2$, which means the growth of the inner and outer regions are in lockstep, which is consistent with previous studies. 

The bottom middle panel, which displays the SFR density ($\Sigma_{SFR}$) as a function of redshift, shows that both inner and outer regions decline in SFR steadily as redshift decreases. The $\Sigma_{SFR}$ at later times is consistent with the measured SFR densities of the MW's inner regions, and solar neighborhood from \cite{Elia:2025}. At higher redshift ($z > 2$), the $\mstar$ of the outer regions is increasing rapidly, while its $\Sigma_{star}$ remains constant, so we should then expect the $\Sigma_{SFR}$ to also remain constant, but that is not the case. The $\Sigma_{SFR}$ of the outer regions is decreasing at $z > 2$. Additionally, there is enhanced sSFR in the outer regions from $1.5<z<3$, as the decline in outer sSFR is much shallower than the inner regions. This is shown more clearly in the bottom-right panel, which shows the difference in dex between the sSFR of the inner and outer regions with respect to the overall sSFR at that redshift bin. This may be an indication that the \textit{observed} SFR may not account for all of the mass assembly of this sample.

The trends in the SFR density and the enhancement of the sSFR of the outer regions may indicate more varied star-formation activity in the outer regions at different epochs of growth, and perhaps indicate enhancement of star formation around cosmic noon. Since this sample is only galaxies with total $S/N \geq 30$, the MWA sample is not mass complete at $z > 3$. Therefore, the stellar mass and especially the SFR trends at $z > 3$ are only representative of the most massive and highly star-forming MWAs at those epochs. However, we have conducted tests to see if the trends found in this section hold with only the top $20\%$ of mass at each redshift epoch apart from the earliest epoch at $4<z<5$. We found that the trends between inner and outer regions discussed in this section still hold. Additionally, hierarchical formation and mergers are expected to play a larger role as redshift increases. Mergers and interactions can enhance SFR as well (though \citealt{Moreno:2021} shows in FIRE-2 that galaxy interactions can both enhance and suppress SFR differently at different radii for both main and secondary galaxies in pairs), so it is important to understand how much stellar mass is formed in-situ versus ex-situ from mergers.

\subsection{Is SFR the main driver of mass assembly?} \label{sec:sfr-mass-compare}
\begin{figure}[t]
    \centering
    \includegraphics[width=\columnwidth]{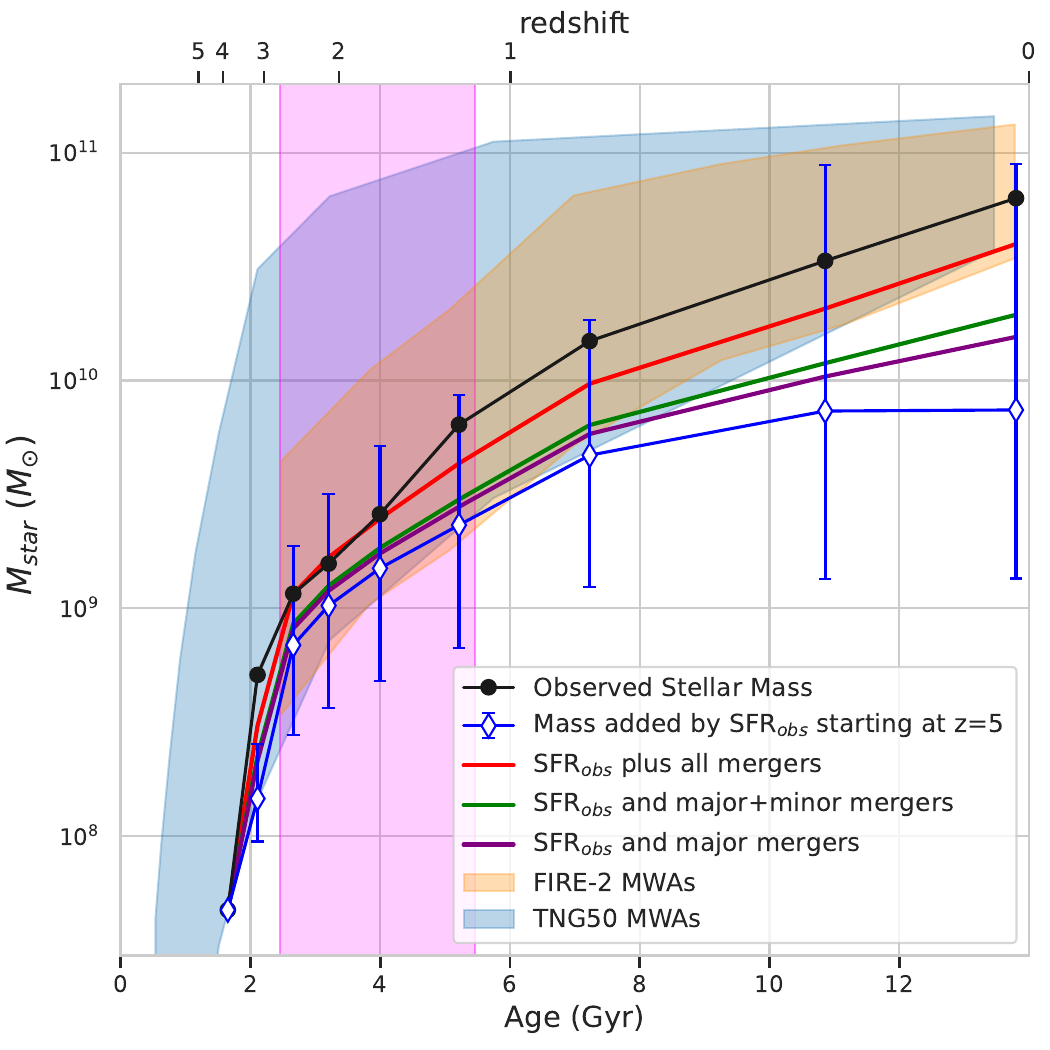}
    \caption{The observed stellar mass evolution over time of the full unresolved sample (solid black circles), and the stellar mass added by integrating the observed unresolved sample's SFR of the previous epoch (blue open diamonds). Error bars are upper and lower limits derived from the $1\sigma$ scatter of the SFR at each epoch. The mass added by both observed SFR and estimated accretion rate (from \citealt{Rodriguez-Gomez:2016}, integrated over the time span of each epoch) improve the agreement when more mergers are included (red line is merger mass ratio $\geq 0.0001$, green line is mass ratio $\geq 0.1$, and purple line is mass ratio $\geq 0.25$). Shaded purple region is the possible time frame at which the GSE merger occurred for the MW \citep{Belokurov:2018, Helmi:2018}}
    \label{fig:mass-growth-z5}
\end{figure}

We seek to understand the mass growth observed over the redshift ranges that are mass-complete, whether it is dominated by in-situ SFR or if galaxy mergers may play a role. For each redshift bin, we have obtained the median total stellar mass, and the median total SFR. To test if the SFR is consistent with the total mass assembled, we first take the total resolved SFR in units of $\msun/$yr at each redshift bin $z_{obs}$, and calculate the total mass formed ($M_{formed}$) by the SFR by integrating over the span of time that has passed.
\begin{equation}
    M_{formed}(z_{obs}) = \int_{t_i}^{t_f} \text{SFR}_{z_{obs}}(t) dt \;,
\end{equation}
where SFR$_{z_{obs}}$ is the median SFR at the redshift epoch $z_{obs}$, $t_i$ is the age in years of the start of $z_{obs}$ and $t_f$ is the age in years of the end of $z_{obs}$. Note that for the median SFR of each redshift range, we only consider star-forming galaxies and ignore quiescent galaxies. This $M_{formed}$ however does not represent the stellar mass, $\mstar$ that was added in that time period. This is because a portion of the total mass formed is returned to the interstellar medium (ISM), and must be accounted for in our calculations. Therefore, we also take into account the ``mass loss" factor, which at its simplest is the percentage of formed mass that becomes stellar mass. 

For every redshift epoch, we define the stellar mass that is \textit{not} returned to the ISM as 
\begin{equation}
    \mstar_{added}(z_{obs}) = \int_{t_i}^{t_f} \text{SFR}_{z_{obs}}(t)f_{ret}(t) dt \;,
\end{equation}
where $f_{ret}(t)$ is the mass-loss curve from FSPS \citep{Conroy:2010} that is implemented in Dense Basis to calculate SFHs from the posteriors. At redshifts of $0.5 < z < 5$, the mass loss rate is roughly $\sim 37\%$ to $\sim 43\%$, with higher mass loss at higher-$z$.

The SFR-derived $\mstar_{added}$ is then combined with the $M_{initial}$ at the starting redshift. The starting redshift point is at $z=3$, therefore $M_{initial} \equiv M_{obs, z=3} = 2.4\times10^9\msun$, and every subsequent redshift epoch:
\begin{align}
    \mstar_{added}(z_{obs}) =  &&\nonumber \\
      M_{obs, z=3}  \quad + && 
      \int_{t_i}^{t_f} \text{SFR}_{z_{obs}}(t)f_{ret}(t) dt \;, 
\end{align}
where the integral represents all of the $\mstar_{added}$ generated from integrating each median SFR at the respective redshift ranges ($t_i$ to $t_f$) that they cover, with the mass loss factor $f_{ret}$ applied. 
The comparison of this added stellar mass growth versus the \textit{observed} stellar mass growth (i.e. the median total resolved stellar mass measured at each redshift bin) is presented in Figure \ref{fig:mass-growth-z5}. 

Although we have resolved SFR measurements, we use the median of the integrated(unresolved) SFR and median integrated \mstar of each galaxy from the full sample in order to mitigate any biases from incompleteness. In Figure \ref{fig:mass-growth-z5}, the resulting mass added line, plotted in blue with white diamonds, consistently underestimates the observed stellar mass of the next redshift epoch, and at $z=0$ would underestimate the growth by almost an order of magnitude. However, the large error bars, which are from the $1\sigma$ scatter of the SFR of the redshift bin, does marginally include the observed stellar mass. This indicates that mass formed ex-situ, or mass accreted from merging galaxies likely play a significant role in mass assembly. In order to assess the contribution of mergers to mass assembly, we would need to integrate the mass accretion rate due to mergers over time.

As the  accretion rate described in \cite{Behroozi:2013b} is based on halo masses, we utilize the \textit{stellar mass accretion rate }from the Illustris simulations, ($M_0\dot{m}_{acc} (M_0, \theta, z) = d^2M_{acc}/d\theta dt$) as described in Table 1 of \cite{Rodriguez-Gomez:2016} \footnote{In \cite{Rodriguez-Gomez:2016}, $\mu$ was used to indicate mass ratio, but we have changed the symbol for the mass ratio to $\theta$ to avoid confusion with the $\mu$ used to describe magnification factors earlier in the paper} . Their merger rate agrees well with work on observational estimates of merger rates from close pairs in \cite{Robotham:2014}. Observational estimates of merger rates have been made up to $z=3$ for massive galaxies (e.g. \citealt{Snyder:2017, Conselice:2022, Fuentealba-Fuentes:2025}). The mass accretion rate from Illustris is used as an estimate of the ``missing ex-situ mass" to demonstrate the possible importance of the role of mergers. As Figure \ref{fig:numdensity} shows, our stellar mass from abundance matching is in very good agreement with the mass range of MWAs in TNG50. The red, green, and purple lines in Figure. \ref{fig:mass-growth-z5} reveal that in order to properly account for the ex-situ mass fraction, mergers with a mass ratio as small as $\theta=10^{-4}$ must be counted to get close to observed values. Only counting major mergers ($\theta\geq 0.25$) or both major and minor mergers ($\theta \geq 0.1$) is insufficient. 

Comparing our results to MWA simulations from FIRE-2 \citep{Garrison-Kimmel:2018} and from TNG50 \citep{Pillepich:2023}, we find that the $M_{\star added} + M_{acc}$ for the stellar mass accreted is consistent with both simulations only when nearly every merger is taken into account. \cite{Sotillo-Ramos:2022} also find that MW/M31 analogs in TNG50 that have experienced $\geq1$ major mergers in the last 5 Gyr have ex-situ stars make up most (up to $90\%$) of their total stellar mass at $z=0$. Furthermore, this addition does not account for how much of the observed SFR at each redshift is due to merger enhanced star formation. A closer look into merger rates is required for better understanding the mass assembly of this sample of MWA progenitors.

\section{Morphological Measurements of MWA progenitors} \label{sec:morphology}

\begin{figure*}[hbt!]
\centering
\includegraphics[width=\textwidth]{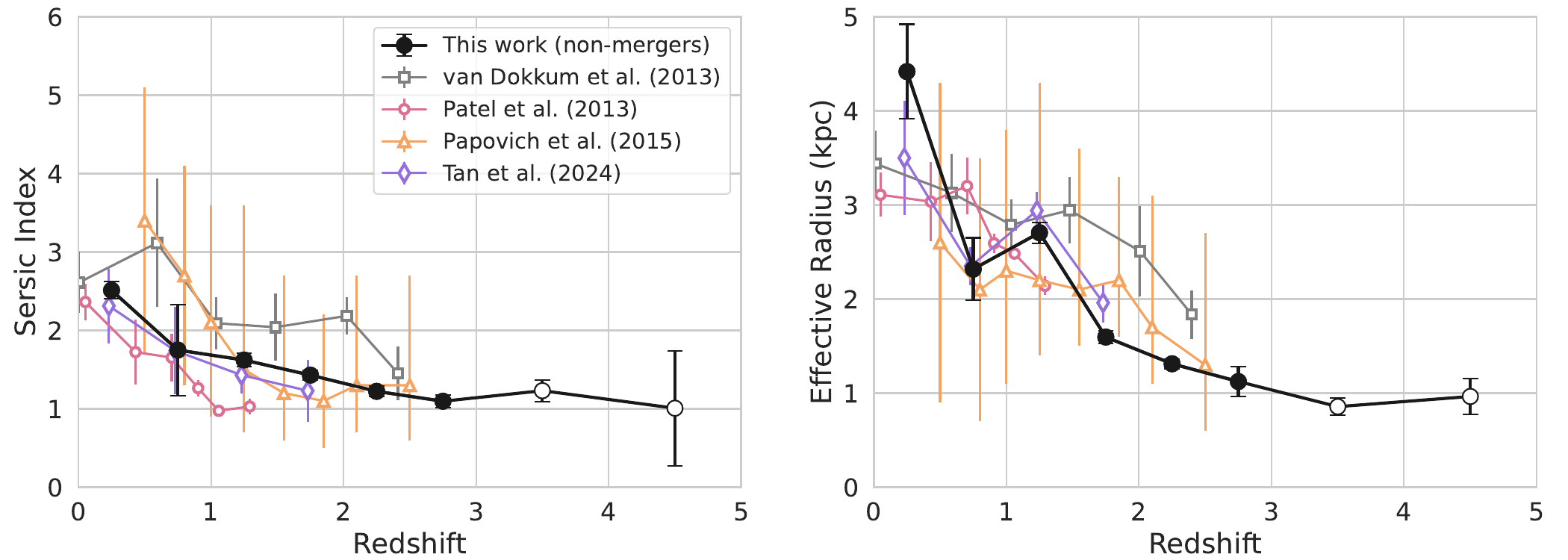}
\caption{\emph{Left: } Median S\'ersic index for each redshift bin. Black circles represent this sample of MWA progenitors from CANUCS with mergers removed, open circles represent values affected by mass completeness. \emph{Right:} Median half-mass radius $R_e$ of each redshift bin in kpc, converted via angular diameter distance from photometric redshifts.  Error bars represent the standard error in each redshift bin. Other lines are S\'ersic parameter measurements from other studies: \cite{vanDokkum:2013} in grey, \cite{Patel:2013} in pink, \cite{Papovich:2015} in orange, and \cite{Tan:2024} in purple.}\label{fig:sersic-params}
\end{figure*}

Examining the morphology of the stellar mass maps can not only strengthen the arguments for inside-out growth of disks at early epochs, but also reveal whether disturbances in structure which resulted from galaxy interactions or mergers have played key roles in mass assembly apart from in-situ star formation. For the morphological measurements, we utilize Statmorph \citep{Rodriguez-Gomez:2019}, a tool developed for asymmetry measurements and clump detection for galaxies. We run Statmorph on the stellar mass maps, the SFR maps, and the F444W maps. 

Resolved maps are prepared for Statmorph in a similar manner as described in \cite{Tan:2022} and \cite{Tan:2024} for morphological fitting.

\begin{itemize}
    \item We take the corresponding segmentation map of each object from the catalog to define the foreground and the background pixels.
    \item Noise is added to the stellar mass and SFR maps according to the FWHM of the background noise present in the F444W cutout, using the segmap of the galaxy from the catalog.
    \item The input PSF is the F444W PSF, since all of the photometry has been PSF-matched to F444W, which means the stellar mass and SFR maps would have the same resolution as F444W.
    \item A high value was set for gain to allow Statmorph to generate its own weight map for the images, as the provided weight map from the catalog would not apply to derived stellar mass and SFR quantities.
\end{itemize}

\subsection{S\'ersic profiles}\label{sec:sersic-profiles}

A S\'ersic profile \citep{Sersic:1963} is a function with two main parameters, $n$ the S\'ersic index, and $R_{e}$ the effective radius, or the radius that contains half of the flux (or in this case, half of the stellar mass). 
\begin{equation}
    I(R) = I_0 \exp\left(-b_n \left[(R/R_e)^{1/n} - 1 \right]\right)
\end{equation}

When $n= 1$, the S\'ersic profile is purely exponential, and indicates that the surface brightness profile of the galaxy is likely an exponential disk. When $n=4$, the S\'ersic profile resembles a de Vaucouleurs profile, and indicates that the galaxy may be bulge-dominated, and elliptical in morphology.  Most galaxies are in-between $n=1$ and $n=4$, as galaxies usually consist of both a bulge and a disk component.

Statmorph utilizes astropy's \texttt{2DSersic} function and scipy's \texttt{curve-fitting} package to fit either single or double S\'ersic profiles to galaxy images. Similar to Galfit \citep{Peng:2010}, it takes in a galaxy image and a PSF kernel as input. When tested on fitting F444W photometry of MWA progenitors from the A370NCF field, Statmorph returned similar S\'ersic profiles to Galfit when fitted to the same galaxy image, but has lower failure rates.
We utilize the fitted S\'ersic profile to the spatially resolved stellar mass maps to study general trends in the changes of the morphology of MWAs over cosmic time.

\subsection{S\'ersic parameters of MWA progenitors since $z=5$}\label{sec:sersic-params}

In Figure \ref{fig:sersic-params}, the median S\'ersic index of each redshift bin is plotted in the left panel, while the median $R_e$ (half-mass radius) is plotted in the right panel. From $3 < z < 5$, the median S\'ersic index remains constant at $n \sim 1$, and from $1.5 <z <3$, the S\'ersic index grows to a value of $n=1.62\pm0.09$. Within the same redshift ranges, the median $R_{e}$, representing the median half-mass radius of the MWA progenitors, is also increasing. In the same time frame of $2 <z< 5$, the $R_e$ increases from $1.0\pm0.2$ kpc at $4 < z < 5$ to $1.84\pm0.16$ kpc at $1.5 <z < 2$.  

As noted in \S \ref{sec:inside-out-growth}, the redshift range $2< z < 5$ is the same epoch where the MWA progenitors exhibit inside-out growth, where more stellar mass is accumulated in the outskirts of the galaxy versus in the inner regions, according to the density profiles in Figure \ref{fig:all_profiles} and in Figure \ref{fig:inner-outer}. This is consistent with growth of progenitors' disk components, because half-mass radius corresponds to size. The median S\'ersic index remaining at $n < 2$ indicates most galaxies at $2<z<5$ are likely disk-dominated systems, but the slight increase in median S\'ersic index is also in line with proportional bulge growth. Since Figure \ref{fig:mass-growth-z5} has shown that mergers are required at all redshifts to match the observed stellar mass build-up, any potential mergers that have occurred tend \textit{to not} morphologically transform these galaxies from disks to pure bulges. However, we note that the individual measurements of S\'ersic index and $R_e$ for individual galaxies have a large scatter, and $16\%$ of non-merger progenitors at $4<z<5$ have $n > 4$. The median values show that in general, our MWA progenitors tend to be disk-dominated at higher redshift. 

The next stage of mass assembly takes place from $z=2$ to $z= 0.3$. As discussed in \S \ref{sec:inside-out-growth}, this is the epoch where the change in mass density is roughly equivalent for both inner and outer regions. Figure \ref{fig:sersic-params} shows the median S\'ersic index at later times increases up to $n = 2.42\pm0.42$ at $0.3 < z < 0.5$. This indicates that the population of MWA progenitors appear to both grow their bulges alongside their disks. While it is impossible to directly measure the S\'ersic index of the MW from an internal vantage point, the bulge-to-total mass ratio of the MW is measured to be $B/T = 0.15$ by \cite{Licquia:2015a}. Using conversions derived by \cite{Brennan:2015} between $B/T$ ratios and S\'ersic indices, if the radius of the bulge to the radius of the disk is $< 0.4$, the S\'ersic index of a galaxy with $B/T =0.15$ is likely to be $n \lesssim 2$. Therefore, our S\'ersic index measurement of  $n\sim 2.4$ for the lowest-$z$ progenitors is consistent, but higher than the value expected for the MW. Furthermore, consensus on the MW's bulge is that its structure is a box/peanut  or ``X-shaped" bulge (see e.g. \citealt{Barbuy:2018,Helmi:2020}), rather than a classical bulge that is described by a de Vaucouleurs profile. As mentioned before, the MW's bulge is small for a disk galaxy of its mass, so any star-forming disk galaxies within our sample at the lowest redshifts will have a stronger bulge component than the MW, which results in a slightly higher S\'ersic index.

Over the same redshift range, the median half-mass radius increases from $R_e = 1.7\pm0.2$ kpc at $1.5<z<2$ to $R_e = 4.4\pm0.3$ kpc at $0.3 < z < 0.5$. This effective radius is comparable to the measurement of the MW's effective radius from \cite{Zhou:2023} at $R_{e} = 4 - 4.5$ kpc, assuming an exponential scale length of the disk of 2.7 kpc from \cite{Licquia:2015a}. This is a doubling in the average size of MWA progenitors from $z=2$ to $z=0.3$ over a time scale of $\sim 7$ Gyr. The median S\'ersic index at the lowest redshift bin is $n\sim2.4$, which means that while star-forming MWA progenitors at $z<0.5$ are still disk-dominated, a number of them likely have substantial bulges.

\subsection{Gini-M20 Statistics}\label{sec:gini-m20}

For our sample of galaxies, fitting a single S\'ersic profile to the main galaxy in each cutout has its limitations. S\'ersic profiles can only reveal the most general information about the flux distribution (or in our case the mass distribution) of the galaxy that it is fit to, and fails to capture the finer structures, such as clumps, tails, or spiral arms. Fitting the unique, finer structures of 877 galaxies is unfeasible, but there are statistical methods to quantify the overall ``clumpiness" of a galaxy image. One such method is via Gini-M20 statistics.

\begin{figure*}
    \centering
    \includegraphics[width=\textwidth]{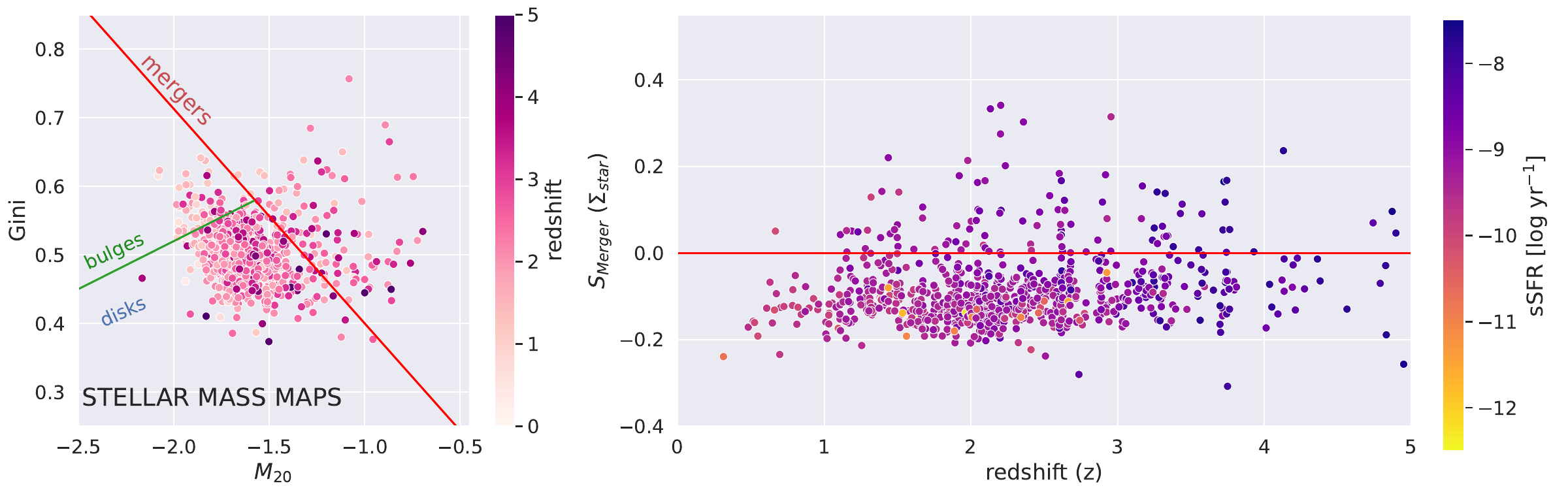}
\caption{\emph{Left Panels:} Gini-M20 plots of the stellar mass distributions of the MWA progenitors. The color of the point represents the redshift of the galaxy. The red line represents the separation between the galaxies with a positive merger statistic, versus the galaxies with a negative merger statistic ($S_{merger}$). The line is defined in Equation \ref{eq:mergerline}. The green line represents the division between bulged-dominated and disk-dominated galaxies.  Although most galaxies high-$z$ galaxies are in the disk region, there are some that reside in the bulge region.
\emph{Right Panels:} The Merger Statistic $S_{merger}$ of the MWA progenitors as a function of redshift, for both the stellar mass maps 
Color represents the sSFR of each galaxy. }
    \label{fig:gini-m20}
\end{figure*}
The Gini-M20 plane \citep{Lotz:2004} is another morphology tool which utilizes the relative flux distribution of galaxy images and the contribution of each pixel to the flux to quantify galaxy morphology. It is exceptionally useful for picking out double peaks which can either indicate pairs of merging galaxies, or star-forming clumps, depending on the filter used.

The Gini coefficient was first used for astronomical image classification by \cite{Abraham:2003}. It is defined for a set of $n$ pixels each with individual flux values $I_i$ as:
\begin{equation}
    G = \frac{1}{\bar{I} n (n-1)} \sum^{n}_{i = 1} (2i - n - 1)I_i ,
\end{equation}
where $\bar{I}$ is the average flux of all the pixels. The Gini coefficient is defined in a way such that the values range from 0 to 1, with $G=0$ assigned to an image with a completely uniform flux, and $G=1$ assigned to a field with a single bright pixel, and all other pixels containing zero flux. One limitation of the Gini coefficient is that the surface brightness of galaxies is highly redshift dependent. So the implementation of the Gini coefficient calculation in \cite{Lotz:2004} and in Statmorph is to construct a Gini segmentation map, which only depends on the Petrosian radius \citep{Petrosian:1976}, and is insensitive to redshift. The Petrosian radius $R_p$ is defined as the radius at which the surface brightness $\mu(r)$ at $R_p$ is equal to the mean surface brightness $\bar{\mu}$ contained within $R_p$, and the ratio of the two values is defined in \cite{Lotz:2004} as:
\begin{equation}
    \eta = \frac{\mu(R_p)}{\bar{\mu}(r < R_p)} 
\end{equation}

The default $\eta$ for the Petrosian radius calculation in Statmorph is set to 0.2.

The $M_{20}$ statistic is the second order moment of a galaxy's brightest pixels, which contains 20\% of the total flux, relative to the total second order central moment $\mu_{tot}$. For an image with $n$ pixels located at positions $(I_i, y_i)$ with $(x_c, y_c)$ defined as the galactic centre, $\mu_{tot}$ is defined as:
\begin{equation}
    \mu_{tot} = \sum_{i=1}^{n} \mu_i \equiv  \sum_{i=1}^{n} I_i\left[ (x_i - x_c)^2 + (y_i - y_c)^2\right],
\end{equation}
where $I_i$ is the flux of an individual pixel. Then, $M_{20}$ is defined as
\begin{equation}
    M_{20} \equiv \log_{10} \left(\frac{\sum_{i} \mu_i}{\mu_{tot}}\right) , \text{while }\sum_{i} I_i< 0.2 I_{tot},
\end{equation}
where $I_{tot}$ is the total flux of the area defined by the segmentation map of the galaxy. Generally, a lower value of $M_{20}$ indicates higher flux concentration. However, unlike the Gini coefficient,  $M_{20}$ is not measured on circular or elliptical apertures, and the centre of a galaxy is left as a free parameter. $M_{20}$ also scales with $r^2$ which makes it more sensitive to detecting multiple spatially resolved bright regions.

Results are presented in the Gini-M20 plane, where different regions of the plane identify roughly the morphology of the galaxy. The Gini-M20 plane for the stellar mass maps of the MWA progenitors are shown in the left half of Figure \ref{fig:gini-m20}. The Gini-M20 plane can also distinguish between disk and bulge-dominated galaxies. The redshift epoch with the highest fraction of bulge-dominated galaxies as determined by Gini-M20 is at $1<z<1.5$, where the bulge morphology fraction is $28\%$. Otherwise, the percentage of bulge-dominated MWA progenitors stays consistent with redshift at $4-8\%$. Overall, the disk/bulge distinction using Gini-M20 agrees well with the trends from S\'ersic profile fitting in Figure \ref{fig:sersic-params}.

One key region of the Gini-M20 plane is the top-right half, where both Gini and $M_{20}$ are high. The galaxies in this region exhibit more than one bright peak in its flux distribution (or in our case, in its stellar mass or SFR distribution), and are thus more likely to be ongoing mergers. However, the interpretation of whether an object in that region is truly a merger is highly dependent on the filter used. Galaxies with satellites, star-forming clumps, and spiral arms can be factors that increase the merger statistic ($S_{merger}$) of a galaxy's image. Since we are using the Gini-M20 plane on a dataset that spans many redshifts, we define a new dividing line from which to measure the merger statistic. This is because the default $S_{merger}$ from Statmorph is based on the data that \cite{Rodriguez-Gomez:2019} were using for their studies, which was comparisons of the Pan-STARRS survey to simulated IllustrisTNG galaxies. We defined a new line for the Gini-M20 plane based on visual inspection that the images classified as having a positive $S_{merger}$ indeed has more than one distinct bright peak in its distribution. The equation that defines the new dividing line is:
\begin{equation}
  S_{merger}(G, M_{20})  =  0.322M_{20} + 1.030G - 0.090 . \label{eq:mergerline}
\end{equation}

\subsection{Mergers and disturbed galaxies at high redshift}\label{sec:asymmetry}

We choose three different proxies for merger rate: $S_{merger}$ from the Gini-M20 plane, CAS asymmetry ($A_{CAS}$), and RMS asymmetry ($A_{RMS}$).  Asymmetry is sensitive to structures such as tidal disturbances, dust lanes, clumpy structures, and double cores. Asymmetry is correlated to mergers, but depending on the stage of the merger, the galaxy may not be highly asymmetric (late stage mergers for instance, do not exhibit great asymmetry). Nevertheless, asymmetry is a useful parameter for detecting disturbances in the process of galaxy evolution, which correlates with merger rate and with galaxy pair interactions.

The most well-known asymmetry parameter comes from the CAS (concentration, asymmetry, smoothness) classification system as described in \citep{Conselice:2003}. Statmorph calculates the asymmetry parameter by subtracting the galaxy's image rotated 180 degrees from its original image: 

\begin{equation}
    A_{CAS} = \frac{\sum_{i,j}|I_{ij} - I_{ij}^{180}|}{\sum_{i,j}|I_{ij}|} - A_{sky} ,
\end{equation}
where $I_{ij}$ and $I_{ij}^{180}$ are pixel flux values of the original and rotated images, and $A_{sky}$ is the average asymmetry of the sky background.

An alternative measurement of asymmetry is RMS asymmetry, which was introduced in \cite{Conselice:2000}.  An updated version of RMS asymmetry from \cite{Sazonova:2024} is currently implemented in the latest version of Statmorph:
\begin{equation}
    A_{RMS}^2 = \frac{\sum_{i,j}(I_{ij} - I_{ij}^{180})^2 - 2\sigma_{sky}^2}{\sum_{i,j}I_{ij}^2 - \sigma_{sky}^2},
\end{equation}
where $\sigma_{sky}$ is the RMS noise of the background. 

\cite{Sazonova:2024} argues that RMS asymmetry is a more robust asymmetry measurement as it is fully independent from noise and aperture size. However, both $A_{CAS}$ and $A_{RMS}$ are correlated to image resolution. Since our study compares galaxies across a wide span of redshifts, it is of the utmost importance that the \textit{physical resolution scale} (i.e. the kpc scale per pixel) is the same. 

\begin{figure*}
    \centering
    \includegraphics[width=\linewidth]{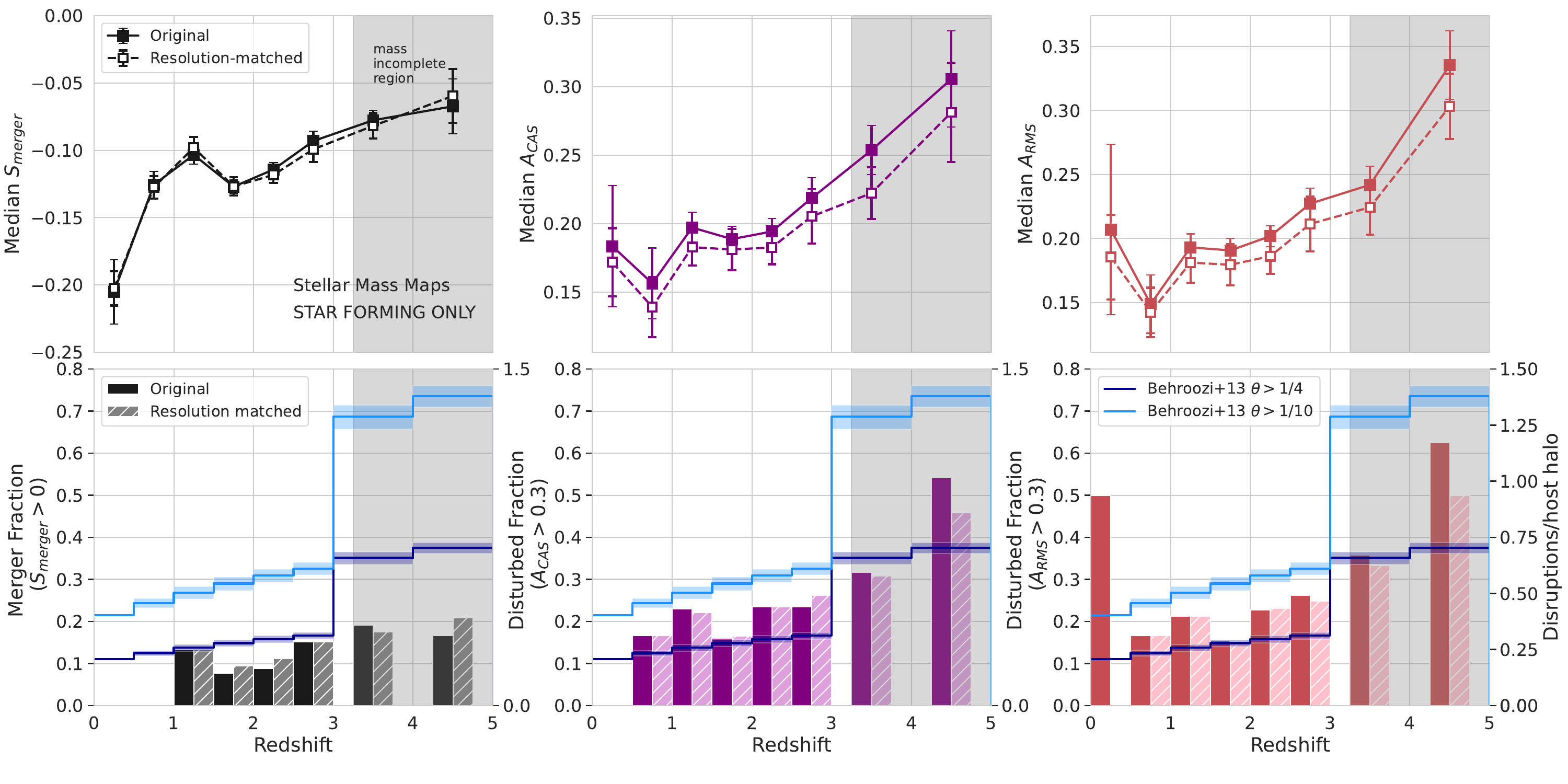}
    \caption{\emph{Top: } From left to right, the median $S_{merger}$, $A_{CAS}$, and $A_{RMS}$ for each redshift bin. In each top panel,  the solid lines with filled squares represent the median of the distribution as measured from the original resolution of the stellar mass maps, and the dashed lines with unfilled squares represent the median from the \textit{resolution matched} stellar mass maps. 
    \emph{Bottom: } Left panel is merger fraction as determined by number of galaxies per redshift that has $S_{merger} > 0$. Center and right panels are ``disturbed" fractions, determined by galaxies with $A_{CAS} > 0.3$ or $A_{RMS} > 0.3$ respectively. In all cases, resolution matching decreases the fraction of disturbed galaxies, but increases the merger fraction at the highest redshift bins. The integrated halo disruption rate from \cite{Behroozi:2013b} is plotted in dark blue for major mergers ($\theta \geq 0.25$) and major and minor mergers ($\theta \geq 0.1$) for comparison.} 
    
    \label{fig:merger-fracs}
\end{figure*}

We compare the merger statistic, CAS asymmetry, and RMS asymmetry for the galaxies' stellar mass maps in both their original resolution (which is the same resolution as the F444W photometry) versus the images that have been matched the same \textit{physical resolution} via the angular diameter distance obtained at each object's redshift. Therefore, the images that have been resolution-matched have one pixel represent the exact same physical extent of the galaxy with the poorest resolution (at $z\sim1.6$, where the angular diameter distance vs. redshift reverses in our cosmology). This resolution is $0.339$ kpc per pixel.

We define the merger fraction morphologically with the stellar mass maps. We use the fraction of galaxies with  $S_{merger} > 0$ in stellar mass to determine the ``merger candidate fraction" \citep{Lotz:2008a}. The merger statistic is a reasonable predictor of ''true" ongoing mergers, as shown from comparisons of the correlation of the value of $S_{merger}$ on mock observation images of galaxy mergers at different stages produced from simulations (i.e. \citealt{Lotz:2008b, Snyder:2015a, Snyder:2015b, Rodriguez-Gomez:2019}). For our purposes, we use merger fraction synonymously with merger candidate fraction, but we note that it is not a true ``merger fraction".

In the top row of Figure \ref{fig:merger-fracs}, we plot the median $S_{merger}$, $A_{CAS}$ and $A_{RMS}$ as a function of redshift, while in the bottom row of the same figure, we plot the merger fractions and disturbed fractions. For all parameters, both the median parameter based on the original stellar mass maps, and the median parameter based on the resolution-matched stellar mass maps, increase with redshift. Though the resolution-matched parameters tend to be slightly lower, with the difference increasing with redshift since angular resolution decreases with redshift up to $z\sim 1.6$ and then increases with redshift for $z > 1.6$). There is a similar trend with the merger fractions (derived from number of galaxies with $S_{merger} > 0$, as seen on the top-right panel of Figure \ref{fig:gini-m20}) and disturbed fractions, derived from either $A_{CAS} > 0.3$ or $A_{RMS}> 0.3$ for each redshift bin. The division line was chosen to be 0.3 because the median $A_{CAS}$ and $A_{RMS}$ are both  $\sim0.3$ at the highest redshift bin. 

The merger fraction on the whole follows a downward trajectory, and the peak $S_{merger}$ is $19.3\pm4.0\%$ at $3 < z < 4$ for the original resolution stellar mass maps, and  $20.8\pm9.3\%$ at $4<z<5$ for the resolution-matched stellar mass maps. This demonstrates that in general, the rate of active, ongoing mergers increases with redshift, and there were more frequent mergers in the past, as we approach the epoch of reionization. The disturbed fraction also increases with redshift, whether the fraction is measured based on $A_{CAS}$ or $A_{RMS}$.  The maximum of the resolution-matched disturbed fraction based on $A_{CAS}$ is $45.8\pm13.8\%$ and the maximum disturbed fraction based on $A_{RMS}$ is $50.0\pm14.4\%$, both found at the highest redshift bin ($4<z<5$).  Although there is a slight upward trend at the lowest redshift bin, the sample size is small, and thus it is still consistent with a decline in asymmetry and disturbed morphology.

For comparison with the implied merger rate from abundance matching with \cite{Behroozi:2013a}, we have obtained the ``subhalo disruption rate per host halo per unit redshift per unit log mass ratio $\log_{10}(\theta)$ ", described in Appendix I of \cite{Behroozi:2013b} which is a run of the same simulation as in \cite{Behroozi:2013a}. We integrate the halo disruption rate over two mass ratios: one at $\theta \geq 1/10$ representing major and minor mergers, and one at $\theta \geq 1/4$, representing major mergers only. We also integrate the function over each redshift step assuming $M_0$ to be the halo mass of the next time step. This becomes the number of disruptions per host halo. We find that the \textit{trend} of the observed merger and disturbed fractions seem overall consistent with the rate of disrupted halos as from \cite{Behroozi:2013b}, however, the merger fraction as defined by positive $S_{merger}$ underestimates the number of mergers at all times, but especially at the two highest redshift bins $3<z<4$ and $4<z<5$. The disturbed fraction using both $A_{CAS} >0.3 $ or $A_{RMS} >0.3$ meanwhile, has slightly better agreement with the major and minor halo disruption rates compared to positive $S_{merger}$.  This shows that our merger and disturbed fractions are more consistent at the redshift ranges where mass completeness is not an issue. Nevertheless, these morphological measurements can still be used to pick out ongoing mergers from observations with relative accuracy compared to simulations, if the sample was selected via abundance matching.

\subsection{Overlap of star-forming regions with stellar mass distribution}\label{sec:sfr-mass-overlap}

\begin{figure*}
\centering
    \begin{subfigure}
       \centering
        \includegraphics[height= 8cm]{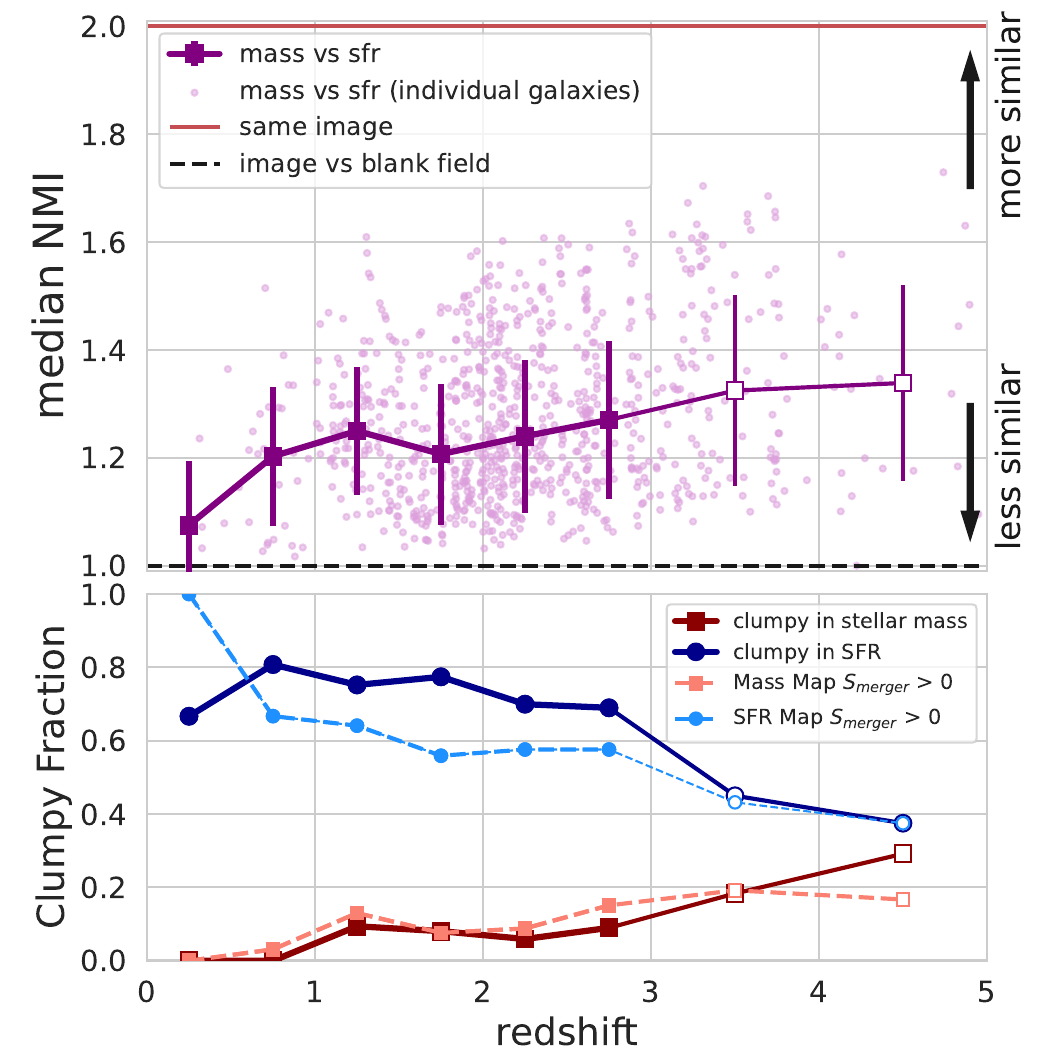}
    \end{subfigure}
    \begin{subfigure}
        \centering
        \includegraphics[height=8cm]{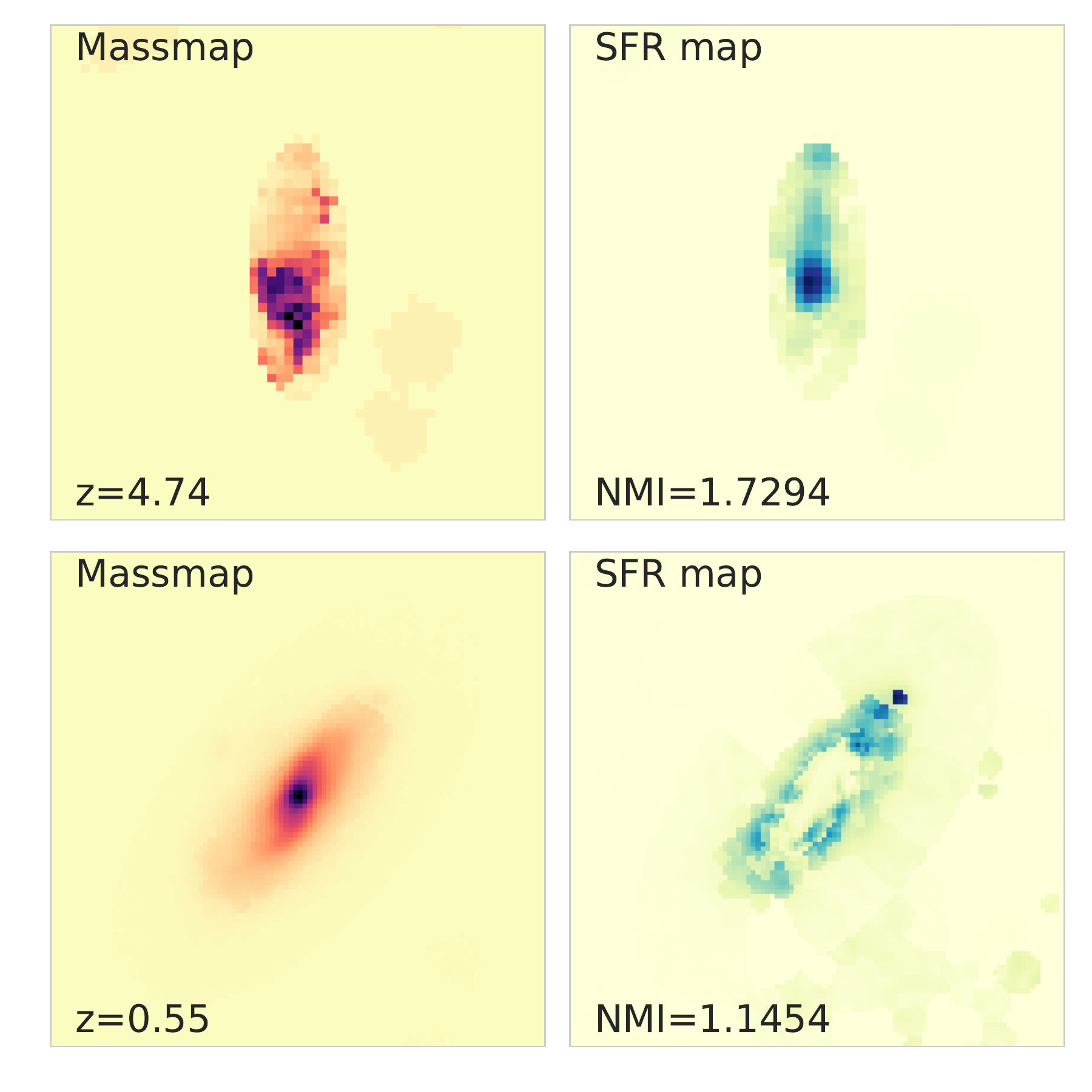}
    \end{subfigure}
    \caption{\textit{Top-Left: }Normalized mutual information (NMI) comparing mass maps to SFR maps for each MWA progenitor (purple points). The median NMI at each redshift bin is plotted as the purple line, with the error bars as $1\sigma$ of the distribution. Red line indicates an NMI of 2, where two images are identical. Dashed black line indicates an NMI of 1, when any non-blank image is compared to a blank field. 
    \textit{Bottom-Left: } Clumpy fraction as a function of redshift, as determined by the clump-finding algorithm from \cite{Sok:2025a} for SFR maps in dark blue, and stellar mass maps in dark red.  The dotted orange line and dotted blue line correspond to $S_{merger}>0$ in mass maps and $S_{merger} > 0$ in SFR maps respectively. 
    \textit{Right Panel:} Example of a high-$z$ galaxy with a high NMI (top) and a low-$z$ galaxy with a low NMI (bottom). The amount of stellar mass covered by the clumpy star-forming regions agree less at lower redshift, resulting in decreasing NMI with decreasing redshift.} \label{fig:img-similarity}
\end{figure*}

As stated earlier in \S \ref{sec:gini-m20},  $S_{merger}$ is, in this case, a measurement of the likelihood that an image contains double or multiple peaks. Therefore, if we use $S_{merger}$ on the SFR maps, then it could be a generalized measurement for whether a galaxy is clumpy in star formation. However, $S_{merger, SFR}$  does not reveal any information about where the clumps are located, or the intensity of the star formation (i.e. the burstiness of the star formation). Due to these limitations, we utilize an alternative method to measure clump fraction for our sample. This is based on the clump detection method outlined in \cite{Wuyts:2012} and \cite{Sok:2022}, which is calibrated to detect clumps in both rest-frame $u$-band and stellar mass maps. In addition, we also measure the similarity between the stellar mass map and the SFR map for each galaxy using the  Normalized Mutual Information index (NMI, \citealt{Studholme:1999}) between the stellar mass maps and SFR maps. The more correlated two images are, the higher the NMI, and the more similar the SFR distribution is to the stellar mass distribution.

In Figure \ref{fig:img-similarity}, we plot both the median NMI between stellar mass and SFR maps in the top-left, and in the bottom left, we plot the clumpiness fractions' for both stellar mass and SFR maps as well as the $S_{merger} > 0$ fraction as a function of redshift. Stellar mass maps are generally clumpier at higher redshift, although there is a slight decreasing trend from $z\sim1$ to $z\sim2$. For the SFR maps however, the clumpiness fraction is lower at higher redshift, though still higher than the clumpiness fraction of stellar mass. Both trends also agree well with the ``clumpiness" fraction derived from the number of galaxies which have $S_{merger} > 0$ in mass maps and SFR maps respectively. 
In \S \ref{sec:mass-growth}, the results show that star formation is much greater at earlier times. As such, we include the NMI measurement to see if the regions where star formation rate is occurring correspond to regions of higher stellar mass. In the median NMI, the correlation between the SFR maps and the stellar mass maps are decreasing as redshift decreases. Furthermore, we show two example galaxies, one at high-$z$ and one at low-$z$, to show that indeed, the SFR maps are more similar to stellar mass maps at earlier times. 

From this, we can infer that at early times, such as the galaxy that is at $z=4.74$ in the top-right figure, regions of bursty star formation correspond more to where the stellar mass is. In that sense, it is not that bursty star formation is less clumpy, but that both SFR \textit{and} stellar mass are \textit{both clumpy} at higher-$z$. But once more stellar mass is set in place, the star-forming regions make up less and less of the galaxy, and eventually only form in the outskirts of the galaxy, which is evidenced by the bottom right images of the galaxy at $z=0.57$, where the SFR, while technically "clumpy", does not correspond to the smooth stellar mass distribution, and is also much lower than at high-$z$.

\section{Discussion}\label{sec:discussion}

\subsection{Star formation, mergers, and morphology}\label{sec:disc-sfr-morphology}

Semi-analytic galaxy simulations (such as \citealt{Benincasa:2020,Grudic:2023,Yu:2023,Pinna:2024,Semenov:2024}) have just begun to explore the question of how exactly the MW's disk formed and evolved, especially as we have evidence that a subset of the stars in the MW's thick disk originated from a major merger that occurred 8-11 Gyr ago known as Gaia-Enceladus (GSE merger). We will consider how the CANUCS sample of observed MWAs compares to previous simulation-based studies of MWA formation.

Although we find strong evidence for inside-out growth as well as disk assembly with our MWA sample, our  mass map derived morphology parameters exhibit both opposing and agreeable trends  to \cite{Yu:2023}. The S\'ersic index of our sample increases with decreasing redshift, showing that our MWAs are becoming more \textit{bulge-dominated} at later times, even after removing quiescent galaxies. In contrast, the circularity of stellar orbits for the FIRE-2 MWAs in \cite{Yu:2023} (which they use as a proxy for disk-like morphology) increases with decreasing redshift, showing their mass distributions are becoming more \textit{disk-dominated} at later times. Galaxies in general become more bulge dominated at lower redshifts (e.g. \citealt{Hashemizadeh:2022}, therefore it is expected that the evolution of MWA progenitors would trend the same way, in spite of the MW itself having a small bulge.

Another simulation-based study with MWAs in TNG50 from \cite{Sotillo-Ramos:2022} tracked both in-situ SFR and ``diskiness" (measured via $D/T$ or disk-to-total mass ratio) over time, and found that increased in-situ SFR is associated with increased diskiness of the MWAs. Their definition of disk is also dependent upon circularity ($\epsilon > 0.7$ for the disks). However, a major merger episode will disrupt diskiness, but still increase in-situ SFR (in a starburst), with the disks reforming afterwards (the orbits of the newly formed stars settling into circular orbits). It can be argued that the generally decreasing trend of the merger and disturbed fractions of the CANUCS MWAs is evidence of this settling down after the starbursts that occurred at $z > 3$. The timescale of the disk reformation according to \cite{Sotillo-Ramos:2022} is anywhere from $<1$ Gyr to $\sim 2$ Gyr. If a redshift bin covers a span of time that is $\gtrsim 1$ Gyr, then it is more likely the median S\'ersic index $n$ in our observed MWA sample will average out all of the morphology of the merging and non-merger galaxies and return a disk-like value close to $n=1$.

\begin{figure}[t]
    \centering
    \includegraphics[width=\columnwidth]{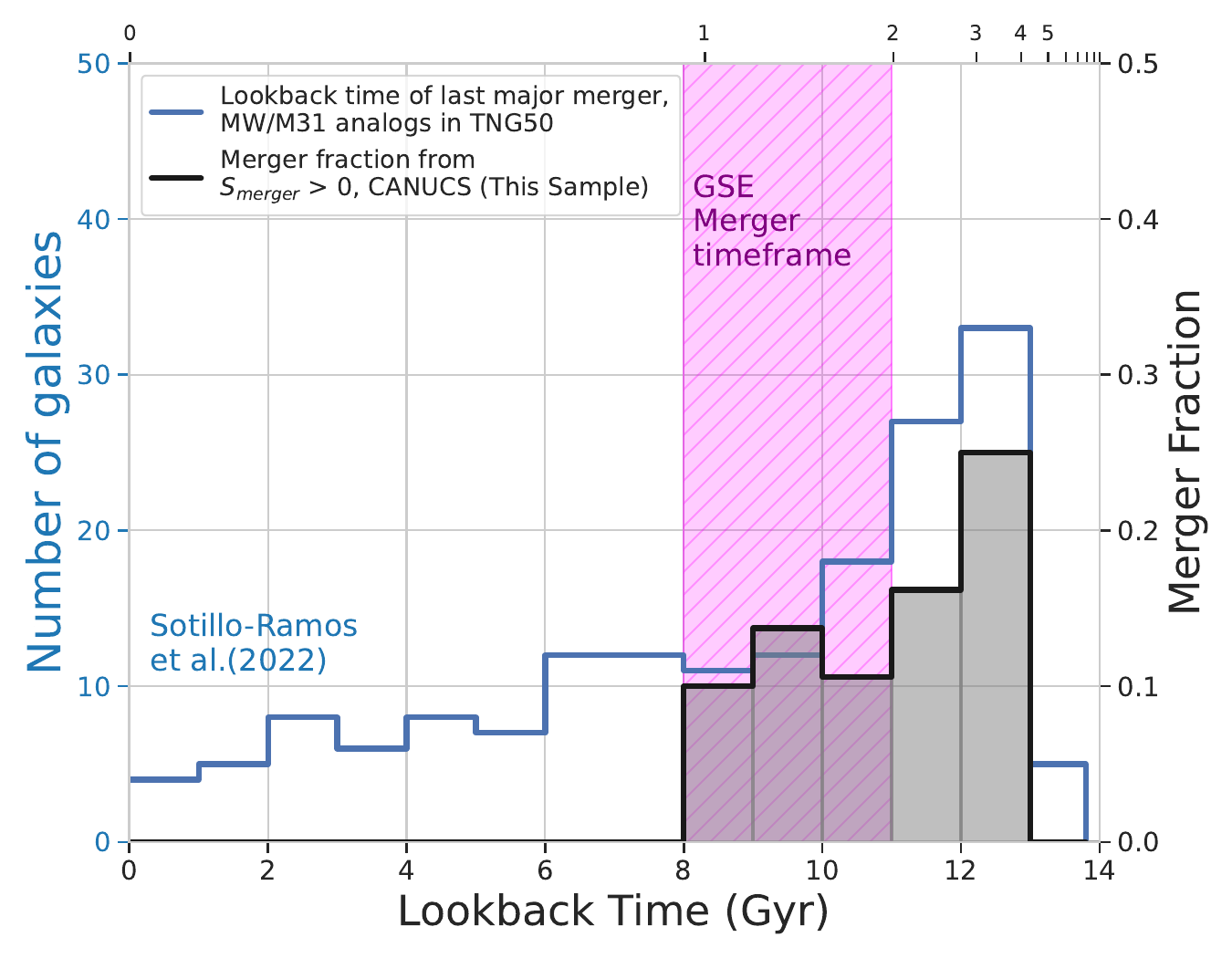}
    \caption{The lookback time of last major merger in TNG50 MWAs (blue line, \citealt{Sotillo-Ramos:2022}) compared to the merger fraction at each lookback time for CANUCS MWAs (black line, grey shading).}
    \label{fig:mergefrac-simcompare}
\end{figure}

On the other hand, $A_{CAS}$, $A_{RMS}$, and $S_{merger}$ of the CANUCS MWA progenitors all show an increasing trend with \textit{increasing} redshift. Asymmetry parameters are correlated with galaxy pair interactions and tidal features, both of which contribute to turbulence and non-circular orbits. Ultimately, S\'ersic profiles are derived from surface brightness, which is not related to orbit circularity. Meanwhile, asymmetry parameters are more closely related to orbital circularity, and thus similar trends appear between observation and simulation.

\subsection{Merger rates over time}



In Figure \ref{fig:mergefrac-simcompare}, we re-bin the merger fraction  defined by $S_{merger} > 0$ of CANCUS MWA progenitors, from the bottom-left panel of Figure \ref{fig:merger-fracs}, by lookback time instead of redshift (black solid line with grey shading). We compare the CANUCS MWA merger fraction with results from TNG50, specifically \cite{Sotillo-Ramos:2022}, as they focused on disk survival rates of MWAs that experienced major mergers. The CANUCS MWA merger fractions agree well with the distribution of last major merger occurrence versus lookback time from the TNG50 MWAs. Note that this is \textit{not} a one-to-one comparison of occurrence rates of mergers. Simulations have access to the full star formation histories of each galaxy, and know when, and how many mergers occurred. Observational results are snapshots of various galaxies at only one point in their evolutionary history. Those galaxies may have had previous mergers, and/or go on to have other major (or minor) mergers as they evolve through cosmic time. But what the agreement between both samples do show is that major mergers occur with more frequency at earlier times/higher redshifts, and that high redshift major mergers \textit{tend to} not alter the disk morphology of  the progenitors of present-day disk galaxies on long time scales. 

This may indicate that systems like the Sunrise arc \citep{Vanzella:2023}, Firefly Sparkle \citep{Mowla:2024}, and Cosmic Gems \citep{Adamo:2024, Bradley:2024} --- glimpses into the highly chaotic formation of the central regions --- can potentially go onto form disky systems at $3<z<5$ that remain disky and become grand design spiral galaxies at $z=0$.


\section{Summary}\label{sec:conclusion}

We used abundance matching on all ten fields of the CANUCS catalogs to find a sample of 877 spatially resolved MWA progenitors since $z = 5$. The stellar mass ranges of these galaxies match well with both simulations and previous observation-based studies of MWA evolution. We then performed resolved SED-fitting with Dense Basis to create stellar mass and SFR maps of our MWA progenitors in order to track the mass assembly, sSFR, and morphology evolution. Our main results are as follows:
\begin{itemize}
\item There is strong evidence of inside-out growth for the selected MWA progenitors before $z =2$. From the mass density profiles (Figures \ref{fig:all_profiles} and \ref{fig:inner-outer}), we find that at $2 < z < 5$, the $\mstar$ in the outer regions ($R > 2$ kpc) grows by 0.8 dex, which is eight times more than the $\mstar$ growth in the inner regions (0.1 dex within $R < 2$ kpc). The maximum sSFR of each stacked and normalized sSFR profile in Figure \ref{fig:all_profiles} from $z\sim4$ to $z \sim 0.5$ is located at or close to the furthest point from the profile's center. The sSFR of the inner regions in Figure \ref{fig:inner-outer} decrease steadily since $z=5$, while the sSFR of the outer regions increase between $2<z<5$, matching the inner regions at $z\sim 3$, and only begin to decrease after cosmic noon ($z \sim 1.75$).
\item Additionally, in Figure \ref{fig:sersic-params}, the median S\'ersic indices of the MWA progenitors demonstrate that up to $z\sim0.5$, these galaxies remain mostly disk-dominated systems. Meanwhile, the change in their median half-mass radius implies that the progenitors are growing rapidly from $z\sim 5$ to $z\sim 3$. The lack of strong evolution in S\'ersic index combined with steady growth in $R_e$ imply the galaxies are building their disks inside-out at earlier times.
\item The merger rates of our MWA progenitors are defined using the position of the galaxy's mass map on the Gini-M20 plane ($S_{merger} > 0$, Figure \ref{fig:gini-m20} ). The most conservative merger fraction at $4<z<5$ is $16.7\pm8.3\%$ (Figure \ref{fig:merger-fracs}). Between $45.8\pm13.8\%$ to $62.5\pm16.1\%$ of MWA progenitors at $4<z<5$ are classified as disturbed, using either CAS or RMS asymmetry. Merger fractions and disturbed fractions have an overall decreasing trend with redshift. The mass growth of early galaxies likely involve frequent interactions and higher merger rates than low-$z$ galaxies.
\item The clumpiness of the stellar mass maps is higher at high-$z$, especially $4 <z <5$, but remains at $<20\%$ at $z < 4$ (Figure \ref{fig:img-similarity}). Although the clumpiness of the SFR maps increases towards lower-$z$, the overall SFR decreases, so the star-forming clumps at lower-z are less bursty than their higher-$z$ counterparts. There is also more mismatch between regions of high SFR and the underlying stellar mass distribution. Highly star-forming regions at higher-$z$ are more likely to be located where there is more stellar mass, whereas the star-forming regions at low-$z$ usually are found in regions with lower mass density. Note however, that this sample of galaxies is affected by mass-incompleteness at $z>3$, and so the $z>3$ sample only represent more massive and more highly star-forming progenitors.

\end{itemize}

MWAs may have evolved from small, disturbed systems with extreme star formation activity to a large disk with pockets of star formation that surround a massive, quiescent bulge. Since both SFR and asymmetry of the CANUCS MWA progenitors all increase as $z$ increases, this matches simulations of MWA evolution where disk galaxies exhibit burstier star formation at earlier times that leads to less circularized/more eccentric and thus more ``asymmetric" stellar orbits. Therefore, the thick disk of MWAs in general may be formed either via bursty episodes of star formation that dynamically heat a less massive disk, or via mergers disrupting the disk or inducing a starburst as the galaxy forms.


\section{Acknowledgments}
The authors wish to thank the anonymous reviewer for their comments and suggestions that greatly improved the manuscript of this paper. This research was supported by grant 18JWST-GTO1 and 23JWGO2A13 from the Canadian Space Agency (CSA), and funding from the Natural Sciences and Engineering Research Council of Canada (NSERC). The MAST DOI for CANUCS is \href{https://doi.org/10.17877/ph4n-6n76}{doi:10.17877/ph4n-6n76}. J. A-D. is supported by the Natural Sciences and Engineering Research Council of Canada (NSERC).  Y.A. is supported by a Research Fellowship for Young Scientists from the Japan Society of the Promotion of Science (JSPS). This research used the Canadian Advanced Network For Astronomy Research (CANFAR) operated in partnership by the Canadian Astronomy Data Centre and The Digital Research Alliance of Canada with support from the National Research Council of Canada the Canadian Space Agency, CANARIE, and the Canadian Foundation for Innovation.

%
\bibliography{paper}

\begin{thebibliography}{}
\expandafter\ifx\csname natexlab\endcsname\relax\def\natexlab#1{#1}\fi
\providecommand{\url}[1]{\href{#1}{#1}}
\providecommand{\dodoi}[1]{doi:~\href{http://doi.org/#1}{\nolinkurl{#1}}}
\providecommand{\doeprint}[1]{\href{http://ascl.net/#1}{\nolinkurl{http://ascl.net/#1}}}
\providecommand{\doarXiv}[1]{\href{https://arxiv.org/abs/#1}{\nolinkurl{https://arxiv.org/abs/#1}}}

\bibitem[{M.~G. {Abadi} {et~al.}(2003){Abadi}, {Navarro}, {Steinmetz}, \& {Eke}}]{Abadi:2003b}
{Abadi}, M.~G., {Navarro}, J.~F., {Steinmetz}, M., \& {Eke}, V.~R. 2003, \bibinfo{title}{{Simulations of Galaxy Formation in a {\ensuremath{\Lambda}} Cold Dark Matter Universe. II. The Fine Structure of Simulated Galactic Disks},} \apj, 597, 21, \dodoi{10.1086/378316}

\bibitem[{R.~G. {Abraham} {et~al.}(2003){Abraham}, {van den Bergh}, \& {Nair}}]{Abraham:2003}
{Abraham}, R.~G., {van den Bergh}, S., \& {Nair}, P. 2003, \bibinfo{title}{{A New Approach to Galaxy Morphology. I. Analysis of the Sloan Digital Sky Survey Early Data Release},} \apj, 588, 218, \dodoi{10.1086/373919}

\bibitem[{A. {Adamo} {et~al.}(2024){Adamo}, {Bradley}, {Vanzella}, {Claeyssens}, {Welch}, {Diego}, {Mahler}, {Oguri}, {Sharon}, {Abdurro'uf}, {Hsiao}, {Xu}, {Messa}, {Lassen}, {Zackrisson}, {Brammer}, {Coe}, {Kokorev}, {Ricotti}, {Zitrin}, {Fujimoto}, {Inoue}, {Resseguier}, {Rigby}, {Jim{\'e}nez-Teja}, {Windhorst}, {Hashimoto}, \& {Tamura}}]{Adamo:2024}
{Adamo}, A., {Bradley}, L.~D., {Vanzella}, E., {et~al.} 2024, \bibinfo{title}{{Bound star clusters observed in a lensed galaxy 460 Myr after the Big Bang},} arXiv e-prints, arXiv:2401.03224, \dodoi{10.48550/arXiv.2401.03224}

\bibitem[{J. {Antwi-Danso} {et~al.}(2023){Antwi-Danso}, {Papovich}, {Leja}, {Marchesini}, {Marsan}, {Martis}, {Labb{\'e}}, {Muzzin}, {Glazebrook}, {Straatman}, \& {Tran}}]{Antwi-Danso:2023}
{Antwi-Danso}, J., {Papovich}, C., {Leja}, J., {et~al.} 2023, \bibinfo{title}{{Beyond UVJ: Color Selection of Galaxies in the JWST Era},} \apj, 943, 166, \dodoi{10.3847/1538-4357/aca294}

\bibitem[{Y. {Asada} {et~al.}(2024{\natexlab{a}}){Asada}, {Sawicki}, {Abraham}, {Brada{\v{c}}}, {Brammer}, {Desprez}, {Estrada-Carpenter}, {Iyer}, {Martis}, {Matharu}, {Mowla}, {Muzzin}, {Noirot}, {Sarrouh}, {Strait}, {Willott}, \& {Harshan}}]{Asada:2024}
{Asada}, Y., {Sawicki}, M., {Abraham}, R., {et~al.} 2024{\natexlab{a}}, \bibinfo{title}{{Bursty star formation and galaxy-galaxy interactions in low-mass galaxies 1 Gyr after the Big Bang},} \mnras, 527, 11372, \dodoi{10.1093/mnras/stad3902}

\bibitem[{Y. {Asada} {et~al.}(2024{\natexlab{b}}){Asada}, {Desprez}, {Willott}, {Sawicki}, {Brada{\v{c}}}, {Brammer}, {Dubath}, {Iyer}, {Martis}, {Muzzin}, {Noirot}, {Paltani}, {Sarrouh}, {Harshan}, \& {Markov}}]{Asada:2024b}
{Asada}, Y., {Desprez}, G., {Willott}, C.~J., {et~al.} 2024{\natexlab{b}}, \bibinfo{title}{{Improving photometric redshifts of Epoch of Reionization galaxies: a new transmission curve with the neutral hydrogen damped Ly$\alpha$ absorption},} arXiv e-prints, arXiv:2410.21543, \dodoi{10.48550/arXiv.2410.21543}

\bibitem[{B. {Barbuy} {et~al.}(2018){Barbuy}, {Chiappini}, \& {Gerhard}}]{Barbuy:2018}
{Barbuy}, B., {Chiappini}, C., \& {Gerhard}, O. 2018, \bibinfo{title}{{Chemodynamical History of the Galactic Bulge},} \araa, 56, 223, \dodoi{10.1146/annurev-astro-081817-051826}

\bibitem[{P.~S. {Behroozi} {et~al.}(2013{\natexlab{a}}){Behroozi}, {Marchesini}, {Wechsler}, {Muzzin}, {Papovich}, \& {Stefanon}}]{Behroozi:2013a}
{Behroozi}, P.~S., {Marchesini}, D., {Wechsler}, R.~H., {et~al.} 2013{\natexlab{a}}, \bibinfo{title}{{Using Cumulative Number Densities to Compare Galaxies across Cosmic Time},} \apjl, 777, L10, \dodoi{10.1088/2041-8205/777/1/L10}

\bibitem[{P.~S. {Behroozi} {et~al.}(2013{\natexlab{b}}){Behroozi}, {Wechsler}, \& {Conroy}}]{Behroozi:2013b}
{Behroozi}, P.~S., {Wechsler}, R.~H., \& {Conroy}, C. 2013{\natexlab{b}}, \bibinfo{title}{{The Average Star Formation Histories of Galaxies in Dark Matter Halos from z = 0-8},} \apj, 770, 57, \dodoi{10.1088/0004-637X/770/1/57}

\bibitem[{S. {Belli} {et~al.}(2019){Belli}, {Newman}, \& {Ellis}}]{Belli:2019}
{Belli}, S., {Newman}, A.~B., \& {Ellis}, R.~S. 2019, \bibinfo{title}{{MOSFIRE Spectroscopy of Quiescent Galaxies at 1.5 < z < 2.5. II. Star Formation Histories and Galaxy Quenching},} \apj, 874, 17, \dodoi{10.3847/1538-4357/ab07af}

\bibitem[{V. {Belokurov} {et~al.}(2018){Belokurov}, {Erkal}, {Evans}, {Koposov}, \& {Deason}}]{Belokurov:2018}
{Belokurov}, V., {Erkal}, D., {Evans}, N.~W., {Koposov}, S.~E., \& {Deason}, A.~J. 2018, \bibinfo{title}{{Co-formation of the disc and the stellar halo},} \mnras, 478, 611, \dodoi{10.1093/mnras/sty982}

\bibitem[{V. {Belokurov} \& A. {Kravtsov}(2022){Belokurov} \& {Kravtsov}}]{Belokurov:2022}
{Belokurov}, V., \& {Kravtsov}, A. 2022, \bibinfo{title}{{From dawn till disc: Milky Way's turbulent youth revealed by the APOGEE+Gaia data},} \mnras, 514, 689, \dodoi{10.1093/mnras/stac1267}

\bibitem[{S.~M. {Benincasa} {et~al.}(2020){Benincasa}, {Loebman}, {Wetzel}, {Hopkins}, {Murray}, {Bellardini}, {Faucher-Gigu{\`e}re}, {Guszejnov}, \& {Orr}}]{Benincasa:2020}
{Benincasa}, S.~M., {Loebman}, S.~R., {Wetzel}, A., {et~al.} 2020, \bibinfo{title}{{Live fast, die young: GMC lifetimes in the FIRE cosmological simulations of Milky Way mass galaxies},} \mnras, 497, 3993, \dodoi{10.1093/mnras/staa2116}

\bibitem[{T. {Bensby} {et~al.}(2003){Bensby}, {Feltzing}, \& {Lundstr{\"o}m}}]{Bensby:2003}
{Bensby}, T., {Feltzing}, S., \& {Lundstr{\"o}m}, I. 2003, \bibinfo{title}{{Elemental abundance trends in the Galactic thin and thick disks as traced by nearby F and G dwarf stars},} \aap, 410, 527, \dodoi{10.1051/0004-6361:20031213}

\bibitem[{E. {Bertin} \& S. {Arnouts}(1996){Bertin} \& {Arnouts}}]{Bertin:1996}
{Bertin}, E., \& {Arnouts}, S. 1996, \bibinfo{title}{{SExtractor: Software for source extraction.},} \aaps, 117, 393, \dodoi{10.1051/aas:1996164}

\bibitem[{L.~A. {Bignone} {et~al.}(2019){Bignone}, {Helmi}, \& {Tissera}}]{Bignone:2019}
{Bignone}, L.~A., {Helmi}, A., \& {Tissera}, P.~B. 2019, \bibinfo{title}{{A Gaia-Enceladus Analog in the EAGLE Simulation: Insights into the Early Evolution of the Milky Way},} \apjl, 883, L5, \dodoi{10.3847/2041-8213/ab3e0e}

\bibitem[{J. {Bland-Hawthorn} \& O. {Gerhard}(2016){Bland-Hawthorn} \& {Gerhard}}]{Bland-Hawthorn:2016}
{Bland-Hawthorn}, J., \& {Gerhard}, O. 2016, \bibinfo{title}{{The Galaxy in Context: Structural, Kinematic, and Integrated Properties},} \araa, 54, 529, \dodoi{10.1146/annurev-astro-081915-023441}

\bibitem[{N. {Boardman} {et~al.}(2020{\natexlab{a}}){Boardman}, {Zasowski}, {Seth}, {Newman}, {Andrews}, {Bershady}, {Bird}, {Chiappini}, {Fielder}, {Fraser-McKelvie}, {Jones}, {Licquia}, {Masters}, {Minchev}, {Schiavon}, {Brownstein}, {Drory}, \& {Lane}}]{Boardman:2020a}
{Boardman}, N., {Zasowski}, G., {Seth}, A., {et~al.} 2020{\natexlab{a}}, \bibinfo{title}{{Milky Way analogues in MaNGA: multiparameter homogeneity and comparison to the Milky Way},} \mnras, 491, 3672, \dodoi{10.1093/mnras/stz3126}

\bibitem[{N. {Boardman} {et~al.}(2020{\natexlab{b}}){Boardman}, {Zasowski}, {Newman}, {Andrews}, {Fielder}, {Bershady}, {Brinkmann}, {Drory}, {Krishnarao}, {Lane}, {Mackereth}, {Masters}, \& {Stringfellow}}]{Boardman:2020b}
{Boardman}, N., {Zasowski}, G., {Newman}, J.~A., {et~al.} 2020{\natexlab{b}}, \bibinfo{title}{{Are the Milky Way and Andromeda unusual? A comparison with Milky Way and Andromeda analogues},} \mnras, 498, 4943, \dodoi{10.1093/mnras/staa2731}

\bibitem[{J. {Bovy} \& H.-W. {Rix}(2013){Bovy} \& {Rix}}]{Bovy:2013}
{Bovy}, J., \& {Rix}, H.-W. 2013, \bibinfo{title}{{A Direct Dynamical Measurement of the Milky Way's Disk Surface Density Profile, Disk Scale Length, and Dark Matter Profile at 4 kpc <\raisebox{-0.5ex}\textasciitilde R <\raisebox{-0.5ex}\textasciitilde 9 kpc},} \apj, 779, 115, \dodoi{10.1088/0004-637X/779/2/115}

\bibitem[{L. {Bradley} {et~al.}(2016){Bradley}, {Sipocz}, {Robitaille}, {Tollerud}, {Deil}, {Vin{\'\i}cius}, {Barbary}, {G{\"u}nther}, {Bostroem}, {Droettboom}, {Bray}, {Bratholm}, {Pickering}, {Craig}, {Pascual}, {Greco}, {Donath}, {Kerzendorf}, {Littlefair}, {Barentsen}, {D'Eugenio}, \& {Weaver}}]{Bradley:2016}
{Bradley}, L., {Sipocz}, B., {Robitaille}, T., {et~al.} 2016, {Photutils: Photometry tools},, Astrophysics Source Code Library, record ascl:1609.011

\bibitem[{L. {Bradley} {et~al.}(2023){Bradley}, {Sip{\H{o}}cz}, {Robitaille}, {Tollerud}, {Vin{\'\i}cius}, {Deil}, {Barbary}, {Wilson}, {Busko}, {Donath}, {G{\"u}nther}, {Cara}, {Lim}, {Me{\ss}linger}, {Conseil}, {Burnett}, {Bostroem}, {Droettboom}, {Bray}, {Andersen Bratholm}, {Jamieson}, {Ginsburg}, {Barentsen}, {Craig}, {Morris}, {Perrin}, {Rathi}, {Pascual}, {Perren}, \& {Georgiev}}]{Bradley:2023}
{Bradley}, L., {Sip{\H{o}}cz}, B., {Robitaille}, T., {et~al.} 2023, {astropy/photutils: 1.10.0}, 1.10.0 Zenodo, \dodoi{10.5281/zenodo.1035865}

\bibitem[{L.~D. {Bradley} {et~al.}(2024){Bradley}, {Adamo}, {Vanzella}, {Sharon}, {Brammer}, {Coe}, {Diego}, {Kokorev}, {Mahler}, {Oguri}, {Abdurro'uf}, {Bhatawdekar}, {Christensen}, {Fujimoto}, {Hashimoto}, {Y. -Y Hsiao}, {Inoue}, {Jim{\'e}nez-Teja}, {Messa}, {Norman}, {Ricotti}, {Tamura}, {Windhorst}, {Xu}, \& {Zitrin}}]{Bradley:2024}
{Bradley}, L.~D., {Adamo}, A., {Vanzella}, E., {et~al.} 2024, \bibinfo{title}{{Unveiling the Cosmic Gems Arc at $z\sim10.2$ with JWST},} arXiv e-prints, arXiv:2404.10770, \dodoi{10.48550/arXiv.2404.10770}

\bibitem[{G.~B. {Brammer} {et~al.}(2008){Brammer}, {van Dokkum}, \& {Coppi}}]{Brammer:2008}
{Brammer}, G.~B., {van Dokkum}, P.~G., \& {Coppi}, P. 2008, \bibinfo{title}{{EAZY: A Fast, Public Photometric Redshift Code},} \apj, 686, 1503, \dodoi{10.1086/591786}

\bibitem[{R. {Brennan} {et~al.}(2015){Brennan}, {Pandya}, {Somerville}, {Barro}, {Taylor}, {Wuyts}, {Bell}, {Dekel}, {Ferguson}, {McIntosh}, {Papovich}, \& {Primack}}]{Brennan:2015}
{Brennan}, R., {Pandya}, V., {Somerville}, R.~S., {et~al.} 2015, \bibinfo{title}{{Quenching and morphological transformation in semi-analytic models and CANDELS},} \mnras, 451, 2933, \dodoi{10.1093/mnras/stv1007}

\bibitem[{T. {Buck} {et~al.}(2020){Buck}, {Obreja}, {Macci{\`o}}, {Minchev}, {Dutton}, \& {Ostriker}}]{Buck:2020}
{Buck}, T., {Obreja}, A., {Macci{\`o}}, A.~V., {et~al.} 2020, \bibinfo{title}{{NIHAO-UHD: the properties of MW-like stellar discs in high-resolution cosmological simulations},} \mnras, 491, 3461, \dodoi{10.1093/mnras/stz3241}

\bibitem[{D. {Calzetti} {et~al.}(2000){Calzetti}, {Armus}, {Bohlin}, {Kinney}, {Koornneef}, \& {Storchi-Bergmann}}]{Calzetti:2000}
{Calzetti}, D., {Armus}, L., {Bohlin}, R.~C., {et~al.} 2000, \bibinfo{title}{{The Dust Content and Opacity of Actively Star-forming Galaxies},} \apj, 533, 682, \dodoi{10.1086/308692}

\bibitem[{M. {Cappellari} \& Y. {Copin}(2003){Cappellari} \& {Copin}}]{Cappellari:2003}
{Cappellari}, M., \& {Copin}, Y. 2003, \bibinfo{title}{{Adaptive spatial binning of integral-field spectroscopic data using Voronoi tessellations},} \mnras, 342, 345, \dodoi{10.1046/j.1365-8711.2003.06541.x}

\bibitem[{G. {Chabrier}(2003){Chabrier}}]{Chabrier:2003}
{Chabrier}, G. 2003, \bibinfo{title}{{Galactic Stellar and Substellar Initial Mass Function},} \pasp, 115, 763, \dodoi{10.1086/376392}

\bibitem[{X. {Cheng} {et~al.}(2024){Cheng}, {Anguiano}, {Majewski}, \& {Arras}}]{Cheng:2024}
{Cheng}, X., {Anguiano}, B., {Majewski}, S.~R., \& {Arras}, P. 2024, \bibinfo{title}{{The surface mass density of the Milky Way: does the traditional K$_{Z}$ approach work in the context of new surveys?},} \mnras, 527, 959, \dodoi{10.1093/mnras/stad3013}

\bibitem[{C. {Conroy} \& J.~E. {Gunn}(2010){Conroy} \& {Gunn}}]{Conroy:2010}
{Conroy}, C., \& {Gunn}, J.~E. 2010, \bibinfo{title}{{The Propagation of Uncertainties in Stellar Population Synthesis Modeling. III. Model Calibration, Comparison, and Evaluation},} \apj, 712, 833, \dodoi{10.1088/0004-637X/712/2/833}

\bibitem[{C.~J. {Conselice}(2003){Conselice}}]{Conselice:2003}
{Conselice}, C.~J. 2003, \bibinfo{title}{{The Relationship between Stellar Light Distributions of Galaxies and Their Formation Histories},} \apjs, 147, 1, \dodoi{10.1086/375001}

\bibitem[{C.~J. {Conselice} {et~al.}(2000){Conselice}, {Bershady}, \& {Jangren}}]{Conselice:2000}
{Conselice}, C.~J., {Bershady}, M.~A., \& {Jangren}, A. 2000, \bibinfo{title}{{The Asymmetry of Galaxies: Physical Morphology for Nearby and High-Redshift Galaxies},} \apj, 529, 886, \dodoi{10.1086/308300}

\bibitem[{C.~J. {Conselice} {et~al.}(2022){Conselice}, {Mundy}, {Ferreira}, \& {Duncan}}]{Conselice:2022}
{Conselice}, C.~J., {Mundy}, C.~J., {Ferreira}, L., \& {Duncan}, K. 2022, \bibinfo{title}{{A Direct Measurement of Galaxy Major and Minor Merger Rates and Stellar Mass Accretion Histories at Z < 3 Using Galaxy Pairs in the REFINE Survey},} \apj, 940, 168, \dodoi{10.3847/1538-4357/ac9b1a}

\bibitem[{L. {Costantin} {et~al.}(2023){Costantin}, {P{\'e}rez-Gonz{\'a}lez}, {Guo}, {Buttitta}, {Jogee}, {Bagley}, {Barro}, {Kartaltepe}, {Koekemoer}, {Cabello}, {Corsini}, {M{\'e}ndez-Abreu}, {de la Vega}, {Iyer}, {Bisigello}, {Cheng}, {Morelli}, {Arrabal Haro}, {Buitrago}, {Cooper}, {Dekel}, {Dickinson}, {Finkelstein}, {Giavalisco}, {Holwerda}, {Huertas-Company}, {Lucas}, {Papovich}, {Pirzkal}, {Seill{\'e}}, {Vega-Ferrero}, {Wuyts}, \& {Yung}}]{Costantin:2023}
{Costantin}, L., {P{\'e}rez-Gonz{\'a}lez}, P.~G., {Guo}, Y., {et~al.} 2023, \bibinfo{title}{{A Milky Way-like barred spiral galaxy at a redshift of 3},} \nat, 623, 499, \dodoi{10.1038/s41586-023-06636-x}

\bibitem[{G. {Desprez} {et~al.}(2024){Desprez}, {Martis}, {Asada}, {Sawicki}, {Willott}, {Muzzin}, {Abraham}, {Brada{\v{c}}}, {Brammer}, {Estrada-Carpenter}, {Iyer}, {Matharu}, {Mowla}, {Noirot}, {Sarrouh}, {Strait}, {Gledhill}, \& {Rihtar{\v{s}}i{\v{c}}}}]{Desprez:2024}
{Desprez}, G., {Martis}, N.~S., {Asada}, Y., {et~al.} 2024, \bibinfo{title}{{{\ensuremath{\Lambda}}CDM not dead yet: massive high-z Balmer break galaxies are less common than previously reported},} \mnras, 530, 2935, \dodoi{10.1093/mnras/stae1084}

\bibitem[{D. {Elia} {et~al.}(2025){Elia}, {Evans}, {Soler}, {Strafella}, {Schisano}, {Molinari}, {Giannetti}, \& {Patra}}]{Elia:2025}
{Elia}, D., {Evans}, N.~J., {Soler}, J.~D., {et~al.} 2025, \bibinfo{title}{{Measuring Star Formation Rates in the Milky Way from Hi-GAL 70 {\ensuremath{\mu}}m Observations},} \apj, 980, 216, \dodoi{10.3847/1538-4357/adaeb2}

\bibitem[{C. {Engler} {et~al.}(2023){Engler}, {Pillepich}, {Joshi}, {Pasquali}, {Nelson}, \& {Grebel}}]{Engler:2023}
{Engler}, C., {Pillepich}, A., {Joshi}, G.~D., {et~al.} 2023, \bibinfo{title}{{Satellites of Milky Way- and M31-like galaxies with TNG50: quenched fractions, gas content, and star formation histories},} \mnras, 522, 5946, \dodoi{10.1093/mnras/stad1357}

\bibitem[{V. {Estrada-Carpenter} {et~al.}(2024){Estrada-Carpenter}, {Sawicki}, {Brammer}, {Desprez}, {Abraham}, {Asada}, {Brada{\v{c}}}, {Iyer}, {Martis}, {Matharu}, {Mowla}, {Muzzin}, {Noirot}, {Sarrouh}, {Strait}, \& {Willott}}]{Estrada-Carpenter:2024}
{Estrada-Carpenter}, V., {Sawicki}, M., {Brammer}, G., {et~al.} 2024, \bibinfo{title}{{When, where, and how star formation happens in a galaxy pair at cosmic noon using CANUCS JWST/NIRISS grism spectroscopy},} \mnras, 532, 577, \dodoi{10.1093/mnras/stae1368}

\bibitem[{A. {Fraser-McKelvie} {et~al.}(2019){Fraser-McKelvie}, {Merrifield}, \& {Arag{\'o}n-Salamanca}}]{Fraser-McKelvie:2019}
{Fraser-McKelvie}, A., {Merrifield}, M., \& {Arag{\'o}n-Salamanca}, A. 2019, \bibinfo{title}{{From the outside looking in: what can Milky Way analogues tell us about the star formation rate of our own galaxy?},} \mnras, 489, 5030, \dodoi{10.1093/mnras/stz2493}

\bibitem[{M.~F. {Fuentealba-Fuentes} {et~al.}(2025){Fuentealba-Fuentes}, {Davies}, {Robotham}, {Cook}, {Bellstedt}, {Lagos}, {Bravo}, \& {Siudek}}]{Fuentealba-Fuentes:2025}
{Fuentealba-Fuentes}, M.~F., {Davies}, L. J.~M., {Robotham}, A. S.~G., {et~al.} 2025, \bibinfo{title}{{Deep Extragalactic VIsible Legacy Survey (DEVILS): new robust merger rates at intermediate redshifts},} \mnras, 539, 1651, \dodoi{10.1093/mnras/staf596}

\bibitem[{K. {Fuhrmann}(2011){Fuhrmann}}]{Fuhrmann:2011}
{Fuhrmann}, K. 2011, \bibinfo{title}{{Nearby stars of the Galactic disc and halo - V},} \mnras, 414, 2893, \dodoi{10.1111/j.1365-2966.2011.18476.x}

\bibitem[{S. {Garrison-Kimmel} {et~al.}(2018){Garrison-Kimmel}, {Hopkins}, {Wetzel}, {El-Badry}, {Sanderson}, {Bullock}, {Ma}, {van de Voort}, {Hafen}, {Faucher-Gigu{\`e}re}, {Hayward}, {Quataert}, {Kere{\v{s}}}, \& {Boylan-Kolchin}}]{Garrison-Kimmel:2018}
{Garrison-Kimmel}, S., {Hopkins}, P.~F., {Wetzel}, A., {et~al.} 2018, \bibinfo{title}{{The origin of the diverse morphologies and kinematics of Milky Way-mass galaxies in the FIRE-2 simulations},} \mnras, 481, 4133, \dodoi{10.1093/mnras/sty2513}

\bibitem[{C. {Gim{\'e}nez-Arteaga} {et~al.}(2023){Gim{\'e}nez-Arteaga}, {Oesch}, {Brammer}, {Valentino}, {Mason}, {Weibel}, {Barrufet}, {Fujimoto}, {Heintz}, {Nelson}, {Strait}, {Suess}, \& {Gibson}}]{Gimenez-Arteaga:2023}
{Gim{\'e}nez-Arteaga}, C., {Oesch}, P.~A., {Brammer}, G.~B., {et~al.} 2023, \bibinfo{title}{{Spatially Resolved Properties of Galaxies at 5 < z < 9 in the SMACS 0723 JWST ERO Field},} \apj, 948, 126, \dodoi{10.3847/1538-4357/acc5ea}

\bibitem[{R. {Gledhill} {et~al.}(2024){Gledhill}, {Strait}, {Desprez}, {Rihtar{\v{s}}i{\v{c}}}, {Brada{\v{c}}}, {Brammer}, {Willott}, {Martis}, {Sawicki}, {Noirot}, {Sarrouh}, \& {Muzzin}}]{Gledhill:2024}
{Gledhill}, R., {Strait}, V., {Desprez}, G., {et~al.} 2024, \bibinfo{title}{{CANUCS: An Updated Mass and Magnification Model of Abell 370 with JWST},} arXiv e-prints, arXiv:2403.07062, \dodoi{10.48550/arXiv.2403.07062}

\bibitem[{D. {Goddard} {et~al.}(2017){Goddard}, {Thomas}, {Maraston}, {Westfall}, {Etherington}, {Riffel}, {Mallmann}, {Zheng}, {Argudo-Fern{\'a}ndez}, {Lian}, {Bershady}, {Bundy}, {Drory}, {Law}, {Yan}, {Wake}, {Weijmans}, {Bizyaev}, {Brownstein}, {Lane}, {Maiolino}, {Masters}, {Merrifield}, {Nitschelm}, {Pan}, {Roman-Lopes}, {Storchi-Bergmann}, \& {Schneider}}]{Goddard:2017}
{Goddard}, D., {Thomas}, D., {Maraston}, C., {et~al.} 2017, \bibinfo{title}{{SDSS-IV MaNGA: Spatially resolved star formation histories in galaxies as a function of galaxy mass and type},} \mnras, 466, 4731, \dodoi{10.1093/mnras/stw3371}

\bibitem[{R.~J.~J. {Grand} {et~al.}(2017){Grand}, {G{\'o}mez}, {Marinacci}, {Pakmor}, {Springel}, {Campbell}, {Frenk}, {Jenkins}, \& {White}}]{Grand:2017}
{Grand}, R. J.~J., {G{\'o}mez}, F.~A., {Marinacci}, F., {et~al.} 2017, \bibinfo{title}{{The Auriga Project: the properties and formation mechanisms of disc galaxies across cosmic time},} \mnras, 467, 179, \dodoi{10.1093/mnras/stx071}

\bibitem[{A. {Grazian} {et~al.}(2015){Grazian}, {Fontana}, {Santini}, {Dunlop}, {Ferguson}, {Castellano}, {Amorin}, {Ashby}, {Barro}, {Behroozi}, {Boutsia}, {Caputi}, {Chary}, {Dekel}, {Dickinson}, {Faber}, {Fazio}, {Finkelstein}, {Galametz}, {Giallongo}, {Giavalisco}, {Grogin}, {Guo}, {Kocevski}, {Koekemoer}, {Koo}, {Lee}, {Lu}, {Merlin}, {Mobasher}, {Nonino}, {Papovich}, {Paris}, {Pentericci}, {Reddy}, {Renzini}, {Salmon}, {Salvato}, {Sommariva}, {Song}, \& {Vanzella}}]{Grazian:2015}
{Grazian}, A., {Fontana}, A., {Santini}, P., {et~al.} 2015, \bibinfo{title}{{The galaxy stellar mass function at 3.5 {\ensuremath{\leq}}z {\ensuremath{\leq}} 7.5 in the CANDELS/UDS, GOODS-South, and HUDF fields},} \aap, 575, A96, \dodoi{10.1051/0004-6361/201424750}

\bibitem[{M.~Y. {Grudi{\'c}} {et~al.}(2023){Grudi{\'c}}, {Hafen}, {Rodriguez}, {Guszejnov}, {Lamberts}, {Wetzel}, {Boylan-Kolchin}, \& {Faucher-Gigu{\`e}re}}]{Grudic:2023}
{Grudi{\'c}}, M.~Y., {Hafen}, Z., {Rodriguez}, C.~L., {et~al.} 2023, \bibinfo{title}{{Great balls of FIRE - I. The formation of star clusters across cosmic time in a Milky Way-mass galaxy},} \mnras, 519, 1366, \dodoi{10.1093/mnras/stac3573}

\bibitem[{J. {Guedes} {et~al.}(2011){Guedes}, {Callegari}, {Madau}, \& {Mayer}}]{Guedes:2011}
{Guedes}, J., {Callegari}, S., {Madau}, P., \& {Mayer}, L. 2011, \bibinfo{title}{{Forming Realistic Late-type Spirals in a {\ensuremath{\Lambda}}CDM Universe: The Eris Simulation},} \apj, 742, 76, \dodoi{10.1088/0004-637X/742/2/76}

\bibitem[{A. {Hashemizadeh} {et~al.}(2022){Hashemizadeh}, {Driver}, {Davies}, {Robotham}, {Bellstedt}, {Foster}, {Holwerda}, {Jarvis}, {Phillipps}, {Siudek}, {Thorne}, {Windhorst}, \& {Wolf}}]{Hashemizadeh:2022}
{Hashemizadeh}, A., {Driver}, S.~P., {Davies}, L. J.~M., {et~al.} 2022, \bibinfo{title}{{Deep extragalactic visible legacy survey (DEVILS): the emergence of bulges and decline of disc growth since z = 1},} \mnras, 515, 1175, \dodoi{10.1093/mnras/stac1195}

\bibitem[{A. {Helmi}(2020){Helmi}}]{Helmi:2020}
{Helmi}, A. 2020, \bibinfo{title}{{Streams, Substructures, and the Early History of the Milky Way},} \araa, 58, 205, \dodoi{10.1146/annurev-astro-032620-021917}

\bibitem[{A. {Helmi} {et~al.}(2018){Helmi}, {Babusiaux}, {Koppelman}, {Massari}, {Veljanoski}, \& {Brown}}]{Helmi:2018}
{Helmi}, A., {Babusiaux}, C., {Koppelman}, H.~H., {et~al.} 2018, \bibinfo{title}{{The merger that led to the formation of the Milky Way's inner stellar halo and thick disk},} \nat, 563, 85, \dodoi{10.1038/s41586-018-0625-x}

\bibitem[{P.~F. {Hopkins}(2015){Hopkins}}]{Hopkins:2015}
{Hopkins}, P.~F. 2015, \bibinfo{title}{{A new class of accurate, mesh-free hydrodynamic simulation methods},} \mnras, 450, 53, \dodoi{10.1093/mnras/stv195}

\bibitem[{P.~F. {Hopkins} {et~al.}(2018){Hopkins}, {Wetzel}, {Kere{\v{s}}}, {Faucher-Gigu{\`e}re}, {Quataert}, {Boylan-Kolchin}, {Murray}, {Hayward}, {Garrison-Kimmel}, {Hummels}, {Feldmann}, {Torrey}, {Ma}, {Angl{\'e}s-Alc{\'a}zar}, {Su}, {Orr}, {Schmitz}, {Escala}, {Sanderson}, {Grudi{\'c}}, {Hafen}, {Kim}, {Fitts}, {Bullock}, {Wheeler}, {Chan}, {Elbert}, \& {Narayanan}}]{Hopkins:2018}
{Hopkins}, P.~F., {Wetzel}, A., {Kere{\v{s}}}, D., {et~al.} 2018, \bibinfo{title}{{FIRE-2 simulations: physics versus numerics in galaxy formation},} \mnras, 480, 800, \dodoi{10.1093/mnras/sty1690}

\bibitem[{D. {Horta} \& R.~P. {Schiavon}(2024){Horta} \& {Schiavon}}]{Horta:2024}
{Horta}, D., \& {Schiavon}, R.~P. 2024, \bibinfo{title}{{On the mass assembly history of the Milky Way: clues from its stellar halo},} arXiv e-prints, arXiv:2404.16939, \dodoi{10.48550/arXiv.2404.16939}

\bibitem[{K. {Iyer} \& E. {Gawiser}(2017){Iyer} \& {Gawiser}}]{Iyer:2017}
{Iyer}, K., \& {Gawiser}, E. 2017, \bibinfo{title}{{Reconstruction of Galaxy Star Formation Histories through SED Fitting:The Dense Basis Approach},} \apj, 838, 127, \dodoi{10.3847/1538-4357/aa63f0}

\bibitem[{K.~G. {Iyer} {et~al.}(2019){Iyer}, {Gawiser}, {Faber}, {Ferguson}, {Kartaltepe}, {Koekemoer}, {Pacifici}, \& {Somerville}}]{Iyer:2019}
{Iyer}, K.~G., {Gawiser}, E., {Faber}, S.~M., {et~al.} 2019, \bibinfo{title}{{Nonparametric Star Formation History Reconstruction with Gaussian Processes. I. Counting Major Episodes of Star Formation},} \apj, 879, 116, \dodoi{10.3847/1538-4357/ab2052}

\bibitem[{M. {Kilic} {et~al.}(2017){Kilic}, {Munn}, {Harris}, {von Hippel}, {Liebert}, {Williams}, {Jeffery}, \& {DeGennaro}}]{Kilic:2017}
{Kilic}, M., {Munn}, J.~A., {Harris}, H.~C., {et~al.} 2017, \bibinfo{title}{{The Ages of the Thin Disk, Thick Disk, and the Halo from Nearby White Dwarfs},} \apj, 837, 162, \dodoi{10.3847/1538-4357/aa62a5}

\bibitem[{J. {Kormendy} \& R. {Bender}(2019){Kormendy} \& {Bender}}]{Kormendy:2019}
{Kormendy}, J., \& {Bender}, R. 2019, \bibinfo{title}{{Structural Analogs of the Milky Way Galaxy: Stellar Populations in the Boxy Bulges of NGC 4565 and NGC 5746},} \apj, 872, 106, \dodoi{10.3847/1538-4357/aafdff}

\bibitem[{Z.~A. {Le Conte} {et~al.}(2024){Le Conte}, {Gadotti}, {Ferreira}, {Conselice}, {de S{\'a}-Freitas}, {Kim}, {Neumann}, {Fragkoudi}, {Athanassoula}, \& {Adams}}]{LeConte:2024}
{Le Conte}, Z.~A., {Gadotti}, D.~A., {Ferreira}, L., {et~al.} 2024, \bibinfo{title}{{A JWST investigation into the bar fraction at redshifts 1 {\ensuremath{\leq}} z {\ensuremath{\leq}} 3},} \mnras, 530, 1984, \dodoi{10.1093/mnras/stae921}

\bibitem[{T.~C. {Licquia} \& J.~A. {Newman}(2015){Licquia} \& {Newman}}]{Licquia:2015a}
{Licquia}, T.~C., \& {Newman}, J.~A. 2015, \bibinfo{title}{{Improved Estimates of the Milky Way's Stellar Mass and Star Formation Rate from Hierarchical Bayesian Meta-Analysis},} \apj, 806, 96, \dodoi{10.1088/0004-637X/806/1/96}

\bibitem[{T.~C. {Licquia} {et~al.}(2016){Licquia}, {Newman}, \& {Bershady}}]{Licquia:2016}
{Licquia}, T.~C., {Newman}, J.~A., \& {Bershady}, M.~A. 2016, \bibinfo{title}{{Does the Milky Way Obey Spiral Galaxy Scaling Relations?},} \apj, 833, 220, \dodoi{10.3847/1538-4357/833/2/220}

\bibitem[{T.~C. {Licquia} {et~al.}(2015){Licquia}, {Newman}, \& {Brinchmann}}]{Licquia:2015b}
{Licquia}, T.~C., {Newman}, J.~A., \& {Brinchmann}, J. 2015, \bibinfo{title}{{Unveiling the Milky Way: A New Technique for Determining the Optical Color and Luminosity of Our Galaxy},} \apj, 809, 96, \dodoi{10.1088/0004-637X/809/1/96}

\bibitem[{J.~M. {Lotz} {et~al.}(2008{\natexlab{a}}){Lotz}, {Jonsson}, {Cox}, \& {Primack}}]{Lotz:2008b}
{Lotz}, J.~M., {Jonsson}, P., {Cox}, T.~J., \& {Primack}, J.~R. 2008{\natexlab{a}}, \bibinfo{title}{{Galaxy merger morphologies and time-scales from simulations of equal-mass gas-rich disc mergers},} \mnras, 391, 1137, \dodoi{10.1111/j.1365-2966.2008.14004.x}

\bibitem[{J.~M. {Lotz} {et~al.}(2004){Lotz}, {Primack}, \& {Madau}}]{Lotz:2004}
{Lotz}, J.~M., {Primack}, J., \& {Madau}, P. 2004, \bibinfo{title}{{A New Nonparametric Approach to Galaxy Morphological Classification},} \aj, 128, 163, \dodoi{10.1086/421849}

\bibitem[{J.~M. {Lotz} {et~al.}(2008{\natexlab{b}}){Lotz}, {Davis}, {Faber}, {Guhathakurta}, {Gwyn}, {Huang}, {Koo}, {Le Floc'h}, {Lin}, {Newman}, {Noeske}, {Papovich}, {Willmer}, {Coil}, {Conselice}, {Cooper}, {Hopkins}, {Metevier}, {Primack}, {Rieke}, \& {Weiner}}]{Lotz:2008a}
{Lotz}, J.~M., {Davis}, M., {Faber}, S.~M., {et~al.} 2008{\natexlab{b}}, \bibinfo{title}{{The Evolution of Galaxy Mergers and Morphology at z < 1.2 in the Extended Groth Strip},} \apj, 672, 177, \dodoi{10.1086/523659}

\bibitem[{J.~M. {Lotz} {et~al.}(2017){Lotz}, {Koekemoer}, {Coe}, {Grogin}, {Capak}, {Mack}, {Anderson}, {Avila}, {Barker}, {Borncamp}, {Brammer}, {Durbin}, {Gunning}, {Hilbert}, {Jenkner}, {Khandrika}, {Levay}, {Lucas}, {MacKenty}, {Ogaz}, {Porterfield}, {Reid}, {Robberto}, {Royle}, {Smith}, {Storrie-Lombardi}, {Sunnquist}, {Surace}, {Taylor}, {Williams}, {Bullock}, {Dickinson}, {Finkelstein}, {Natarajan}, {Richard}, {Robertson}, {Tumlinson}, {Zitrin}, {Flanagan}, {Sembach}, {Soifer}, \& {Mountain}}]{Lotz:2017}
{Lotz}, J.~M., {Koekemoer}, A., {Coe}, D., {et~al.} 2017, \bibinfo{title}{{The Frontier Fields: Survey Design and Initial Results},} \apj, 837, 97, \dodoi{10.3847/1538-4357/837/1/97}

\bibitem[{D.~J. {McLeod} {et~al.}(2021){McLeod}, {McLure}, {Dunlop}, {Cullen}, {Carnall}, \& {Duncan}}]{McLeod:2021}
{McLeod}, D.~J., {McLure}, R.~J., {Dunlop}, J.~S., {et~al.} 2021, \bibinfo{title}{{The evolution of the galaxy stellar-mass function over the last 12 billion years from a combination of ground-based and HST surveys},} \mnras, 503, 4413, \dodoi{10.1093/mnras/stab731}

\bibitem[{J. {Moreno} {et~al.}(2021){Moreno}, {Torrey}, {Ellison}, {Patton}, {Bottrell}, {Bluck}, {Hani}, {Hayward}, {Bullock}, {Hopkins}, \& {Hernquist}}]{Moreno:2021}
{Moreno}, J., {Torrey}, P., {Ellison}, S.~L., {et~al.} 2021, \bibinfo{title}{{Spatially resolved star formation and fuelling in galaxy interactions},} \mnras, 503, 3113, \dodoi{10.1093/mnras/staa2952}

\bibitem[{B.~P. {Moster} {et~al.}(2018){Moster}, {Naab}, \& {White}}]{Moster:2018}
{Moster}, B.~P., {Naab}, T., \& {White}, S. D.~M. 2018, \bibinfo{title}{{EMERGE - an empirical model for the formation of galaxies since z {\ensuremath{\sim}} 10},} \mnras, 477, 1822, \dodoi{10.1093/mnras/sty655}

\bibitem[{L. {Mowla} {et~al.}(2024){Mowla}, {Iyer}, {Asada}, {Desprez}, {Tan}, {Martis}, {Sarrouh}, {Strait}, {Abraham}, {Brada{\v{c}}}, {Brammer}, {Muzzin}, {Pacifici}, {Ravindranath}, {Sawicki}, {Willott}, {Estrada-Carpenter}, {Jahan}, {Noirot}, {Matharu}, {Rihtar{\v{s}}i{\v{c}}}, \& {Zabl}}]{Mowla:2024}
{Mowla}, L., {Iyer}, K., {Asada}, Y., {et~al.} 2024, \bibinfo{title}{{The Firefly Sparkle: The Earliest Stages of the Assembly of A Milky Way-type Galaxy in a 600 Myr Old Universe},} arXiv e-prints, arXiv:2402.08696, \dodoi{10.48550/arXiv.2402.08696}

\bibitem[{J.~C. {Mu{\~n}oz-Mateos} {et~al.}(2007){Mu{\~n}oz-Mateos}, {Gil de Paz}, {Boissier}, {Zamorano}, {Jarrett}, {Gallego}, \& {Madore}}]{Munoz-Mateos:2007}
{Mu{\~n}oz-Mateos}, J.~C., {Gil de Paz}, A., {Boissier}, S., {et~al.} 2007, \bibinfo{title}{{Specific Star Formation Rate Profiles in Nearby Spiral Galaxies: Quantifying the Inside-Out Formation of Disks},} \apj, 658, 1006, \dodoi{10.1086/511812}

\bibitem[{S.~J. {Mutch} {et~al.}(2011){Mutch}, {Croton}, \& {Poole}}]{Mutch:2011}
{Mutch}, S.~J., {Croton}, D.~J., \& {Poole}, G.~B. 2011, \bibinfo{title}{{The Mid-life Crisis of the Milky Way and M31},} \apj, 736, 84, \dodoi{10.1088/0004-637X/736/2/84}

\bibitem[{A. {Muzzin} {et~al.}(2013){Muzzin}, {Marchesini}, {Stefanon}, {Franx}, {McCracken}, {Milvang-Jensen}, {Dunlop}, {Fynbo}, {Brammer}, {Labb{\'e}}, \& {van Dokkum}}]{Muzzin:2013b}
{Muzzin}, A., {Marchesini}, D., {Stefanon}, M., {et~al.} 2013, \bibinfo{title}{{The Evolution of the Stellar Mass Functions of Star-forming and Quiescent Galaxies to z = 4 from the COSMOS/UltraVISTA Survey},} \apj, 777, 18, \dodoi{10.1088/0004-637X/777/1/18}

\bibitem[{G. {Noirot} {et~al.}(2023){Noirot}, {Desprez}, {Asada}, {Sawicki}, {Estrada-Carpenter}, {Martis}, {Sarrouh}, {Strait}, {Abraham}, {Brada{\v{c}}}, {Brammer}, {Iyer}, {MacFarland}, {Matharu}, {Mowla}, {Muzzin}, {Pacifici}, {Ravindranath}, {Willott}, {Albert}, {Doyon}, {Hutchings}, \& {Rowlands}}]{Noirot:2023}
{Noirot}, G., {Desprez}, G., {Asada}, Y., {et~al.} 2023, \bibinfo{title}{{The first large catalogue of spectroscopic redshifts in Webb's first deep field, SMACS J0723.3-7327},} \mnras, 525, 1867, \dodoi{10.1093/mnras/stad1019}

\bibitem[{J. {O{\~n}orbe} {et~al.}(2015){O{\~n}orbe}, {Boylan-Kolchin}, {Bullock}, {Hopkins}, {Kere{\v{s}}}, {Faucher-Gigu{\`e}re}, {Quataert}, \& {Murray}}]{Onorbe:2015}
{O{\~n}orbe}, J., {Boylan-Kolchin}, M., {Bullock}, J.~S., {et~al.} 2015, \bibinfo{title}{{Forged in FIRE: cusps, cores and baryons in low-mass dwarf galaxies},} \mnras, 454, 2092, \dodoi{10.1093/mnras/stv2072}

\bibitem[{C. {Papovich} {et~al.}(2015){Papovich}, {Labb{\'e}}, {Quadri}, {Tilvi}, {Behroozi}, {Bell}, {Glazebrook}, {Spitler}, {Straatman}, {Tran}, {Cowley}, {Dav{\'e}}, {Dekel}, {Dickinson}, {Ferguson}, {Finkelstein}, {Gawiser}, {Inami}, {Faber}, {Kacprzak}, {Kawinwanichakij}, {Kocevski}, {Koekemoer}, {Koo}, {Kurczynski}, {Lotz}, {Lu}, {Lucas}, {McIntosh}, {Mehrtens}, {Mobasher}, {Monson}, {Morrison}, {Nanayakkara}, {Persson}, {Salmon}, {Simons}, {Tomczak}, {van Dokkum}, {Weiner}, \& {Willner}}]{Papovich:2015}
{Papovich}, C., {Labb{\'e}}, I., {Quadri}, R., {et~al.} 2015, \bibinfo{title}{{ZFOURGE/CANDELS: On the Evolution of M* Galaxy Progenitors from z = 3 to 0.5},} \apj, 803, 26, \dodoi{10.1088/0004-637X/803/1/26}

\bibitem[{S.~G. {Patel} {et~al.}(2013){Patel}, {Fumagalli}, {Franx}, {van Dokkum}, {van der Wel}, {Leja}, {Labb{\'e}}, {Brammer}, {Skelton}, {Momcheva}, {Whitaker}, {Lundgren}, {Muzzin}, {Quadri}, {Nelson}, {Wake}, \& {Rix}}]{Patel:2013}
{Patel}, S.~G., {Fumagalli}, M., {Franx}, M., {et~al.} 2013, \bibinfo{title}{{The Structural Evolution of Milky-Way-like Star-forming Galaxies since z \raisebox{-0.5ex}\textasciitilde 1.3},} \apj, 778, 115, \dodoi{10.1088/0004-637X/778/2/115}

\bibitem[{C.~Y. {Peng} {et~al.}(2010){Peng}, {Ho}, {Impey}, \& {Rix}}]{Peng:2010}
{Peng}, C.~Y., {Ho}, L.~C., {Impey}, C.~D., \& {Rix}, H.-W. 2010, \bibinfo{title}{{Detailed Decomposition of Galaxy Images. II. Beyond Axisymmetric Models},} \aj, 139, 2097, \dodoi{10.1088/0004-6256/139/6/2097}

\bibitem[{V. {Petrosian}(1976){Petrosian}}]{Petrosian:1976}
{Petrosian}, V. 1976, \bibinfo{title}{{Surface Brightness and Evolution of Galaxies},} \apjl, 210, L53, \dodoi{10.1086/18230110.1086/182253}

\bibitem[{A. {Pillepich} {et~al.}(2019){Pillepich}, {Nelson}, {Springel}, {Pakmor}, {Torrey}, {Weinberger}, {Vogelsberger}, {Marinacci}, {Genel}, {van der Wel}, \& {Hernquist}}]{Pillepich:2019}
{Pillepich}, A., {Nelson}, D., {Springel}, V., {et~al.} 2019, \bibinfo{title}{{First results from the TNG50 simulation: the evolution of stellar and gaseous discs across cosmic time},} \mnras, 490, 3196, \dodoi{10.1093/mnras/stz2338}

\bibitem[{A. {Pillepich} {et~al.}(2023){Pillepich}, {Sotillo-Ramos}, {Ramesh}, {Nelson}, {Engler}, {Rodriguez-Gomez}, {Fournier}, {Donnari}, {Springel}, \& {Hernquist}}]{Pillepich:2023}
{Pillepich}, A., {Sotillo-Ramos}, D., {Ramesh}, R., {et~al.} 2023, \bibinfo{title}{{Milky Way and Andromeda analogs from the TNG50 simulation},} arXiv e-prints, arXiv:2303.16217, \dodoi{10.48550/arXiv.2303.16217}

\bibitem[{F. {Pinna} {et~al.}(2024){Pinna}, {Grand}, {Martig}, \& {Fragkoudi}}]{Pinna:2024}
{Pinna}, F., {Grand}, R. J.~J., {Martig}, M., \& {Fragkoudi}, F. 2024, \bibinfo{title}{{Recovering chemical bimodalities in observed edge-on stellar disks: Insights from AURIGA simulations},} \aap, 691, A61, \dodoi{10.1051/0004-6361/202450843}

\bibitem[{M. {Portail} {et~al.}(2017){Portail}, {Gerhard}, {Wegg}, \& {Ness}}]{Portail:2017}
{Portail}, M., {Gerhard}, O., {Wegg}, C., \& {Ness}, M. 2017, \bibinfo{title}{{Dynamical modelling of the galactic bulge and bar: the Milky Way's pattern speed, stellar and dark matter mass distribution},} \mnras, 465, 1621, \dodoi{10.1093/mnras/stw2819}

\bibitem[{G. {Rihtar{\v{s}}i{\v{c}}} {et~al.}(2025){Rihtar{\v{s}}i{\v{c}}}, {Brada{\v{c}}}, {Desprez}, {Harshan}, {Noirot}, {Estrada-Carpenter}, {Martis}, {Abraham}, {Asada}, {Brammer}, {Iyer}, {Matharu}, {Mowla}, {Muzzin}, {Sarrouh}, {Sawicki}, {Strait}, {Willott}, {Gledhill}, {Markov}, \& {Tripodi}}]{Rihtarsic:2025}
{Rihtar{\v{s}}i{\v{c}}}, G., {Brada{\v{c}}}, M., {Desprez}, G., {et~al.} 2025, \bibinfo{title}{{CANUCS: Constraining the MACS J0416.1-2403 strong lensing model with JWST NIRISS, NIRSpec, and NIRCam},} \aap, 696, A15, \dodoi{10.1051/0004-6361/202451117}

\bibitem[{A.~S.~G. {Robotham} {et~al.}(2014){Robotham}, {Driver}, {Davies}, {Hopkins}, {Baldry}, {Agius}, {Bauer}, {Bland-Hawthorn}, {Brough}, {Brown}, {Cluver}, {De Propris}, {Drinkwater}, {Holwerda}, {Kelvin}, {Lara-Lopez}, {Liske}, {L{\'o}pez-S{\'a}nchez}, {Loveday}, {Mahajan}, {McNaught-Roberts}, {Moffett}, {Norberg}, {Obreschkow}, {Owers}, {Penny}, {Pimbblet}, {Prescott}, {Taylor}, {van Kampen}, \& {Wilkins}}]{Robotham:2014}
{Robotham}, A.~S.~G., {Driver}, S.~P., {Davies}, L.~J.~M., {et~al.} 2014, \bibinfo{title}{{Galaxy And Mass Assembly (GAMA): galaxy close pairs, mergers and the future fate of stellar mass},} \mnras, 444, 3986, \dodoi{10.1093/mnras/stu1604}

\bibitem[{V. {Rodriguez-Gomez} {et~al.}(2016){Rodriguez-Gomez}, {Pillepich}, {Sales}, {Genel}, {Vogelsberger}, {Zhu}, {Wellons}, {Nelson}, {Torrey}, {Springel}, {Ma}, \& {Hernquist}}]{Rodriguez-Gomez:2016}
{Rodriguez-Gomez}, V., {Pillepich}, A., {Sales}, L.~V., {et~al.} 2016, \bibinfo{title}{{The stellar mass assembly of galaxies in the Illustris simulation: growth by mergers and the spatial distribution of accreted stars},} \mnras, 458, 2371, \dodoi{10.1093/mnras/stw456}

\bibitem[{V. {Rodriguez-Gomez} {et~al.}(2019){Rodriguez-Gomez}, {Snyder}, {Lotz}, {Nelson}, {Pillepich}, {Springel}, {Genel}, {Weinberger}, {Tacchella}, {Pakmor}, {Torrey}, {Marinacci}, {Vogelsberger}, {Hernquist}, \& {Thilker}}]{Rodriguez-Gomez:2019}
{Rodriguez-Gomez}, V., {Snyder}, G.~F., {Lotz}, J.~M., {et~al.} 2019, \bibinfo{title}{{The optical morphologies of galaxies in the IllustrisTNG simulation: a comparison to Pan-STARRS observations},} \mnras, 483, 4140, \dodoi{10.1093/mnras/sty3345}

\bibitem[{E. {Rusta} {et~al.}(2024){Rusta}, {Salvadori}, {Gelli}, {Koutsouridou}, \& {Marconi}}]{Rusta:2024}
{Rusta}, E., {Salvadori}, S., {Gelli}, V., {Koutsouridou}, I., \& {Marconi}, A. 2024, \bibinfo{title}{{Linking high-z and low-z: Are We Observing the Progenitors of the Milky Way with JWST?},} arXiv e-prints, arXiv:2407.06255, \dodoi{10.48550/arXiv.2407.06255}

\bibitem[{S. {Salvadori} {et~al.}(2010){Salvadori}, {Ferrara}, {Schneider}, {Scannapieco}, \& {Kawata}}]{Salvadori:2010}
{Salvadori}, S., {Ferrara}, A., {Schneider}, R., {Scannapieco}, E., \& {Kawata}, D. 2010, \bibinfo{title}{{Mining the Galactic halo for very metal-poor stars},} \mnras, 401, L5, \dodoi{10.1111/j.1745-3933.2009.00772.x}

\bibitem[{P. {Santini} {et~al.}(2017){Santini}, {Fontana}, {Castellano}, {Di Criscienzo}, {Merlin}, {Amorin}, {Cullen}, {Daddi}, {Dickinson}, {Dunlop}, {Grazian}, {Lamastra}, {McLure}, {Micha{\l}owski}, {Pentericci}, \& {Shu}}]{Santini:2017}
{Santini}, P., {Fontana}, A., {Castellano}, M., {et~al.} 2017, \bibinfo{title}{{The Star Formation Main Sequence in the Hubble Space Telescope Frontier Fields},} \apj, 847, 76, \dodoi{10.3847/1538-4357/aa8874}

\bibitem[{G.~T. {Sarrouh} {et~al.}(2024){Sarrouh}, {Muzzin}, {Iyer}, {Mowla}, {Abraham}, {Asada}, {Bradac}, {Brammer}, {Desprez}, {Martis}, {Matharu}, {Noirot}, {Sawicki}, {Strait}, {Willott}, \& {Zabl}}]{Sarrouh:2024}
{Sarrouh}, G.~T., {Muzzin}, A., {Iyer}, K.~G., {et~al.} 2024, \bibinfo{title}{{Exposing Line Emission: A First Look At The Systematic Differences of Measuring Stellar Masses With JWST NIRCam Medium Versus Wide Band Photometry},} arXiv e-prints, arXiv:2401.08781, \dodoi{10.48550/arXiv.2401.08781}

\bibitem[{G.~T.~E. {Sarrouh} {et~al.}(2025){Sarrouh}, {Asada}, {Martis}, {Willott}, {Iyer}, {Noirot}, {Muzzin}, {Sawicki}, {Brammer}, {Desprez}, {Rihtar{\v{s}}i{\v{c}}}, {Zabl}, {Abraham}, {Brada{\v{c}}}, {Doyon}, {Antwi-Danso}, {Berek}, {Brown}, {Estrada-Carpenter}, {Favaro}, {Felicioni}, {Forrest}, {Gaspar}, {Gould}, {Gledhill}, {Harshan}, {Jahan}, {Jagga}, {Jude{\v{z}}}, {Marchesini}, {Markov}, {Matharu}, {MacFarland}, {Merchant}, {M{\'e}rida}, {Mowla}, {Myers}, {Omori}, {Pacifici}, {Ravindranath}, {Robbins}, {Strait}, {Sok}, {Tan}, {Tripodi}, {Wilson}, \& {Withers}}]{Sarrouh:2025}
{Sarrouh}, G. T.~E., {Asada}, Y., {Martis}, N.~S., {et~al.} 2025, \bibinfo{title}{{CANUCS/Technicolor Data Release 1: Imaging, Photometry, Slit Spectroscopy, and Stellar Population Parameters},} arXiv e-prints, arXiv:2506.21685, \dodoi{10.48550/arXiv.2506.21685}

\bibitem[{E. {Sazonova} {et~al.}(2024){Sazonova}, {Morgan}, {Balogh}, {Alatalo}, {Benavides}, {Bluck}, {Brough}, {Busa}, {Demarco}, {Donevski}, {Figueira}, {Martin}, {Rodriguez-Gomez}, {Rom{\'a}n}, \& {Rowlands}}]{Sazonova:2024}
{Sazonova}, E., {Morgan}, C., {Balogh}, M., {et~al.} 2024, \bibinfo{title}{{RMS asymmetry: a robust metric of galaxy shapes in images with varied depth and resolution},} arXiv e-prints, arXiv:2404.05792, \dodoi{10.48550/arXiv.2404.05792}

\bibitem[{J. {Schaye} {et~al.}(2015){Schaye}, {Crain}, {Bower}, {Furlong}, {Schaller}, {Theuns}, {Dalla Vecchia}, {Frenk}, {McCarthy}, {Helly}, {Jenkins}, {Rosas-Guevara}, {White}, {Baes}, {Booth}, {Camps}, {Navarro}, {Qu}, {Rahmati}, {Sawala}, {Thomas}, \& {Trayford}}]{Schaye:2015}
{Schaye}, J., {Crain}, R.~A., {Bower}, R.~G., {et~al.} 2015, \bibinfo{title}{{The EAGLE project: simulating the evolution and assembly of galaxies and their environments},} \mnras, 446, 521, \dodoi{10.1093/mnras/stu2058}

\bibitem[{C. {Schreiber} {et~al.}(2015){Schreiber}, {Pannella}, {Elbaz}, {B{\'e}thermin}, {Inami}, {Dickinson}, {Magnelli}, {Wang}, {Aussel}, {Daddi}, {Juneau}, {Shu}, {Sargent}, {Buat}, {Faber}, {Ferguson}, {Giavalisco}, {Koekemoer}, {Magdis}, {Morrison}, {Papovich}, {Santini}, \& {Scott}}]{Schreiber:2015}
{Schreiber}, C., {Pannella}, M., {Elbaz}, D., {et~al.} 2015, \bibinfo{title}{{The Herschel view of the dominant mode of galaxy growth from z = 4 to the present day},} \aap, 575, A74, \dodoi{10.1051/0004-6361/201425017}

\bibitem[{V.~A. {Semenov} {et~al.}(2024){Semenov}, {Conroy}, {Smith}, {Puchwein}, \& {Hernquist}}]{Semenov:2024}
{Semenov}, V.~A., {Conroy}, C., {Smith}, A., {Puchwein}, E., \& {Hernquist}, L. 2024, \bibinfo{title}{{How Early Could the Milky Way's Disk Form?},} arXiv e-prints, arXiv:2409.18173, \dodoi{10.48550/arXiv.2409.18173}

\bibitem[{J.~L. {S{\'e}rsic}(1963){S{\'e}rsic}}]{Sersic:1963}
{S{\'e}rsic}, J.~L. 1963, \bibinfo{title}{{Influence of the atmospheric and instrumental dispersion on the brightness distribution in a galaxy},} Boletin de la Asociacion Argentina de Astronomia La Plata Argentina, 6, 41

\bibitem[{J. {Shen} {et~al.}(2010){Shen}, {Rich}, {Kormendy}, {Howard}, {De Propris}, \& {Kunder}}]{Shen:2010}
{Shen}, J., {Rich}, R.~M., {Kormendy}, J., {et~al.} 2010, \bibinfo{title}{{Our Milky Way as a Pure-disk Galaxy{\textemdash}A Challenge for Galaxy Formation},} \apjl, 720, L72, \dodoi{10.1088/2041-8205/720/1/L72}

\bibitem[{G.~F. {Snyder} {et~al.}(2015{\natexlab{a}}){Snyder}, {Lotz}, {Moody}, {Peth}, {Freeman}, {Ceverino}, {Primack}, \& {Dekel}}]{Snyder:2015a}
{Snyder}, G.~F., {Lotz}, J., {Moody}, C., {et~al.} 2015{\natexlab{a}}, \bibinfo{title}{{Diverse structural evolution at z > 1 in cosmologically simulated galaxies},} \mnras, 451, 4290, \dodoi{10.1093/mnras/stv1231}

\bibitem[{G.~F. {Snyder} {et~al.}(2017){Snyder}, {Lotz}, {Rodriguez-Gomez}, {Guimar{\~a}es}, {Torrey}, \& {Hernquist}}]{Snyder:2017}
{Snyder}, G.~F., {Lotz}, J.~M., {Rodriguez-Gomez}, V., {et~al.} 2017, \bibinfo{title}{{Massive close pairs measure rapid galaxy assembly in mergers at high redshift},} \mnras, 468, 207, \dodoi{10.1093/mnras/stx487}

\bibitem[{G.~F. {Snyder} {et~al.}(2015{\natexlab{b}}){Snyder}, {Torrey}, {Lotz}, {Genel}, {McBride}, {Vogelsberger}, {Pillepich}, {Nelson}, {Sales}, {Sijacki}, {Hernquist}, \& {Springel}}]{Snyder:2015b}
{Snyder}, G.~F., {Torrey}, P., {Lotz}, J.~M., {et~al.} 2015{\natexlab{b}}, \bibinfo{title}{{Galaxy morphology and star formation in the Illustris Simulation at z = 0},} \mnras, 454, 1886, \dodoi{10.1093/mnras/stv2078}

\bibitem[{V. {Sok} {et~al.}(2022){Sok}, {Muzzin}, {Jablonka}, {Marsan}, {Tan}, {Alcorn}, {Marchesini}, \& {Stefanon}}]{Sok:2022}
{Sok}, V., {Muzzin}, A., {Jablonka}, P., {et~al.} 2022, \bibinfo{title}{{Finite-resolution Deconvolution of Multiwavelength Imaging of 20,000 Galaxies in the COSMOS Field: The Evolution of Clumpy Galaxies over Cosmic Time},} \apj, 924, 7, \dodoi{10.3847/1538-4357/ac2f40}

\bibitem[{V. {Sok} {et~al.}(2025){Sok}, {Muzzin}, {Jablonka}, {Tan}, {Marsan}, {Marchesini}, {Wilson}, \& {Alcorn}}]{Sok:2025a}
{Sok}, V., {Muzzin}, A., {Jablonka}, P., {et~al.} 2025, \bibinfo{title}{{An Indication of Gas Inflow in Clumpy Star-forming Galaxies near z {\ensuremath{\sim}} 1: Lower Gas-phase Metallicities in Clumpy Galaxies Compared to Nonclumpy Galaxies},} \apj, 979, 14, \dodoi{10.3847/1538-4357/ad96bb}

\bibitem[{D. {Sotillo-Ramos} {et~al.}(2022){Sotillo-Ramos}, {Pillepich}, {Donnari}, {Nelson}, {Eisert}, {Rodriguez-Gomez}, {Joshi}, {Vogelsberger}, \& {Hernquist}}]{Sotillo-Ramos:2022}
{Sotillo-Ramos}, D., {Pillepich}, A., {Donnari}, M., {et~al.} 2022, \bibinfo{title}{{The merger and assembly histories of Milky Way- and M31-like galaxies with TNG50: disc survival through mergers},} \mnras, 516, 5404, \dodoi{10.1093/mnras/stac2586}

\bibitem[{J.~S. {Speagle} {et~al.}(2014){Speagle}, {Steinhardt}, {Capak}, \& {Silverman}}]{Speagle:2014}
{Speagle}, J.~S., {Steinhardt}, C.~L., {Capak}, P.~L., \& {Silverman}, J.~D. 2014, \bibinfo{title}{{A Highly Consistent Framework for the Evolution of the Star-Forming ``Main Sequence'' from z \raisebox{-0.5ex}\textasciitilde 0-6},} \apjs, 214, 15, \dodoi{10.1088/0067-0049/214/2/15}

\bibitem[{C. {Studholme} {et~al.}(1999){Studholme}, {Hill}, \& {Hawkes}}]{Studholme:1999}
{Studholme}, C., {Hill}, D., \& {Hawkes}, D. 1999, \bibinfo{title}{An overlap invariant entropy measure of 3D medical image alignment,} Pattern Recognition, 32, 71, \dodoi{https://doi.org/10.1016/S0031-3203(98)00091-0}

\bibitem[{V.~Y.~Y. {Tan} {et~al.}(2024){Tan}, {Muzzin}, {Marchesini}, {Sok}, {Sarrouh}, \& {Marsan}}]{Tan:2024}
{Tan}, V. Y.~Y., {Muzzin}, A., {Marchesini}, D., {et~al.} 2024, \bibinfo{title}{{A Measurement of the Assembly of Milky Way Analogues at Redshifts $0.5 < z < 2$ with Resolved Stellar Mass and Star-Formation Rate Profiles},} arXiv e-prints, arXiv:2402.12433, \dodoi{10.48550/arXiv.2402.12433}

\bibitem[{V.~Y.~Y. {Tan} {et~al.}(2022){Tan}, {Muzzin}, {Marsan}, {Sok}, {Alcorn}, {Matharu}, {Shipley}, {Marchesini}, {Nedkova}, {Martis}, {van der Wel}, \& {Whitaker}}]{Tan:2022}
{Tan}, V. Y.~Y., {Muzzin}, A., {Marsan}, Z.~C., {et~al.} 2022, \bibinfo{title}{{Resolved Stellar Mass Maps of Galaxies in the Hubble Frontier Fields: Evidence for Mass Dependency in Environmental Quenching},} \apj, 933, 30, \dodoi{10.3847/1538-4357/ac7051}

\bibitem[{R.~B. {Tully} \& J.~R. {Fisher}(1977){Tully} \& {Fisher}}]{Tully:1977}
{Tully}, R.~B., \& {Fisher}, J.~R. 1977, \bibinfo{title}{{A new method of determining distances to galaxies.},} \aap, 54, 661

\bibitem[{J. {Tumlinson}(2010){Tumlinson}}]{Tumlinson:2010}
{Tumlinson}, J. 2010, \bibinfo{title}{{Chemical Evolution in Hierarchical Models of Cosmic Structure. II. The Formation of the Milky Way Stellar Halo and the Distribution of the Oldest Stars},} \apj, 708, 1398, \dodoi{10.1088/0004-637X/708/2/1398}

\bibitem[{F.~C. {van den Bosch}(1998){van den Bosch}}]{vandenBosch:1998}
{van den Bosch}, F.~C. 1998, \bibinfo{title}{{The Formation of Disk-Bulge-Halo Systems and the Origin of the Hubble Sequence},} \apj, 507, 601, \dodoi{10.1086/306354}

\bibitem[{P.~G. {van Dokkum} {et~al.}(2013){van Dokkum}, {Leja}, {Nelson}, {Patel}, {Skelton}, {Momcheva}, {Brammer}, {Whitaker}, {Lundgren}, {Fumagalli}, {Conroy}, {F{\"o}rster Schreiber}, {Franx}, {Kriek}, {Labb{\'e}}, {Marchesini}, {Rix}, {van der Wel}, \& {Wuyts}}]{vanDokkum:2013}
{van Dokkum}, P.~G., {Leja}, J., {Nelson}, E.~J., {et~al.} 2013, \bibinfo{title}{{The Assembly of Milky-Way-like Galaxies Since z \raisebox{-0.5ex}\textasciitilde 2.5},} \apjl, 771, L35, \dodoi{10.1088/2041-8205/771/2/L35}

\bibitem[{E. {Vanzella} {et~al.}(2023){Vanzella}, {Claeyssens}, {Welch}, {Adamo}, {Coe}, {Diego}, {Mahler}, {Khullar}, {Kokorev}, {Oguri}, {Ravindranath}, {Furtak}, {Hsiao}, {Abdurro'uf}, {Mandelker}, {Brammer}, {Bradley}, {Brada{\v{c}}}, {Conselice}, {Dayal}, {Nonino}, {Andrade-Santos}, {Windhorst}, {Pirzkal}, {Sharon}, {de Mink}, {Fujimoto}, {Zitrin}, {Eldridge}, \& {Norman}}]{Vanzella:2023}
{Vanzella}, E., {Claeyssens}, A., {Welch}, B., {et~al.} 2023, \bibinfo{title}{{JWST/NIRCam Probes Young Star Clusters in the Reionization Era Sunrise Arc},} \apj, 945, 53, \dodoi{10.3847/1538-4357/acb59a}

\bibitem[{J. {Wang} {et~al.}(2011){Wang}, {Kauffmann}, {Overzier}, {Catinella}, {Schiminovich}, {Heckman}, {Moran}, {Haynes}, {Giovanelli}, \& {Kong}}]{Wang:2011}
{Wang}, J., {Kauffmann}, G., {Overzier}, R., {et~al.} 2011, \bibinfo{title}{{The GALEX Arecibo SDSS survey - III. Evidence for the inside-out formation of Galactic discs},} \mnras, 412, 1081, \dodoi{10.1111/j.1365-2966.2010.17962.x}

\bibitem[{C. {Wegg} \& O. {Gerhard}(2013){Wegg} \& {Gerhard}}]{Wegg:2013}
{Wegg}, C., \& {Gerhard}, O. 2013, \bibinfo{title}{{Mapping the three-dimensional density of the Galactic bulge with VVV red clump stars},} \mnras, 435, 1874, \dodoi{10.1093/mnras/stt1376}

\bibitem[{C. {Wegg} {et~al.}(2015){Wegg}, {Gerhard}, \& {Portail}}]{Wegg:2015}
{Wegg}, C., {Gerhard}, O., \& {Portail}, M. 2015, \bibinfo{title}{{The structure of the Milky Way's bar outside the bulge},} \mnras, 450, 4050, \dodoi{10.1093/mnras/stv745}

\bibitem[{A.~R. {Wetzel} {et~al.}(2016){Wetzel}, {Hopkins}, {Kim}, {Faucher-Gigu{\`e}re}, {Kere{\v{s}}}, \& {Quataert}}]{Wetzel:2016}
{Wetzel}, A.~R., {Hopkins}, P.~F., {Kim}, J.-h., {et~al.} 2016, \bibinfo{title}{{Reconciling Dwarf Galaxies with {\ensuremath{\Lambda}}CDM Cosmology: Simulating a Realistic Population of Satellites around a Milky Way-mass Galaxy},} \apjl, 827, L23, \dodoi{10.3847/2041-8205/827/2/L23}

\bibitem[{S.~D.~M. {White} \& M.~J. {Rees}(1978){White} \& {Rees}}]{White:1978}
{White}, S.~D.~M., \& {Rees}, M.~J. 1978, \bibinfo{title}{{Core condensation in heavy halos: a two-stage theory for galaxy formation and clustering.},} \mnras, 183, 341, \dodoi{10.1093/mnras/183.3.341}

\bibitem[{V. {Wild} {et~al.}(2020){Wild}, {Taj Aldeen}, {Carnall}, {Maltby}, {Almaini}, {Werle}, {Wilkinson}, {Rowlands}, {Bolzonella}, {Castellano}, {Gargiulo}, {McLure}, {Pentericci}, \& {Pozzetti}}]{Wild:2020}
{Wild}, V., {Taj Aldeen}, L., {Carnall}, A., {et~al.} 2020, \bibinfo{title}{{The star formation histories of z {\ensuremath{\sim}} 1 post-starburst galaxies},} \mnras, 494, 529, \dodoi{10.1093/mnras/staa674}

\bibitem[{R.~J. {Williams} {et~al.}(2009){Williams}, {Quadri}, {Franx}, {van Dokkum}, \& {Labb{\'e}}}]{Williams:2009}
{Williams}, R.~J., {Quadri}, R.~F., {Franx}, M., {van Dokkum}, P., \& {Labb{\'e}}, I. 2009, \bibinfo{title}{{Detection of Quiescent Galaxies in a Bicolor Sequence from Z = 0-2},} \apj, 691, 1879, \dodoi{10.1088/0004-637X/691/2/1879}

\bibitem[{C.~J. {Willott} {et~al.}(2022){Willott}, {Doyon}, {Albert}, {Brammer}, {Dixon}, {Muzic}, {Ravindranath}, {Scholz}, {Abraham}, {Artigau}, {Brada{\v{c}}}, {Goudfrooij}, {Hutchings}, {Iyer}, {Jayawardhana}, {LaMassa}, {Martis}, {Meyer}, {Morishita}, {Mowla}, {Muzzin}, {Noirot}, {Pacifici}, {Rowlands}, {Sarrouh}, {Sawicki}, {Taylor}, {Volk}, \& {Zabl}}]{Willott:2022}
{Willott}, C.~J., {Doyon}, R., {Albert}, L., {et~al.} 2022, \bibinfo{title}{{The Near-infrared Imager and Slitless Spectrograph for the James Webb Space Telescope. II. Wide Field Slitless Spectroscopy},} \pasp, 134, 025002, \dodoi{10.1088/1538-3873/ac5158}

\bibitem[{C.~J. {Willott} {et~al.}(2024){Willott}, {Desprez}, {Asada}, {Sarrouh}, {Abraham}, {Brada{\v{c}}}, {Brammer}, {Estrada-Carpenter}, {Iyer}, {Martis}, {Matharu}, {Mowla}, {Muzzin}, {Noirot}, {Sawicki}, {Strait}, {Rihtar{\v{s}}i{\v{c}}}, \& {Withers}}]{Willott:2024}
{Willott}, C.~J., {Desprez}, G., {Asada}, Y., {et~al.} 2024, \bibinfo{title}{{A Steep Decline in the Galaxy Space Density beyond Redshift 9 in the CANUCS UV Luminosity Function},} \apj, 966, 74, \dodoi{10.3847/1538-4357/ad35bc}

\bibitem[{S. {Wuyts} {et~al.}(2012){Wuyts}, {F{\"o}rster Schreiber}, {Genzel}, {Guo}, {Barro}, {Bell}, {Dekel}, {Faber}, {Ferguson}, {Giavalisco}, {Grogin}, {Hathi}, {Huang}, {Kocevski}, {Koekemoer}, {Koo}, {Lotz}, {Lutz}, {McGrath}, {Newman}, {Rosario}, {Saintonge}, {Tacconi}, {Weiner}, \& {van der Wel}}]{Wuyts:2012}
{Wuyts}, S., {F{\"o}rster Schreiber}, N.~M., {Genzel}, R., {et~al.} 2012, \bibinfo{title}{{Smooth(er) Stellar Mass Maps in CANDELS: Constraints on the Longevity of Clumps in High-redshift Star-forming Galaxies},} \apj, 753, 114, \dodoi{10.1088/0004-637X/753/2/114}

\bibitem[{M. {Xiang} \& H.-W. {Rix}(2022){Xiang} \& {Rix}}]{Xiang:2022}
{Xiang}, M., \& {Rix}, H.-W. 2022, \bibinfo{title}{{A time-resolved picture of our Milky Way's early formation history},} \nat, 603, 599, \dodoi{10.1038/s41586-022-04496-5}

\bibitem[{S. {Yu} {et~al.}(2023){Yu}, {Bullock}, {Gurvich}, {Hafen}, {Stern}, {Boylan-Kolchin}, {Faucher-Gigu{\`e}re}, {Wetzel}, {Hopkins}, \& {Moreno}}]{Yu:2023}
{Yu}, S., {Bullock}, J.~S., {Gurvich}, A.~B., {et~al.} 2023, \bibinfo{title}{{Born this way: thin disc, thick disc, and isotropic spheroid formation in FIRE-2 Milky Way-mass galaxy simulations},} \mnras, 523, 6220, \dodoi{10.1093/mnras/stad1806}

\bibitem[{S. {Zhou} {et~al.}(2023){Zhou}, {Arag{\'o}n-Salamanca}, {Merrifield}, {Andrews}, {Drory}, \& {Lane}}]{Zhou:2023}
{Zhou}, S., {Arag{\'o}n-Salamanca}, A., {Merrifield}, M., {et~al.} 2023, \bibinfo{title}{{Are Milky-Way-like galaxies like the Milky Way? A view from SDSS-IV/MaNGA},} \mnras, 521, 5810, \dodoi{10.1093/mnras/stad853}

\end{thebibliography}
\bibliographystyle{aasjournalv7}

\appendix
\section{UVJ color calibration for this dataset}\label{app:uvj}
In \S\ref{sec:q-frac-uvj}, the UVJ diagram is used to determine the quiescent fraction of potential MWA progenitors and remove them from the analysis. In order to have unbiased UVJ boundaries, we cannot use the MWA progenitor sample itself to determine the boundaries due to the low quiescent fraction and lack of bimodality. Instead, a sample of 9516 galaxies from the cluster fields is used to fit the UVJ boundaries. The slope of the red sequence determines the slope of the calibrated line, and the local minimum of the distribution determines the $y$-intercept. 

\begin{figure*}
   \centering
   \includegraphics[width=\textwidth]{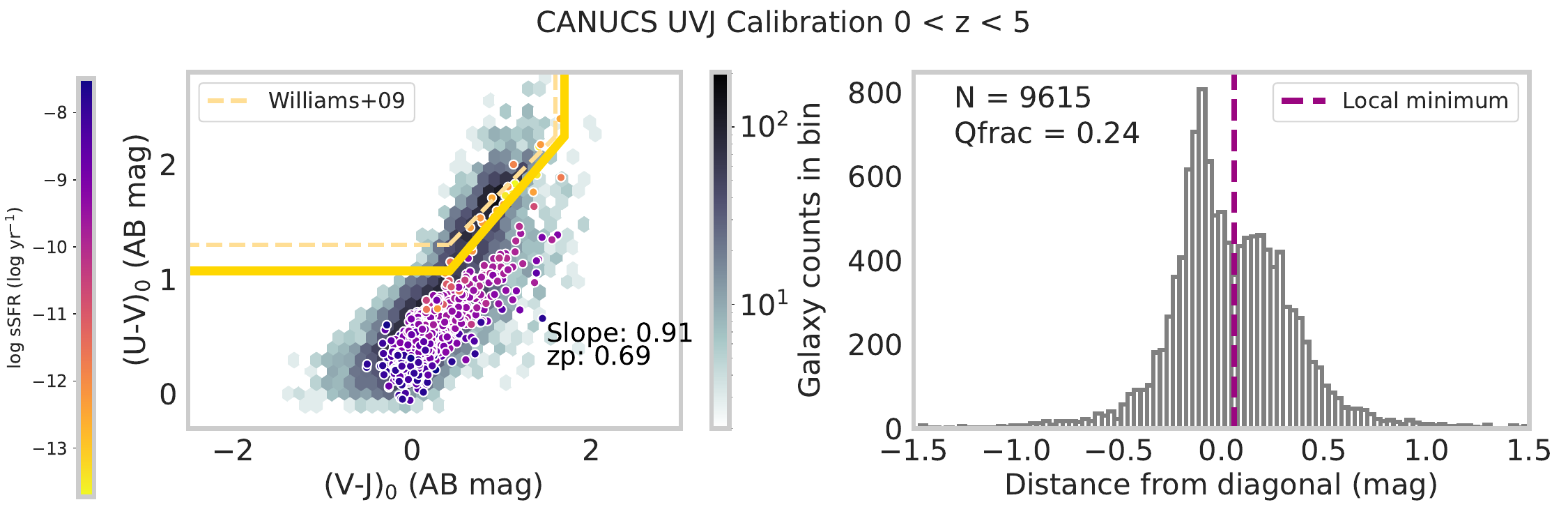}
\caption{\emph{Left Panel:} Rest-frame UVJ diagram of 9516 galaxies from the CANUCS CLU fields with redshifts $0 < z < 5$ shown as the grey bins. Darker bins indicate higher number count of galaxies. The MWA progenitors are plotted on top, with each circle representing one galaxy, and the color indicating the integrated sSFR. The UVJ boundaries used in this work are shown in yellow with dashed yellow line indicating the UVJ boundaries from \cite{Williams:2009}.
\emph{Right Panel:} Histogram distribution of the same 9516 galaxies and their distance from the diagonal UVJ boundary line. Purple dashed line indicates local minimum, and is analogous to the UVJ diagonal line in the left panel.}\label{fig:color-calib}
\end{figure*}

In Figure \ref{fig:color-calib}, we plot the full sample of galaxies that are used to calibrate the UVJ lines. The original color cuts were generated for every galaxy present in the CLU fields from $0.5 < z <6$, which we have adjusted to $0<z<5$ to match the redshift range of the MWA progenitor sample in this work. The reason the cluster fields are used but not the NCF fields is because there are not enough quiescent galaxies in the NCF fields for a clear bimodal distribution in UVJ color space. The quiescent fraction of the CANUCS sample at $0<z<5$ in the NCF fields is only $0.07\%$, while in the CLU fields it is $24\%$ in the same redshift range due to the presence of the clusters.

\section{Creating 1-D profiles}\label{app:profiles-extra}
\begin{figure*}
\centering
    \begin{subfigure}
    \centering
        \includegraphics[width=\textwidth]{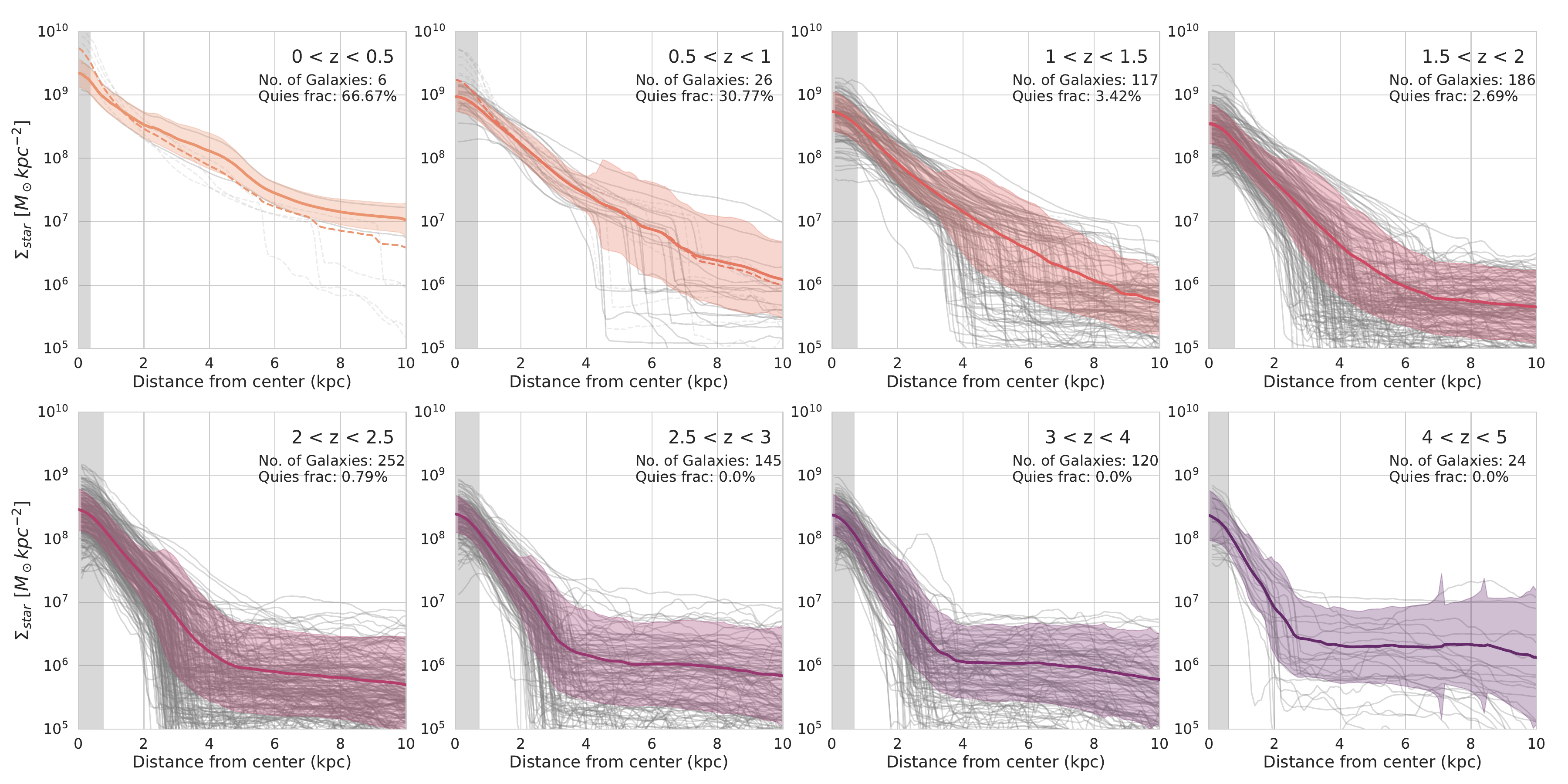}
    \end{subfigure}
    \begin{subfigure}
    \centering 
        \includegraphics[width=\textwidth]{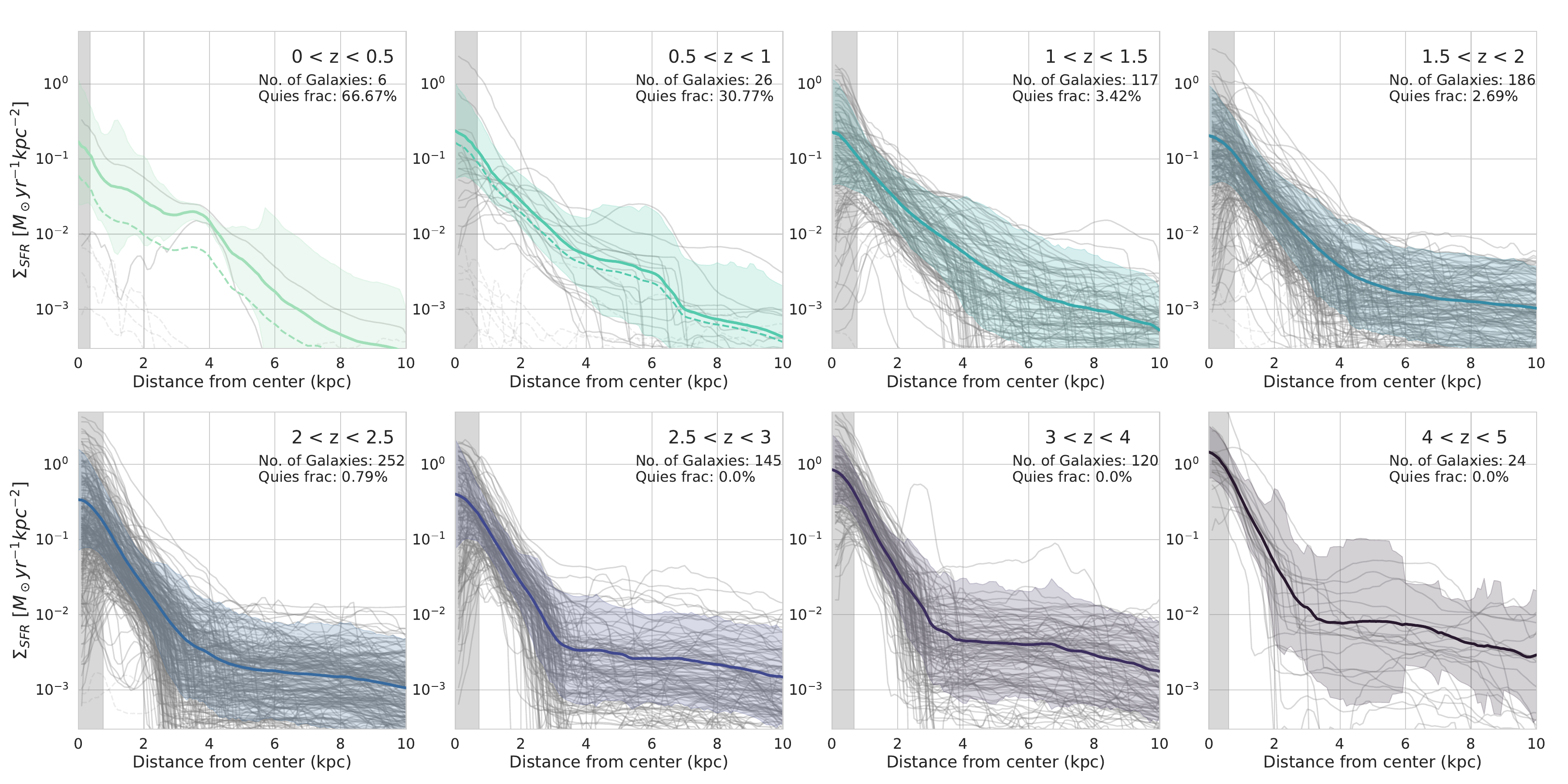}
    \end{subfigure}
    \caption{\emph{Top Panel:} 1-D stellar mass density profiles for each MWA, plotted in grey. Colored lines are the stacked and normalized profile of all the galaxies in each redshift bin. Shaded regions represent the 1-sigma (68th percentile) region. The vertical grey shaded region near $R = 0$kpc represents the average physical size of the PSF FHWM at that redshift bin.
    \emph{Bottom Panel:} Similar to the plot above but for SFR density profiles. The total number of galaxies per redshift bin, and the quiescent fraction is also displayed in the top-right corner.} 
    \label{fig:all_profiles2}
\end{figure*}

As summarized in \S \ref{sec:density-profiles}, we created 1-D stellar mass and SFR density profiles by placing down elliptical annuli with width 0.1kpc centered on the stellar mass maps and SFR maps of each respective galaxy. The stacking is performed along the semi-major axis of each galaxy for consistency. Figure \ref{fig:all_profiles2} is an expanded version of Figure \ref{fig:all_profiles} from \S \ref{sec:density-profiles}. Displayed are each individual galaxy's density profile divided into eight subplots by redshift bin. The stacked and normalized profile is shown in the same color as in Figure \ref{fig:all_profiles}. The dashed grey lines represent profiles of quiescent galaxies while solid lines are the star-forming galaxies. The dashed colored lines represent the stacked and normalized profile for each redshift bin including both star-forming and quiescents, while the solid colored lines only include star-forming galaxies. The profiles are also not truncated by the noise floor. One difference between Figure \ref{fig:all_profiles} and  Figure \ref{fig:all_profiles2} is that the bottom panel of Figure \ref{fig:all_profiles2} shows the SFR density instead of the sSFR density. Indeed the sSFR density profiles were created via dividing each stacked and normalized SFR density profile at each redshift by its corresponding mass density profile. 

\end{document}